\documentclass[aps,prd,onecolumn,superscriptaddress,nofootinbib]{revtex4-1}
\usepackage[margin=1in]{geometry}
\usepackage{graphicx}
\usepackage{subcaption}
\usepackage{float}
\usepackage{amsmath}
\usepackage{booktabs}
\usepackage{multirow}
\usepackage{amsmath,amssymb}
\usepackage{graphicx}
\usepackage[colorlinks=true, linkcolor=blue, citecolor=blue, urlcolor=blue]{hyperref}
\usepackage{xcolor}
\usepackage{bm}
\usepackage{orcidlink}
\usepackage{booktabs}
\usepackage{siunitx}

\newcommand{\Mc}{\mathcal{M}}
\newcommand{\be}{\begin{equation}}
\newcommand{\ee}{\end{equation}}

\begin{document}

\title{Inspiralling Binary merger parameter reconstruction with lunar and satellite Laser Ranging}

% \author{G.~Franciolini\orcidlink{0000-0002-6892-9145}}
% \affiliation{Dipartimento di Fisica e Astronomia “G. Galilei” and INFN, Sezione di Padova, 
% via Marzolo 8, I-35131 Padova, Italy}

% \author{Y. Gouttenoire\orcidlink{0000-0003-2225-6704}}
% \affiliation{Institut d’Astrophysique de Paris (IAP), CNRS, Sorbonne Université, FR-75014, France}

% \author{A.~J.~Iovino\orcidlink{0000-0002-8531-5962}}
% \affiliation{Center for Astrophysics and Space Science (CASS), New York University Abu Dhabi, PO Box 129188, Abu Dhabi, UAE}

% \author{S.~Trifinopoulos\orcidlink{0000-0002-0492-1144}}
% \affiliation{Physik-Institut, Universit\"at Z\"urich, Z\"urich, Switzerland}

\author{Miguel~Vanvlasselaer\orcidlink{0000-0002-8527-7011}}
\affiliation{Departament de F\'{\i}sica Qu\`{a}ntica i Astrof\'{\i}sica and Institut de Ci\`{e}ncies del Cosmos (ICC), Universitat de Barcelona, Mart\'{\i} i Franqu\`{e}s 1, Barcelona, Spain}

\date{\today}

\begin{abstract}
Precision tracking of Earth--Moon and Earth--satellite ranges offers a novel, low-frequency avenue for detecting gravitational waves from inspiralling massive black hole binaries, complementary to space-based interferometers. In this paper, we propose a first study of the reconstruction of the binary parameters from a synthetic set of range observations. \textit{Set-up}: Assuming an ideal unperturbed Keplerian motion for the earth-satellite binary, we develop a framework for reconstructing the chirp mass $\mathcal M_c$ and luminosity distance $D_L$ of chirping sources from the perturbations they induce on ranging observables, using both Lunar Laser Ranging (LLR) and a two-satellite ranging configuration (GUEST) as concrete case studies.
  \textit{Result}: We find that chirp-mass recovery tightly converges as a power law in SNR ($\sigma_{M_c}/M_c \propto \mathrm{SNR}^{\alpha}$, $\alpha \simeq -1$ to $-1.8$) for both LLR and GUEST. In contrast, the luminosity distance suffers from a severe, SNR-independent degeneracy with the sky-orientation angles in the LLR configuration, yielding fractional biases and uncertainties of tens to over a hundred percent regardless of signal strength. The two-satellite GUEST configuration breaks this degeneracy and restores a convergent, SNR-dependent recovery of $D_L$ comparable in scaling to that of the chirp mass. These results demonstrate that multi-baseline ranging architectures are essential for extracting luminosity-distance information from ranging-based gravitational wave searches, while chirp-mass measurements remain reliably accessible even with a single ranging baseline such as the Earth--Moon system.
\end{abstract}

\maketitle

\section{Introduction and motivation}
Probing and constraining the population of astrophysical black holes is one of the principal scientific drivers of current and future gravitational-wave observatories. Population-level inferences --- on the black-hole mass spectrum, the spin distribution, the merger-rate evolution with redshift, and the relative contribution of distinct formation channels --- are ultimately built upon the parameter estimates of individual events. The fidelity of any astrophysical conclusion therefore rests on the accuracy and robustness of single-event inference. 

In this goal of constraining the black hole population and binary merger rate, Gravitational Waves (GW) observations recently rose as a prominent tool, namely with LVK\cite{LIGOScientific:2018jsj, LIGOScientific:2020kqk, KAGRA:2021duu, Gair:2022fsj}, and soon LISA\cite{Gair:2010bx, Toubiana:2026yml}, Cosmic Explorer\cite{Reitze:2019iox}, TianQin\cite{Fan:2024nnp} and Einstein telescope\cite{ET:2025xjr, DeRenzis:2024dvx}. The $\mu$Hz frequency band, between the PTA and LISA~\cite{LISA:2017pwj,LISACosmologyWorkingGroup:2022jok} windows, is particularly relevant in this context. Indeed, the ISCO (Innermost Stable Circular Orbit) typical frequency of the GW emitted by a merging event is related to the total mass $M$ of the binary,
$
f_{\mathrm{ISCO}} \approx \frac{4400 \, \mathrm{Hz}}{M / M_{\odot}}
$. This places binaries with total mass $M \in [10^7, 10^9] M_{\odot}$ in the mHz$-\mu$Hz range. 

Several concepts targeting this band have been proposed, including space-based interferometers~\cite{Sesana:2019vho}, asteroid ranging~\cite{Fedderke:2021kuy}, binary pulsar timing~\cite{Blas:2021mqw}, survey concepts with Gaia and Roman~\cite{Moore:2017ity,Wang:2022sxn}, and spacecraft Doppler tracking~\cite{Armstrong:2006}.  In this respect, two detector concepts have been revived recently: Lunar ranging\cite{Benderarticle,Dickey:1994zz, Murphy:2013qya} and satellite ranging\cite{Blas:2026xol, Blas:2021mqw}. In particular,  Lunar Laser Ranging (LLR)~\cite{Foster:2025csl, Foster:2025nzf, Blas:2021mqw, Turyshev:2026qus} already provides an operational probe of GWs in this range, and allows to constraint. LLR measures the Earth--Moon distance with (sub-)centimeter precision by timing laser pulses reflected off retroreflectors on the lunar surface. A passing GW tidally perturbs the lunar orbit; when the GW frequency matches a harmonic of the lunar orbital frequency ($f_{\rm GW}= n/P_{\rm lunar}$, $n\in\mathbb{N}$), the ranging displacement grows resonantly with observation time~\cite{Hui:2012yp,Blas:2021mqw,Foster:2025csl}. With more than 60 years of accumulated tracking data~\cite{Murphy:2013qya}, LLR provides a uniquely sensitive and long-baseline probe of the $\mu$Hz GW spectrum, enabling constraints on the population of massive black-hole binaries across cosmic time.

Despite its rather long history\cite{1981ApJ...246..569M, 1979ApJ...233..685T, Bertotti:1980pg}, binary resonance searches were introduced in their modern form by
Refs.~\cite{Blas:2021mqw,Blas:2021mpc}, where the secular, orbit-averaged
evolution of the six orbital elements under a stochastic gravitational-wave
background was derived within a Kramers--Moyal/Fokker--Planck formalism, and
first forecasts for lunar laser ranging, satellite laser ranging and binary
pulsar timing were obtained. Ref.~\cite{Foster:2025csl, Foster:2025nzf} subsequently relaxed the
secular approximation and worked directly with the time-resolved ranging and
timing residuals, showing that the resonant response is coherent rather than
diffusive.

On the other hand, the GUEST project\cite{Blas:2026xol} (Gravitational-wave Universe Exploration via Satellite Tracking) is a
proposed ESA mission concept. The
central idea is to monitor the orbits of two satellites. The
centre-of-mass positions of the spacecraft are reconstructed to sub-centimeter
precision from the ranging data, which are averaged into \emph{normal points}
(NPs). This SLR technique has been continuously
developed since the 1970s, with the LAGEOS and LARES missions being leading
examples.

One of the main target of research of the LLR/SLR experiments is the GW produced by a BH binary merger. Interestingly, due to their deterministic oscillatory nature, these events leave a recognizable imprint in the moon/satellite orbit which is in principle free of degeneracies, as opposed to, for example, the coherent drift induced by a stochastic GW signal. Constraining the BH population would then imply to be able to reconstruct the distance and the chirp mass of a noisy event. The noise will be a complicated superposition of  stochastic and deterministic noise, whose study are still in their early phase. The quality of the recovery of the parameters $(\mathcal{M}_c, D_L)$ directly depends on the loudness of the signal (the signal to noise ratio). Similar analysis have been performed for LVK detectors \cite{Christensen:2022bxb} and are foreseen for LISA \cite{Deng:2025qhx}. On the top of this, beside BH binaries, LLR and GUEST concepts have be shown to be able to constraint planetary mass PBH\cite{Blas:2026dsm}, scenario of ultra-light dark matter\cite{Foster:2025nzf, Foster:2025csl}, first order phase transitions in the early universe and other topological defect\cite{Blas:2025lzc}.

In short, the question that we plan to address in this paper is the following: \textit{assuming that we are able detect a binary merger event in SLR or LLR, and considering a gaussian white noise on the top of the deterministic GW signal, how well can we hope to reconstruct the luminosity distance and the chirp mass ?} As a first study in a nascent subfield, our study will propose a first step in the elucidation of the complicated problem, as  the references \cite{Rover:2006bb, Rover:2006ni} were for the LVK observatories. 

\paragraph{A glimpse at the results and assumptions}

In this study, we assume a Keplerian ideal motion for the earth-satellite system and generate a synthetic response to GW perturbations of an inspiralling compact BH binary using the leading order 0PN GW signal. To model the uncertain noise in the LLR and SLR probes, we add white gaussian noise on the synthetic data and then run a bayesian analysis over this noisy data. 
Let us already reveal our main results: for the LLR concept, the detectable binary mergers are concentrated in the range $\mathcal M_c \in [10^8, 10^{9}] M_{\odot}$ over a distance smaller than few Gpc. For all detectable signal, the chirp mass can be reconstructed with a precision and accuracy typically smaller than $10$ percents (a detectable event with $5\sigma-$ significance typically shows precision and accuracy of order 10 percents). The accuracy and precision show a clear pattern of convergence for louder (larger SNR) signals, with error bars dropping like $\sim \text{SNR}^{-1}$ (where the \text{SNR} measure is defined from the cumulative $\chi^2$ presented in Eq.\eqref{eq:chi2}). On the other hand, the luminosity distance, even for very loud signal, is plagued by errors of several tens of percents and similar error bands. Louder signal typically does not improve it. This pattern can be traced back to the degeneracy of the distance with the angles characterizing the binary event, namely the position in the sky, the inclination and the polarisation. 

For the GUEST experiment, detectable events are concentrated in the range $\mathcal M_c \in [\text{few} \times 10^6, 10^{7}] M_{\odot}$ also over a distance smaller than few Gpc. Even though the recovery of the chirp mass is quite similar in LLR and GUEST for equal SNR, showing closely similar patterns of convergence for the accuracy and the precision, the recovery of the luminosity distance, on the other hand, converges fast with larger SNR.  This is a direct reflection of the expected result that a constellation of satellites offers better sky resolution and thus moderate the degeneracy with the luminosity distance.  

To be self-contained, this study should include a study of the \emph{population} of binary black holes and their merger rate close enough to be observable by LLR and SLR. This study will be provided in a parallel paper\cite{Franciollini}. 

\paragraph{Remainder of the paper}
The remainder of this paper is organized as follows: In section \ref{app:framework}, we reproduce the framework of the response of the earth-moon satellite to an incoming GW forcing signal. In section \ref{sec:recovery}, we present all the machinery of the bayesian analysis used in this paper. In section \ref{sec:2d_analysis}, we then present the result of the parameter estimation for the LLR case and provide an in depth analysis of the quality of the chirp mass and luminosity distance for the LLR. In section \ref{sec:recovery_guest}, we reproduce a similar analysis for the satellite ranging experiment GUEST. Finally, in section \ref{sec:conclusion}, we conclude, list the limitations of this first step study and present several directions of improvement.

\section{Theoretical framework: GW signal and orbital response}
\label{app:framework}

\subsection{The idea behind satellite ranging}

Let us start by spelling concisely the idea of detecting gravitational wave with satellite ranging. We assume that a gravitational wave (GW) with strain $h_{ij}(t)$ induces a tidal acceleration
$a_i^{\rm GW}=\ddot{h}_{ij}\,r^j/2$ on the Earth--Satellite separation $r^i$ (where in principle, the satellite is any object orbiting the earth).
The observable is the perturbation $\delta r_{\rm sat}(t)$ (where ``sat'' refers to the satellite) of the
measured Earth--Satellite distance and it is sampled by shooting, on a daily basis, a laser on retroreflectors attached to the satellite and measuring the time of round-travel. To compute it, we use the osculating orbital
element formalism~\cite{Poisson_Will_2014}, in which the six
Keplerian elements $\{p,e,\iota,\Omega,\omega,f\}$ evolve under the GW force
according to the Gauss planetary equations. The range perturbation $\delta r_{\rm sat}(t)$ (see \ref{app:llr} for more details) is then
obtained by projecting the perturbed orbit onto the Earth--Satellite separation
measured by LLR. Tracking the oscillations and the perturbations in the $\delta r_{\rm sat}(t)$ allows to detect the presence of GWs. As a rule of thumb, 1 mm of range corresponds to 6.7 ps of round-trip timing, so millimetre ranging genuinely requires picosecond-class timing systems and event timers. Further details on the LLR experiment and a discussion of the different sources of noise are presented in Appendix \ref{app:llr}.  

% The key physical effect is a \emph{growth hierarchy} for
% on-resonance forcing~\cite{Foster:2025csl}: the semi-latus rectum and
% eccentricity perturbations, $p_1$ and $e_1$, grow linearly with the number of
% orbits, while the true-anomaly perturbation $f_1$ grows \emph{quadratically},
% fed by the linear modes through the fundamental matrix. As a result, the measured range perturbation grows secularly as
% $\delta r_{\rm sat}\propto t^2$ on resonance, so the cumulative sensitivity improves
% rapidly with observing time.

In the remainder of this section, following the lines of the authors of \cite{Blas:2026dsm, Foster:2025csl, Foster:2025nzf} we present a general framework of response of a satellite (natural like the moon, or artificial like GUEST) to a GW perturbation. 

\subsection{Toy model with constant deterministic GW signal}

We assess the sensitivity of both LLR and GUEST/SLR to monochromatic gravitational waves 
with constant amplitude $h_0$ and frequency $f_{\rm GW}$ over an observation period of 
15 years. The injected signals are characterized by:
\begin{itemize}
    \item \textbf{Signal model:} The gravitational wave enters the equations of motion through the geodesic 
deviation equation, producing a tidal acceleration on the orbiting body. The 
transverse-traceless metric perturbation for a monochromatic plane wave is:

\begin{equation}
h_{ij}(t, \mathbf{r}) = h_0\big( A_+ \, e^+_{ij} \cos\Phi_{\rm ret} 
                      + A_\times \, e^\times_{ij} \sin\Phi_{\rm ret} \big)
\end{equation}

\noindent
with the retarded phase

\begin{equation}
\Phi_{\rm ret}(t, \mathbf{r}) = 2\pi f_{\rm GW}\, t  \, . 
\end{equation}
\noindent
Here $\hat{n}$ is the propagation direction, $e^+_{ij}$ and $e^\times_{ij}$ are 
the polarization tensors of the incoming GW, and the polarization amplitudes, $A_+, A_\times$, are set by the 
inclination of the source. For our preliminary discussion, we take the simplifying $A_+ = A_\times = 1/2$.
The resulting tidal acceleration on the relative separation vector $\mathbf{r}$ is:

\begin{equation}
a^{\rm GW}_i(t) = \frac{1}{2} \ddot h_{ij}r^j  = -\frac{1}{2}\,\omega_{\rm GW}^2\, 
                 \sum_j h_{ij}(t,\mathbf{r})\, r_j,
\label{eq:gw_accel_code}
\end{equation}

\noindent
which is added to the Newtonian (and variational) acceleration and integrated 
over the orbit.

Crucially, the actual observable collected by ranging of the Moon or a 
satellite is the round-trip light travel time, which allows the Earth--Moon 
separation to be reconstructed to sub-centimeter precision once the 
propagation delays are modelled (a summary of the precision forecast will be presented in Appendix \ref{app:llr}). Since the laser pulse probes only the 
component of the separation along the Earth-station-to-reflector line, the 
ranging data are sensitive not to the full displacement vector 
$\delta\mathbf{r}$ but to its projection onto the line of sight. The 
observable is therefore the projection of the induced displacement 
$\delta\mathbf{r}$ onto the line of sight (radial direction):

\begin{equation}
\delta r_{\rm sat}(t) = \frac{\delta\mathbf{r}(t)\cdot\mathbf{r}(t)}{|\mathbf{r}(t)|}.
\end{equation}

Let us emphasize that within our simplified framework, $\delta r_{\rm sat}(t)$ is the deviation from the \emph{Keplerian} unperturbed orbit. In a more realistic setting, $\delta r_{\rm sat}(t)$ should already contain all the non-Keplerian corrections to the orbit of the satellite, each of them coming with their own uncertainty. This would imply to marginalise over all the parameters controlling these (non-GW) deviations from the Keplerian orbit. We plan to include these marginalisations in a future study. 

    \item \textbf{Detection criterion:} We quantify detectability through the cumulative $\chi^2$ statistic~\cite{Foster:2025csl},
\be
\chi^2(t)=\sum_{t_i<t}\!\left[\frac{\delta r(t_i)}{\sigma}\right]^2,
\label{eq:chi2}
\ee
where $\sigma$ is the single-measurement distance precision and the sum runs over LLR measurements occurring at the times $\{t_i\}$ before the $t$.  A threshold $\chi^2_{\rm th}=25(9)$ corresponds to $5(3)=\sigma$ detection. This allows to determine the minimum detectable strain amplitude 
    $h_0^{\rm min}(f_{\rm GW})$ at each frequency.

\item \textbf{Gauss planetary equations}: The Satellite's orbit is parametrised by the osculating Keplerian elements
$\mathbf X_{\rm orb}=\{p,e,i,\Omega,\omega\}$ and true anomaly $f$. The GW induces a tidal acceleration on the Earth–Satellite separation,
\begin{equation}
\mathbf a_{\rm GW}=\tfrac12\,\ddot{\mathbf h}\!\cdot\!\mathbf r
\, ,
\end{equation}
in the monochromatic approximation $\omega_{\rm GW}=2\pi f_{\rm GW}$.

\begin{figure}[t]
\centering

\includegraphics[width=0.8\columnwidth]{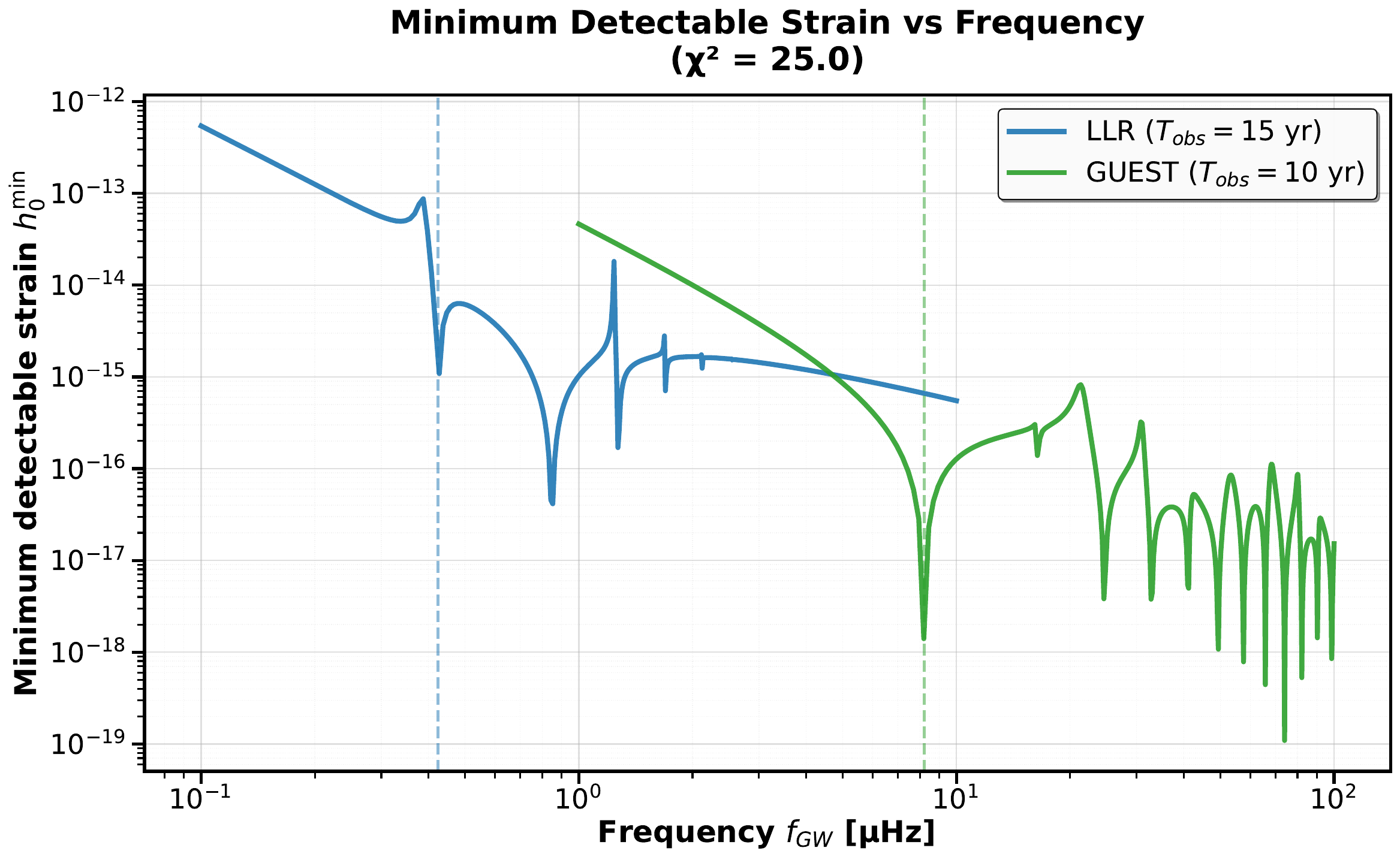}
\caption{Minimum detectable strain amplitude $h_0^{\rm min}$ versus gravitational wave frequency 
for monochromatic signals detected via $\chi^2$ test (threshold = 25). Blue (LLR, $\sigma_{\rm LLR} = 3$mm, Moon, over 15 years of 
observations, see section \ref{sec:recovery} for discussion) and green (GUEST/SLR, $\sigma_{\rm GUEST} = 10$mm,  over 10 years of 
observations, see section \ref{sec:recovery_guest} for discussion) curves show sensitivity 
for constant-amplitude signals. Dashed vertical lines mark the orbital frequencies of 
the Moon and GUEST. LLR excels at low frequencies while GUEST provides superior 
sensitivity at higher frequencies. Notice that the sensitivity of the GUEST experiment is overestimated with respect to the sensitivity reported in \cite{Blas:2026xol}, this difference originates from the assumption of the unperturbed Keplerian orbit that we adopt in this paper. }
\label{fig:stoc_sens}
\end{figure}

To interpret orbital perturbations physically, we decompose the motion in the
co-moving RSW frame centered on the orbiting body (here, the Moon relative to the
Earth). The radial unit vector $\hat{\bm{r}}$ points from the Earth's center straight
out to the Moon, tracking whether the body is pushed higher or lower in its
orbit. The along-track vector $\hat{\bm\theta}$ lies in the orbital plane, perpendicular
to $\hat{\bm{r}}$, and points broadly along the direction of travel; it coincides
with the velocity only for a perfectly circular orbit, so in an eccentric orbit
it reveals how perturbations advance or retard the body's phase. Finally, the
cross-track vector $\hat{\bm{z}}$ is normal to the orbital plane and completes the
right-handed triad ($\hat{\bm{z}} = \hat{\bm{r}} \times \hat{\bm\theta}$), isolating out-of-plane
deflections that tilt or twist the orbit. Because $\mathcal{R}$ and $\mathcal{S}$ span
the instantaneous orbital plane while $\hat{\bm{z}}$ measures departures from it,
this frame cleanly separates changes in orbital size and shape (in-plane) from
changes in orientation (out-of-plane), making it the natural basis for
expressing the perturbing forces acting on the orbit. In this bases the perturbation around the ideal Keplerian orbit takes the form $\bm{a}=\mathcal{R}\,\hat{\bm{r}}+\mathcal{S}\,\hat{\bm\theta}+\mathcal{W}\,\hat{\bm{z}}$. 
 In terms
of the six orbital elements, these unit vectors are given by~\cite{Poisson_Will_2014}:

\begin{widetext}
\begin{align}
\dot{p} &= 2\sqrt{\frac{p^3}{GM}}\;\frac{\mathcal{S}}{1+e\cos f}\,,
\notag\\[4pt]
\dot{e} &= \sqrt{\frac{p}{GM}}\left[\sin f\;\mathcal{R}
  +\frac{2\cos f+e(1+\cos^2\!f)}{1+e\cos f}\;\mathcal{S}\right],
\notag\\[4pt]
\dot{\iota} &= \sqrt{\frac{p}{GM}}\;\frac{\cos(\omega+f)}{1+e\cos f}\;\mathcal{W}\,,
\notag\\[4pt]
\dot{\Omega} &= \csc\iota\;\sqrt{\frac{p}{GM}}\;\frac{\sin(\omega+f)}{1+e\cos f}\;\mathcal{W}\,,
\notag\\[4pt]
\dot{\omega} &= \frac{1}{e}\sqrt{\frac{p}{GM}}\left[-\cos f\;\mathcal{R}
  +\frac{2+e\cos f}{1+e\cos f}\sin f\;\mathcal{S}\right]
 -\cot\iota\;\sqrt{\frac{p}{GM}}\;\frac{\sin(\omega+f)}{1+e\cos f}\;\mathcal{W}\,,
\notag\\[4pt]
\dot{f} &= \sqrt{\frac{GM}{p^3}}(1+e\cos f)^2
+ \frac{1}{e}\sqrt{\frac{p}{GM}}
\times\left[\cos f\;\mathcal{R}
  -\frac{2+e\cos f}{1+e\cos f}\sin f\;\mathcal{S}\right].
\label{eq:gauss_sm}
\end{align}
\end{widetext}
\end{itemize}

Assuming a static Keplerian orbit as a leading order baseline for the system Earth-Satellite, Figure \ref{fig:stoc_sens} presents the resulting sensitivity curves, with $\sigma_{\rm LLR} = 3$mm and  $\sigma_{\rm GUEST} = 10$mm, showing that the 
LLR configuration achieves superior sensitivity at low frequencies (below $\sim 1$ µHz), 
where the Moon's orbital dynamics dominate the signal response. Conversely, the GUEST 
constellation leverages the shorter orbital period of the satellites to access higher frequencies 
(above $\sim 10$ µHz), where it progressively outperforms LLR. These complementary frequency 
ranges suggest that a combined analysis of both datasets could provide comprehensive 
gravitational wave sensitivity across several decades of frequency space.  We provide the details about the numerical solver in Appendix \ref{app:solver}.

The ideal Keplerian unperturbed orbit is a rough approximation and several other parameters enter the global fit of the Earth-Satellite system. In practice, to go beyond the ideal Keplerian motion, what is done is to fit for the GW amplitude with all other parameters being fitted in orbit reconstruction. One then can use the global fit as a measure of sensitivity.

\subsection{GW signal from a chirping binary}

We now turn to the more realistic case of the gravitational waves emitted by a inspiral of a compact binary. 

\paragraph{Inspiral phase:}
A quasi-circular binary of component masses $M_{1,2}$ is described by its
chirp mass
\begin{equation}
\mathcal{M}_c = \frac{(M_1 M_2)^{3/5}}{(M_1+M_2)^{1/5}} .
\end{equation}
At leading (0PN, quadrupole-driven) order, the time to coalescence
$\tau(t)=\tau_0-t$ controls the frequency, phase, and amplitude evolution\cite{Maggiore:2007ulw}:
    \begin{align}
\dot{f} = \frac{96}{5}\,\pi^{8/3}\!\left(\frac{G\Mc}{c^3}\right)^{\!5/3}f^{11/3}\, \qquad \Rightarrow \qquad f_{\rm GW}(\tau) = \frac{1}{\pi}\!\bigg[\frac{256}{5}\frac{(G\mathcal{M}_c)^{5/3}}{c^5}\,\tau\bigg]^{-3/8}\!, \\
\label{eq:fdot}
\end{align}
and the evolution of the phase and the amplitude of the signal are given by
\begin{align}
\Phi_{\rm GW}(\tau) &= \phi_0 - 2\!\left[\frac{5}{256}\frac{c^5\tau}{(G\mathcal{M}_c)^{5/3}}\right]^{5/8}\!, \\
h_0(t) &= \frac{4\,(G\mathcal{M}_c)^{5/3}\,[\pi f_{\rm GW}(t)]^{2/3}}{c^4\,D_L} \, ,
\end{align}

where $\phi_0$ is the coalescence or the reference phase.

For a chirping source located in direction $\hat{\mathbf r}(\theta,\phi)$, we use the orthonormal triad $(\hat{\mathbf u},\hat{\mathbf v},\hat{\mathbf r})$ adapted to the sky position. 
To account for an arbitrary polarization orientation of the source, we rotate this basis by the polarization angle $\psi$:
\begin{equation}
\hat{\mathbf p}=\cos\psi\,\hat{\mathbf u}+\sin\psi\,\hat{\mathbf v},\quad
\hat{\mathbf q}=-\sin\psi\,\hat{\mathbf u}+\cos\psi\,\hat{\mathbf v}.
\end{equation}

This rotation in the plane perpendicular to $\hat{\mathbf r}$ changes the plus and cross polarization tensors to
\begin{equation}
\mathbf{e}_+ = \hat{\mathbf p}\!\otimes\!\hat{\mathbf p}-\hat{\mathbf q}\!\otimes\!\hat{\mathbf q},\quad
\mathbf{e}_\times = \hat{\mathbf p}\!\otimes\!\hat{\mathbf q}+\hat{\mathbf q}\!\otimes\!\hat{\mathbf p},
\end{equation}
which are symmetric, traceless, and transverse to $\hat{\mathbf k}$. Under a rotation of the transverse basis by an angle $\psi$, the polarization tensors transform with twice this angle. This reflects the spin-2 nature of gravitational waves.

More precisely, denoting by $(\mathbf e_+^{\,0}, \mathbf e_\times^{\,0})$ the polarization tensors defined in the unrotated basis $(\hat{\mathbf u}, \hat{\mathbf v})$, the tensors in the rotated basis $(\hat{\mathbf p}, \hat{\mathbf q})$ satisfy
\begin{align}
\mathbf e_+(\psi) = \cos(2\psi)\,\mathbf e_+^{\,0} + \sin(2\psi)\,\mathbf e_\times^{\,0},
\\
\mathbf e_\times(\psi) = -\sin(2\psi)\,\mathbf e_+^{\,0} + \cos(2\psi)\,\mathbf e_\times^{\,0} \, . 
\end{align}

The dependence on the binary inclination angle $\mathcal{I}$ of the binary enters through the standard amplitude factors
\begin{equation}
A_+=\tfrac{1+\cos^2\mathcal{I}}{2},\qquad A_\times=\cos\mathcal{I} ,
\end{equation}
where $\mathcal{I}$ is the angle between the orbital angular momentum vector 
$\hat{\mathbf{L}}$ of the source binary and the line of sight from the source 
to the observer, i.e.\ the propagation direction $\hat{n}$ of the emitted 
wave, so that $\cos\mathcal{I}=\hat{\mathbf{L}}\cdot\hat{n}$. It is a property 
of the \emph{source} geometry alone and should not be confused with the 
inclination $\iota_0$ of the lunar orbit relative to the ecliptic, which 
enters through the initial conditions of the Earth--Moon system. For a 
face-on binary ($\mathcal{I}=0$ or $\pi$) the orbital plane is seen 
perpendicular to the line of sight, $A_+=A_\times=1$, and the emitted 
radiation is circularly polarised with maximal amplitude. For an edge-on 
binary ($\mathcal{I}=\pi/2$) the orbital plane contains the line of sight, 
$A_\times=0$ and $A_+=1/2$, and the wave is linearly polarised in the $+$ 
mode with a factor-of-two reduction in amplitude. The simplifying choice 
$A_+=A_\times=1/2$ adopted above therefore does not correspond to any single 
physical inclination, but serves to fix the overall normalisation while both 
polarisations are retained.
Consequently, the metric perturbation at the detector is
\begin{equation}
h_{ij}(t)=h_0(t)\!\left[
A_+\, e^+_{ij}\cos\Phi_{\rm GW}(t)
+
A_\times\, e^\times_{ij}\sin\Phi_{\rm GW}(t)
\right].
\end{equation}
\begin{figure}[h]
\center{
\includegraphics[width=0.48\textwidth]{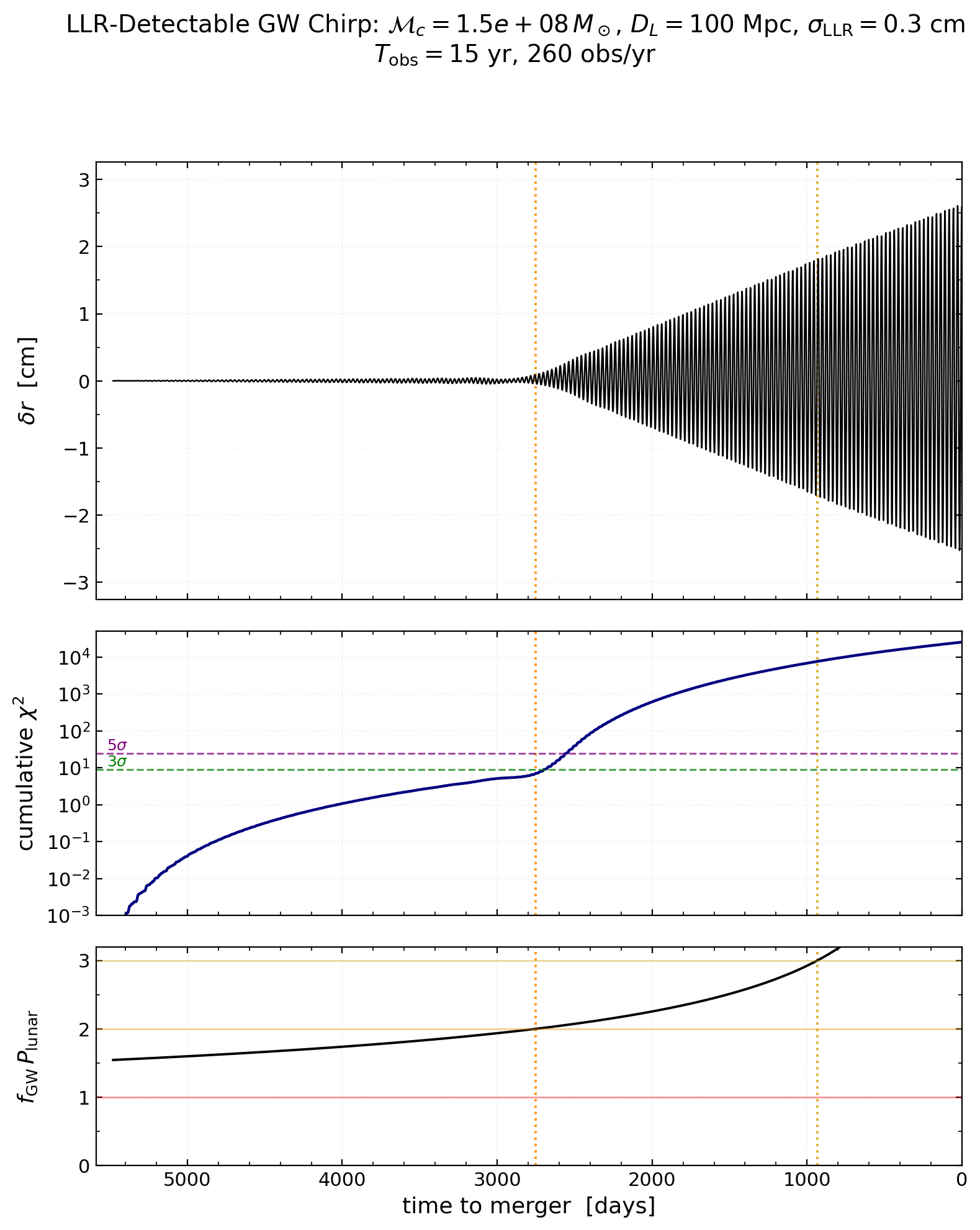}
\includegraphics[width=0.48\textwidth]{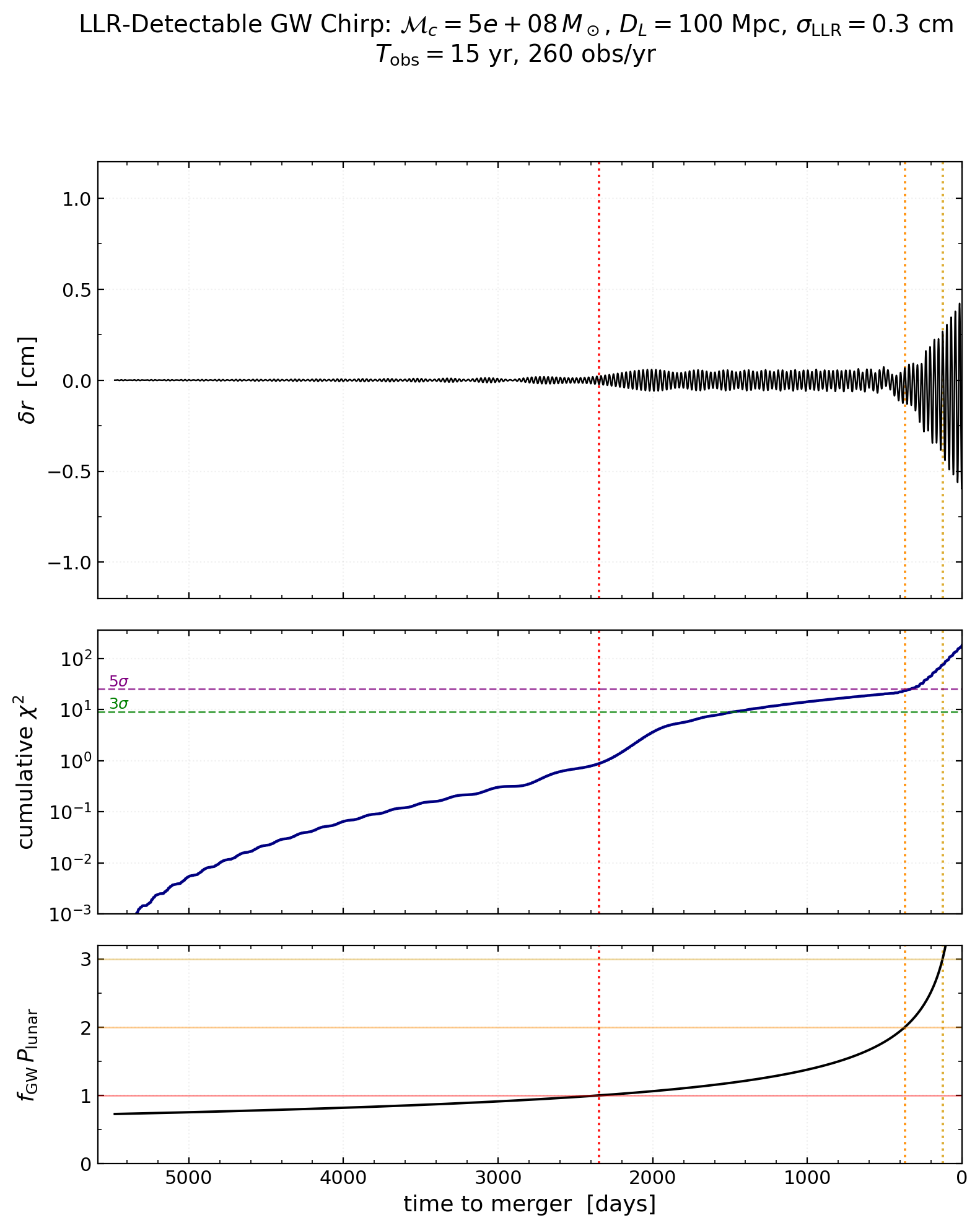}
}
\caption{
\textbf{Evolution of the deviation $\delta r$ around the ideal Keplerian orbit for GW from an inspiralling compact binary}: Two examples of the evolution of the observable $\delta r$ under the perturbation of a binary inspiral situated at $100$ Mpc and with chirp mass $\mathcal M_c = 1.5\times 10^8 M_{\odot}$ (left) and $\mathcal M_c = 5\times 10^8 M_{\odot}$ (right). The upper panels present $\delta r$ across the binary inspiral, the middle panels present the evolution of the $\chi^2(t)$ and finally in the lower panels, we present the evolution of the frequency of the binary. The vertical dashed orange lines show the frequencies corresponding to an harmonic of the moon-earth system frequency $P_{\rm lunar}$. We provide the details about the numerical solver used to compute these trajectories in Appendix \ref{app:solver}. }
\label{fig:delta_r_evolution}
\end{figure}

For example, the special case
$
(\theta,\phi,\psi,\mathcal{I}) = (0,0,0,0),
$
for which
$
\hat{\mathbf n} = (0,0,1),\quad
\hat{\mathbf u} = (1,0,0),\quad
\hat{\mathbf v} = (0,1,0) \, ,
$ corresponds to a gravitational wave propagating along the $+z$ direction (zenith), with no polarization rotation ($\psi=0$) and a face-on binary ($\mathcal I =0$), yielding maximal signal amplitude and a polarization basis aligned with the Cartesian axes.

\paragraph{The merger and the ringdown phase:}

The inspiral phase is followed by the merger and the ringdown phases, which also source gravitational waves. Theses are characterized by numerical or semi-analytical waveforms which are challenging to incorporate directly into a bayesian analysis of the signal. A proper analysis would require a \emph{matched filtering technique}, similar to the ones performed in LVK searches. Namely, it would require the generation of a bank of $\delta r$ signal over which to run the bayesian analysis. We keep this analysis for a further study. 

As an estimate of the importance of this portion of the signal, we can mention that the typical duration of the merger and ringdown phase is of order: \begin{equation}
t_{\rm merger+RD} \sim\mathcal{O}(1) 
\left(\frac{\mathcal{M}_c}{10^8\,M_\odot}\right) \, \times \text{days}.
\end{equation} 

% This means that for masses smaller $\mathcal{M}_c$ than $10^9\,M_\odot$ (for LLR) and $10^8\,M_\odot$ (for SLR, GUEST), the merger occurs on a time scale much shorter than the resonance period (which is of the order of the month (day) respectively), and it has likely a moderate impact. In this respect, we will cut the inspiral signal $1$ day before the divergence. 

It has however been shown (see \cite{Blas:2026xol}) that the merger and ringdown phases can contribute significantly to the SNR of the signal. We will discard them in our study.

\subsection{Detectability from LLR}

On Fig.\ref{fig:delta_r_evolution}, we present the evolution of the observable $\delta r_{\rm sat}$, related to the time of round trip of the signal, under the GW perturbation due to the inspiral phase of two typical binaries with chirp mass of order $\mathcal{M}_c \sim 10^8 M_{\odot}$. We observe that the change in the pattern of evolution of the envelope, starting at the growth of the instability, occurs when the binary inspiral hits a harmonic in the earth-satellite system. In the present case of the earth-moon system, the harmonics are all the multiples of $P_0 \sim 27$ days.

\begin{figure}[h]
\center{
\includegraphics[width=0.48\textwidth]{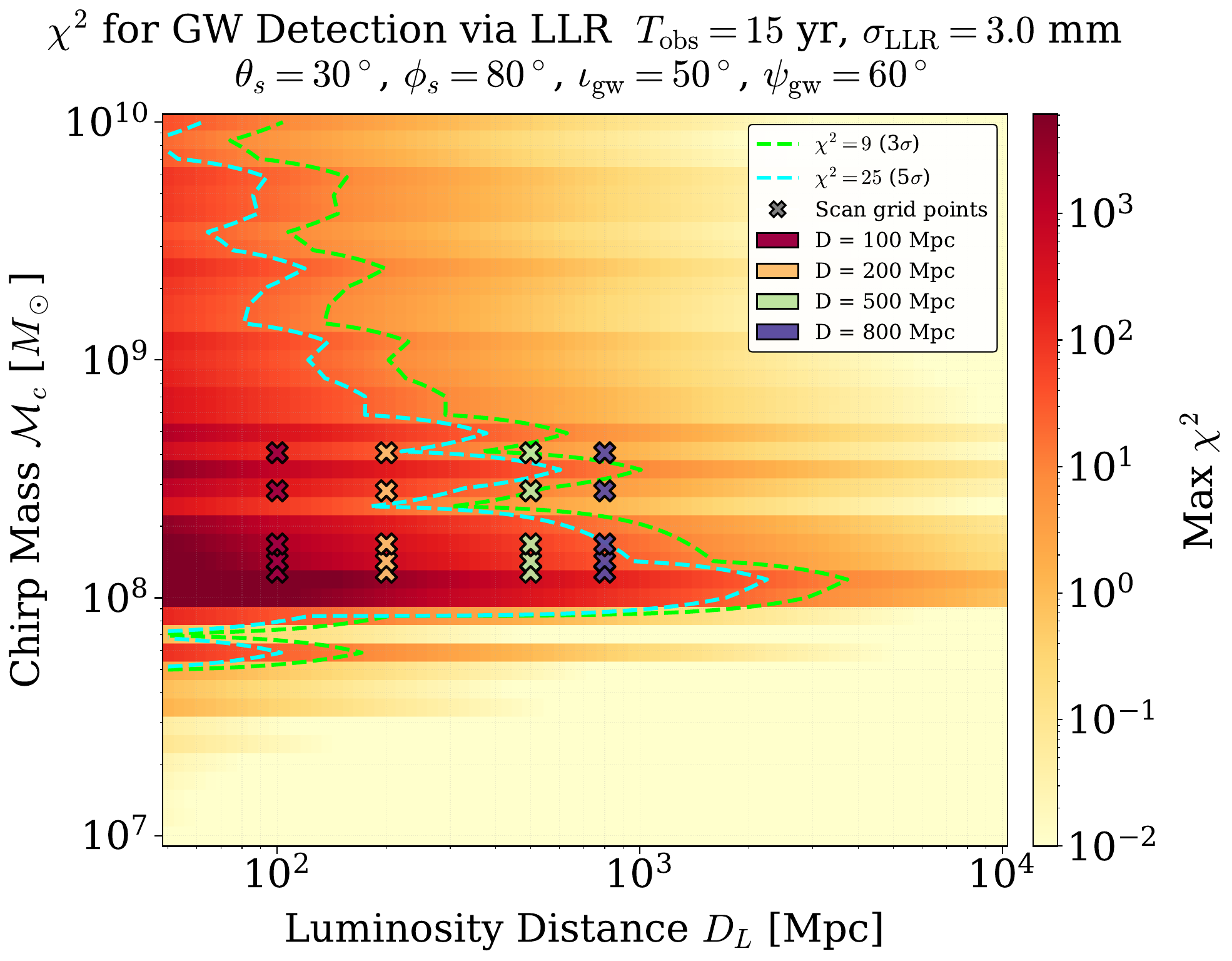}
\includegraphics[width=0.48\textwidth]{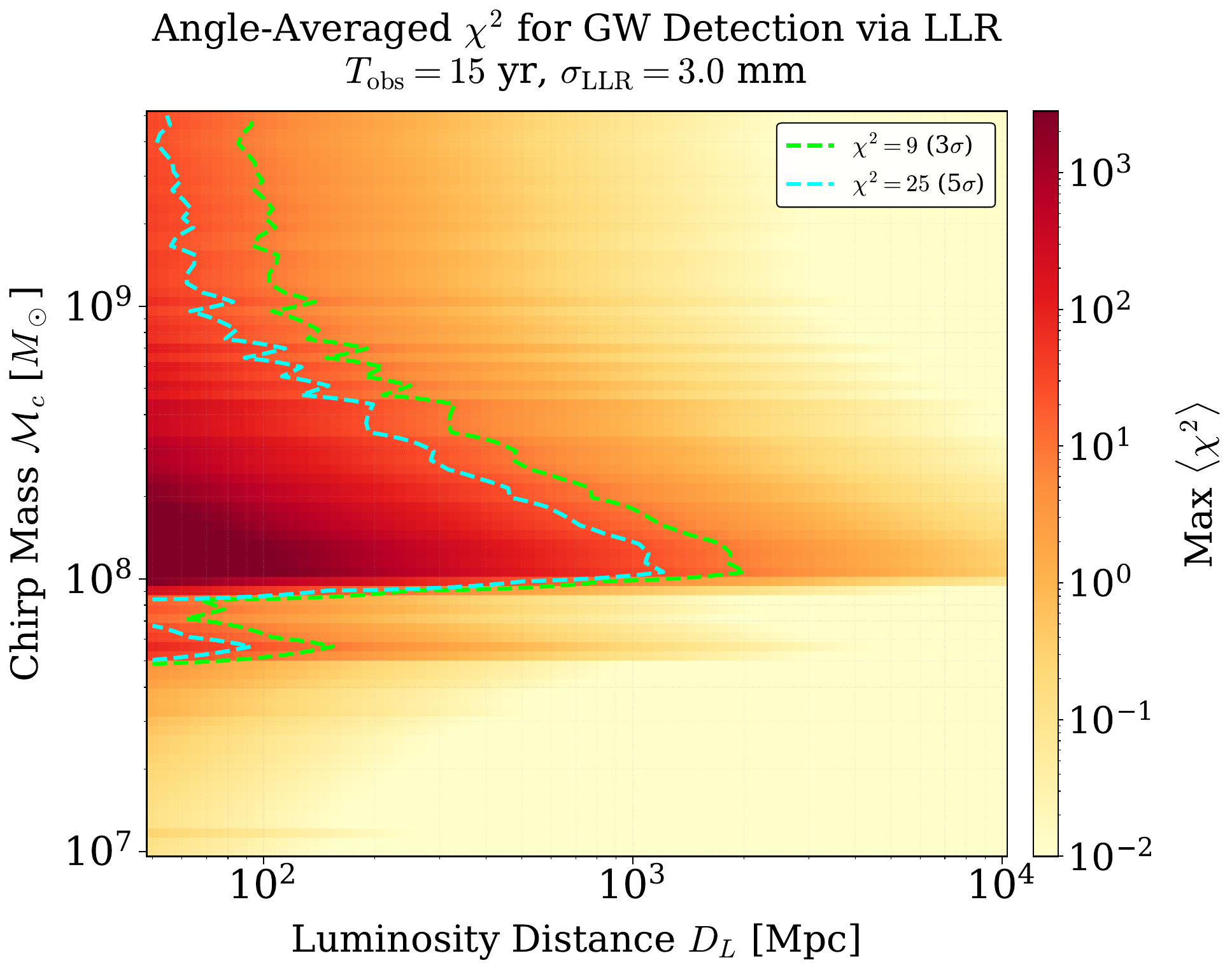}
}
\caption{
\textbf{Detectability heatmap in the $D_L-\mathcal{M}_c$ (Inspiral only)}. Value of the accumulated $\chi^2(t)$ at the merger (Max $\chi^2$). \textbf{Left panel:} The four angles are set to a fixed value. The green (blue) boundary represents the values Max $\chi^2 = 25(9)$ corresponding to a $5(3)$ sigma detection for a chosen set of angles. The crosses represent the samples of binaries that will be statistically studied in Section \ref{sec:2d_analysis} ($D_{\mathrm{true}} \in \{100, 200, 400, 500, 800\}\,\mathrm{Mpc}$
and primary masses $\mathcal{M}_c \in \{9.4, 14.0, 17.0, 32.0, 46.8\}\times 10^{7}\,M_\odot$. 
\textbf{Right Panel:} Same analysis but averaging over the angles. We emphasize that these plots will receive corrections when we include the merger and the ringdown.}
\label{fig:waveformheatmap}
\end{figure}

On Fig.\ref{fig:waveformheatmap}, we present a heatmap of the cumulative $\chi^2$ evaluated at the merger (which we call the Max $\chi^2$) in the space $\mathcal{M}_c-D_L$, for an observation time of $T_{\rm obs}$ and a gaussian with $\sigma_{\rm LLR} = 3 \times 10^{-3}$ mm, and using only the inspiral period. We present in blue (green) the boundary on which $\chi^2 = 25 (9)$ which designates $5(3)-\sigma$ confidence detection. 
 The crosses represent the samples of binaries that will be statistically studied in Section \ref{sec:2d_analysis} ($D_{\mathrm{true}} \in \{100, 200, 400, 500, 800\}\,\mathrm{Mpc}$
and chirp mass $\mathcal{M}_c \in \{9.4, 14.0, 17.0, 32.0, 46.8\}\times 10^{7}\,M_\odot$). The upper panel assumed a fixed direction of the signal (fixed angles) while the lower panel does an average over the four angles. 

% EXPLAIN THE PATTERN: Because the tidal GW acceleration scales as
% $\delta r \propto \mathcal{M}_c^{5/3} f_{\rm gw}^{2/3}/D_L$, the accumulated
% $\chi^2$ increases steeply toward high chirp masses and small distances, as we can directly observe on fig.\ref{fig:chi2_heatmap_guest}

\subsection{Noise detector realisation}

The data collection is plagued by a series of errors originating from atmospheric laser propagation, ground deformation, tidal dissipation, detector systematic errors, some of which are discussed in the Appendix \ref{app:llr}.  
As a minimal modelisation of the impact of these noises, we assume that each measurement is corrupted by an independent Gaussian noise of the form
\begin{equation}
\label{eq:noise_model}
\delta r_{\text{obs}}(t_i, \bm\theta_{\rm true}) = \delta r_{\text{true}}(t_i,\bm\theta_{\rm true}) + \mathcal{N}(0, \sigma_{\text{LLR}}^2),
\end{equation}
where $\delta r_{\text{true}}(t_i)$ is the true GW-induced signal at time $t_i$, 
and $\sigma_{\text{LLR}}$ is the measurement noise standard deviation. 
We generate $N_{\text{obs}}$ independent noise realizations to simulate realistic observational 
data across the $N_{\text{obs}}$ measurement epochs spanning the $T_{\text{obs}}$ year baseline.

%-------------------------------------
%------------------------------------------------------------------

%============================================================
\section{Parameter Recovery analysis for LLR}
\label{sec:recovery}
%============================================================

\subsection{Problem setup}

The response of the satellite, measured by $\delta r_{\rm sat}$, is controlled by a set of 8 parameters 
\begin{equation}
    \bigl(M_1,\; M_2,\; D_L,\;
\theta_s,\; \phi_s,\; \mathcal I,\; \psi, \; \phi_0, \; t_c \bigr) \, . 
\end{equation}

Let us now discuss which of these parameters are the most crucial for our numerical sampling: 

\begin{itemize}
    \item \textbf{The masses $M_1, M_2$ and the luminosity distance $D_L$}: The two masses and the luminosity distance are the main parameters of physical interest to us, over which we will sample. Looking back at the 0PN chirping GW signal, we observe that the signal only depends on the chirp mass and not on the individual masses, implying that the individual masses are not directly resolvable. This reduces the parameters to the chirping mass $\mathcal M_c$ and the luminosity distance $D_L$.
    \item \textbf{The angles}: In our analysis, the angles are not crucial components, as they do not directly provide information on the population of binaries. Consequently, we treat the four angular parameters
$(\theta_s,\; \phi_s,\; \mathcal I,\; \psi)$ --- describing, respectively, the sky
location, orbital inclination, and polarization angle of the source --- as
nuisance parameters in our analysis. We
numerically marginalize by integrating out their posterior dependence. This
approach allows us to account for the uncertainty and potential correlations
these angles introduce into the recovered signal amplitude and phase, at the cost of additional computational overhead in the sampling procedure.

\item \textbf{The reference phase $\phi_0$}: In the present analysis we fix the reference phase to $\phi_0 = 0$, both in the injected and in the bayesian analysis, rather than
including it as a free parameter in the sampling. At the current, leading-order
stage of LLR waveform reconstruction, the template enters the likelihood as a
simple oscillatory term in which the reference phase acts primarily as an overall
offset rather than as a parameter that is strongly informed by the data\cite{Rover:2006ni}\footnote{In principle, since the template is linear in
$\cos\phi_0$ and $\sin\phi_0$, the reference phase could be analytically
marginalized in closed form via the standard quadrature approach used in
matched-filtering and phase-marginalized gravitational-wave likelihoods, yielding
a marginal likelihood proportional to $I_0(R)$, where $R$ is the
quadrature-combined matched-filter statistic. We defer this refinement to future
work and instead adopt the simpler choice of fixing $\phi_0 = 0$.}.  We nonetheless perform some bayesian analysis including the marginalisation over the coalescence phase $\phi_0$  for few selected points in the analysis  and compare the posteriors with the results of the bayesian analysis on same points with fixed determined $\phi_0$. We present the results of this comparison in Appendix \ref{app:phi0}. We find that the posteriors differ by only a 5 to 10\% percents. More specifically, the best fit recovered values (for the luminosity distance and the chirp mass) and the half-width ranges move by  5 to 10\%.
%, consistent with the broader scope-limiting strategy of this paper, and analogous to the phased introduction of parameters seen in the early development of LVK Bayesian inference pipelines\cite{Rover:2006bb, Rover:2006ni}.

\item \textbf{The coalescence time}: We similarly fix the coalescence time $t_c$ rather than sampling over it. As in
standard gravitational-wave matched-filtering pipelines, we expect $t_c$ to be first
identified with high precision through a preliminary matched-filter search of
the LLR data, prior to the Bayesian parameter estimation stage. Since this
quantity is already well constrained by the data, fixing it to its best-fit
value does not discard meaningful information; rather, it avoids introducing
an additional, redundant dimension into the sampling of the parameters of
primary interest ($M_1$, $M_2$, $D_L$).
\end{itemize}

In summary, our numerical sampling is performed over the two-parameter space
$(\mathcal{M}_c, D_L)$, with the four angular parameters
$(\theta_s,\phi_s,\mathcal I,\psi)$ marginalized numerically, and with $\phi_0$
and $t_c$ fixed to their fiducial values as discussed above. We now turn to the systematic analysis of the recovery of the chirp mass and the luminosity distance.

With these assumptions in mind, we construct a synthetic data stream representing the response of an
idealized low-frequency GW detector to a single inspiralling massive
black-hole binary. The source is parametrized by
\begin{equation}
\bm{\theta} \;=\; \bigl(\log_{10}\mathcal{M}_c,\; \log_{10}D_L,\;
\cos\theta_s,\; \phi_s,\; \cos\mathcal I,\; \psi\bigr),
\end{equation}
with fiducial (``truth'') values $\bm{\theta}_{\rm true}$ chosen in the
regime relevant for
the LLR observations, as shown with crosses in Fig.\ref{fig:waveformheatmap}. The data are sampled uniformly at $N_{\rm obs}$ points per year over a
total baseline $T_{\rm obs}$, giving a time
series $\{t_i, d_i\}_{i=1}^{N}$. The noise model is extracted from Eq.\eqref{eq:noise_model}. 

% This choice deliberately isolates the information-theoretic properties of the inference problem --- parameter degeneracies, prior dependence, multi-modality --- from the additional complications of coloured or non-stationary detector noise, which we defer to future work.

\begin{table}[h]
\centering
\caption{Prior distributions used in the 2D $(\mathcal{M}_c, D_L)$ nested
sampling analysis. Sky/orientation angles $(\theta_s, \phi_s, \mathcal I_{\rm
gw}, \psi_{\rm gw})$ are not sampled directly: they are marginalised
deterministically via a fixed Sobol quasi-Monte-Carlo point set drawn once
from the isotropic prior below.}
\label{tab:priors_2d}
\begin{tabular}{lccc}
\hline\hline
Parameter & Prior & Range & Notes \\
\hline
$\mathcal{M}_c$ & Log-uniform & $[10^{7},\, 10^{9}]\,M_\odot$ & sampled in $\log_{10}\mathcal{M}_c$ \\
$D_L$           & Uniform     & $[50,\,1000]\,\mathrm{Mpc}$   & sampled in $D_L$\\
\hline
$\cos\theta_s$        & Uniform     & $[-1,\,1]$                    & marginalised via Sobol set \\
$\phi_s$              & Uniform     & $[0,\,2\pi]$                  & marginalised via Sobol set \\
$\cos\mathcal I_{\rm gw}$  & Uniform     & $[-1,\,1]$                    & marginalised via Sobol set \\
$\psi_{\rm gw}$       & Uniform     & $[0,\,\pi]$                   & marginalised via Sobol set \\
\hline\hline
\end{tabular}
\end{table}

We infer the source parameters from the simulated LLR range residuals
$\delta r(t_i)$ within a Bayesian framework. Assuming independent Gaussian
measurement noise of standard deviation $\sigma_{\rm LLR}$, the loglikelihood
for a model $\delta r_{\rm model}$ of the noisy signal $\delta r_{\rm obs}$ is
\begin{equation}
\ln \mathcal{L}(\bm\theta)
= -\frac{1}{2\sigma_{\rm LLR}^2}
  \sum_{i}\bigl[\delta r_{\rm obs}(t_i)
  - \delta r_{\rm model}(t_i;\bm\theta)\bigr]^2 ,
\end{equation}
up to an additive constant. The sum is performed over all the observations $T_{\rm obs} \times N_{\rm obs}$. The full parameter set comprises the intrinsic
parameters $\{\mathcal{M}_c, D_L\}$ (chirp mass and luminosity distance) and
the four extrinsic angles $\{\theta_s,\phi_s,\mathcal I_{\rm gw},\psi_{\rm gw}\}$.

However, as we mentioned, these four extrinsic angles which control the satellite antenna response and the
orbital/orientation geometry of the source, are numerically marginalised
at the level of the likelihood,
\begin{equation}
    \mathcal{L}_{\rm marg}(\mathcal{M}_c, D_L)
    = \int \mathcal{L}(\mathcal{M}_c, D_L, \Omega)\, \pi(\Omega)\, d\Omega ,
\end{equation}
evaluated over the isotropic angle prior $\pi(\Omega)$,
\begin{equation}
    \cos\theta_s \sim \mathcal{U}(-1,1), \qquad
    \phi_s \sim \mathcal{U}(0, 2\pi), \qquad
    \cos\mathcal I_{\rm gw} \sim \mathcal{U}(-1,1), \qquad
    \psi_{\rm gw} \sim \mathcal{U}(0, \pi).
\end{equation}
In practice, we draw $N_{\rm Sobol}$ points from a Sobol sequence and map
them onto these physical priors. Because the points are drawn directly from
$\pi(\Omega)$, the integral reduces to an average of the
likelihood over the Sobol point set, with no importance-sampling correction
required:
\begin{equation}
    \ln\mathcal{L}_{\rm marg}(\mathcal{M}_c, D_L)
    = \log\!\left[\frac{1}{N_{\rm Sobol}}\sum_{k=1}^{N_{\rm Sobol}}
    \mathcal{L}(\mathcal{M}_c, D_L, \Omega_k)
    \right]
    = \mathrm{logsumexp}_k\!\left(\ln\mathcal{L}_k\right) - \ln N_{\rm Sobol} ,
\end{equation}
where $\ln\mathcal{L}_k$ is the log-likelihood
evaluated at the $k$-th Sobol angle tuple. The quasi-Monte-Carlo convergence
of this average is monitored through the effective sample size
$N_{\rm eff} = (\sum_k w_k)^2 / \sum_k w_k^2$, with $w_k = \mathcal{L}_k$,
ensuring the marginalisation remains well resolved throughout the sampled
parameter space. We explain in further details the procedure and study its convergence in Appendix \ref{app:bias_study}.

\begin{figure*}[t]
    \centering
    \includegraphics[width=0.48\linewidth]{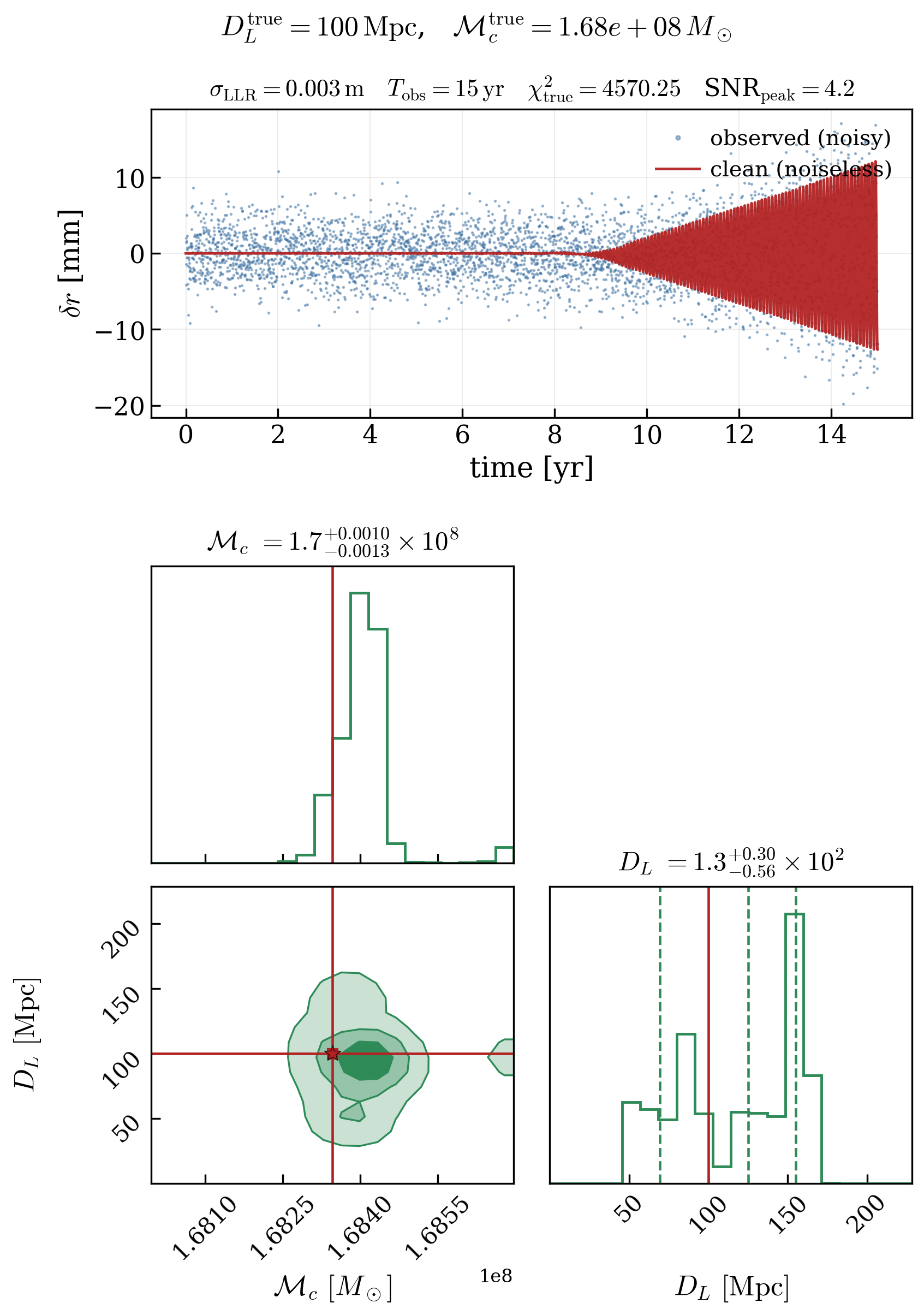}
    \includegraphics[width=0.48\linewidth]{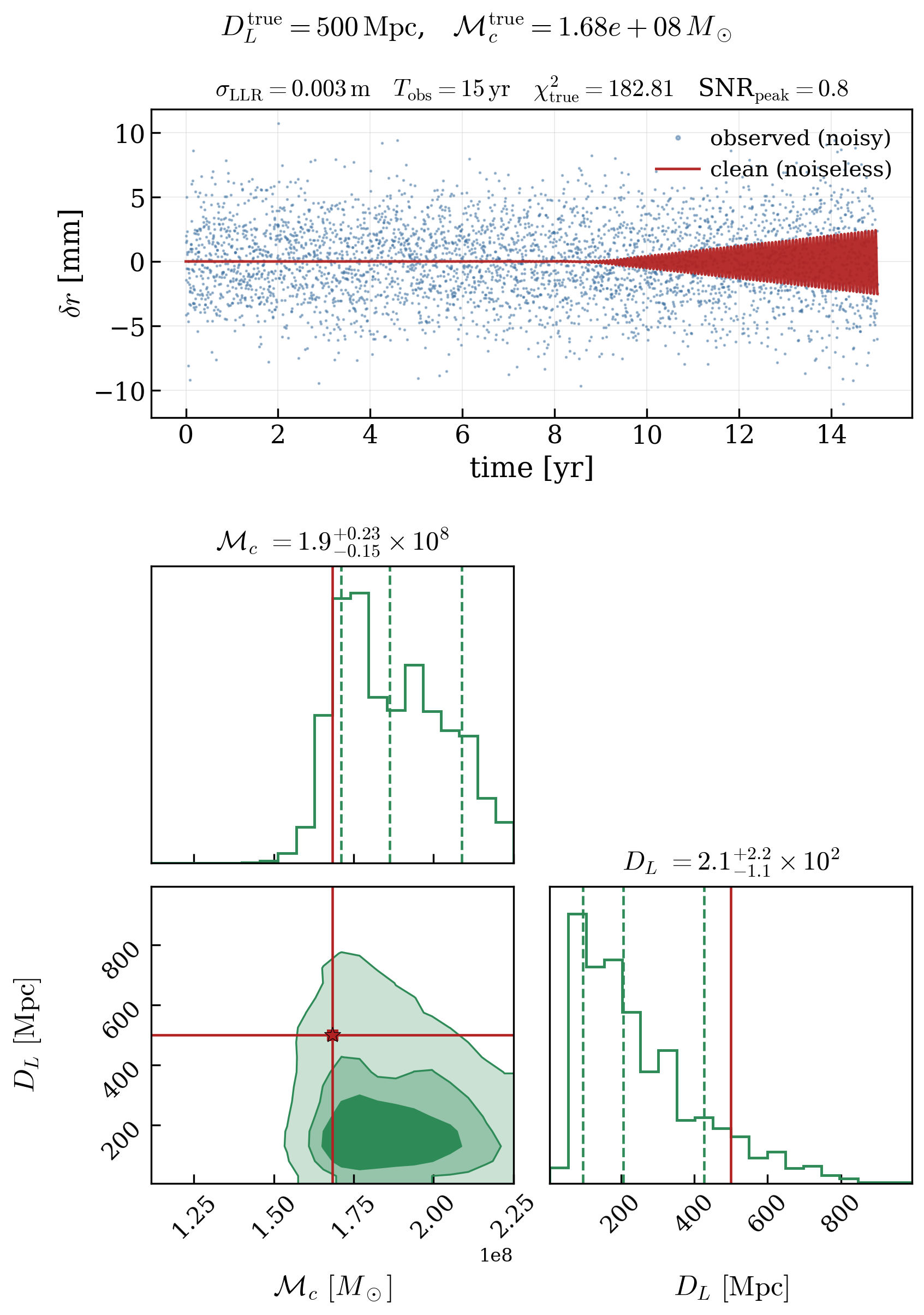}
    \caption{\textbf{Posteriors of the bayesian analysis}:
        Corner plot of the 2-dimensional posterior obtained with 
        \textsc{Dynesty}\cite{Speagle:2019ivv} for a representative injection. Diagonal panels show 
        the marginalised one-dimensional posteriors; off-diagonal panels show 
        the two-dimensional marginalised posteriors with $1\sigma$ and 
        $2\sigma$ contours. Green shaded regions mark the 16th, 50th, and 84th 
        percentiles respectively; the red lines and dots indicate the true injected values. 
        Parameter values and 68\% credible intervals are quoted above each 
        diagonal panel.
    }
    \label{fig:corner_example}
\end{figure*}

Priors on the two sampled parameters are

\begin{itemize}
    \item \textbf{Chirp mass $\mathcal{M}_c$}: log-uniform over
    $\mathcal{M}_c \in [10^{7}, 10^{9}]\,M_\odot$.

   \item \textbf{Luminosity distance $D_L$}: uniform in \emph{linear}
distance over $D_L \in [50, 1000]\,\mathrm{Mpc}$, i.e.\ $p(D_L) =
\mathrm{const}$. We verified that adopting a log-uniform prior for $D_{\rm L}$ (i.e., $p(\log D_{\rm L}) = \text{const}$) 
instead of a linear prior systematically shifts the posterior median toward smaller distances by 20--30\%, 
reflecting the prior's preference for lower distance scales in logarithmic space.
% simplicity, rather than a log-uniform prior (as used for $\mathcal{M}_c$)
% or the Euclidean-volume prior $p(D_L)\propto D_L^2$ common in
% gravitational-wave parameter estimation. We note that none of these three
% choices is uniquely "non-informative": a Jeffreys-type argument for the
% amplitude-scaling parameter $D_L$ (which enters the waveform only through
% an overall $1/D_L$ factor) would instead favour a log-uniform prior,
% similar to $\mathcal{M}_c$. The choice of prior on $D_L$ therefore reflects
% an explicit modelling decision rather than an automatic non-informative
% default, and we caution that at low SNR the recovered $D_L$ posterior can
% be prior-dominated (see Appendix~\ref{app:distance}); we verify this
% sensitivity by repeating a representative case under a log-uniform $D_L$
% prior in Appendix~\ref{app:distance}.
\end{itemize}

They are also summarised in Table~\ref{tab:priors_2d}. With this machinery in hands, we sample the two-dimensional $(\mathcal{M}_c, D_L)$
posterior using the nested sampling algorithm
\textsc{dynesty}~\cite{Speagle:2019ivv} (with $N_{\mathrm{live}}=400$, multi-ellipsoid
 bounds, random-walk sampling, $\Delta\ln Z = 0.1$), which simultaneously returns the
Bayesian evidence $\ln\mathcal{Z}$. As an example of typical posterio we obtain from the analysis, we present in Fig.\ref{fig:corner_example} the corner plot of the runs $(\mathcal M_c, D_L)=(1.68 \times 10^8, 100)$, as an example a very loud signal ($\chi^2 \sim 4000$), and $(\mathcal M_c, D_L)=(1.68 \times 10^8, 500)$, as an example of a weaker signal ($\chi^2 \sim 200$).

\begin{figure*}[t]
    \centering
    \includegraphics[width=0.8\linewidth]{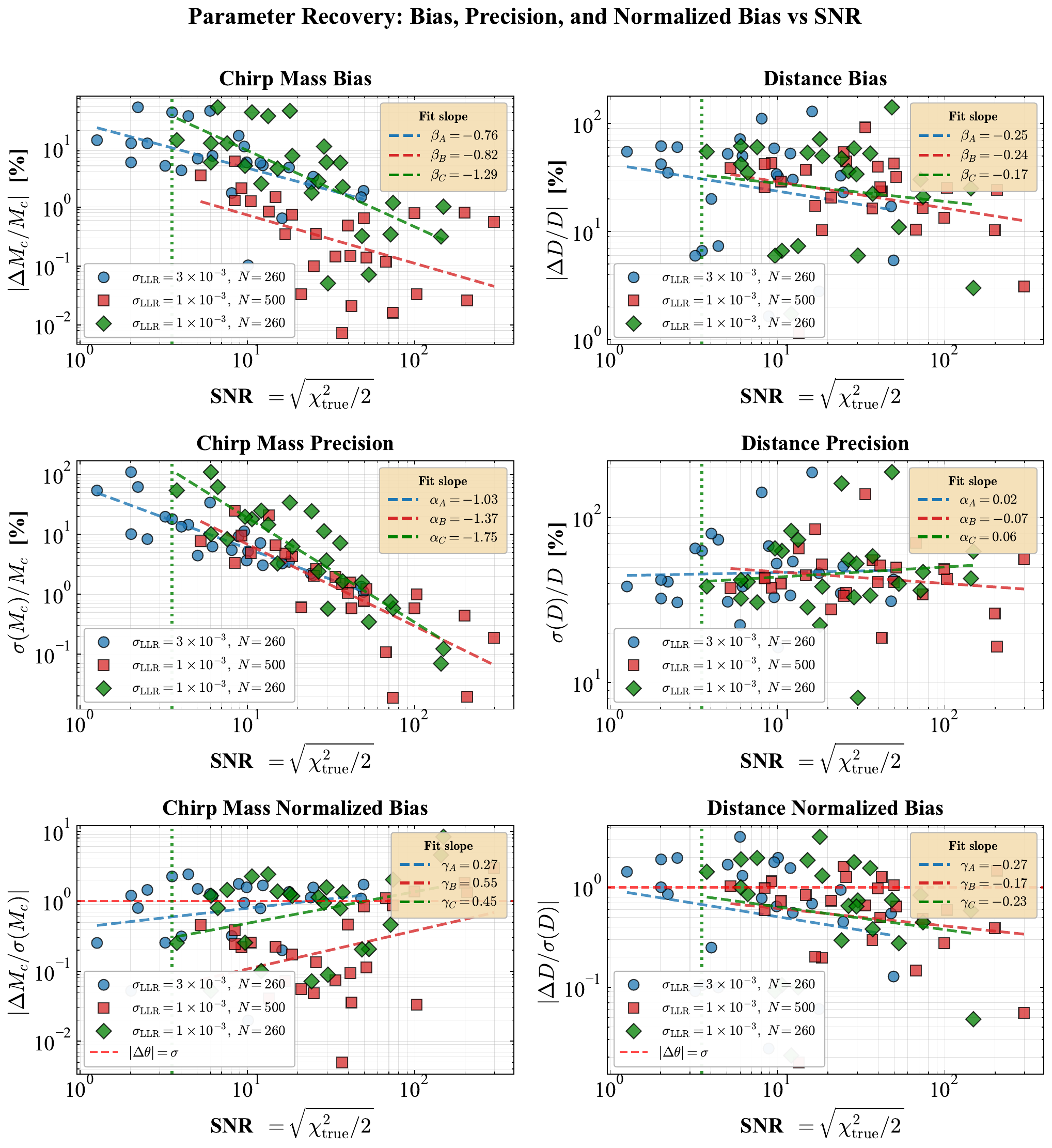}
    \caption{
        \textbf{Quality of the parameter recovery with LLR.} 
This figures shows the recovery of the chirp mass and the luminosity distance as a function of SNR$=\sqrt{\chi^2/2}$. The left panel shows the normalised fractional error (the best estimate compared the injected parameter) and the right panel shows the normalised uncertainty bands. The meaning of the y-axis are explained in Eq.\eqref{eq:bias_frac} and around it. 
The \textbf{lower row} presents the ratio of the two first panels. 
The scan covers a grid of injected source configurations spanning true
distances $D_{\mathrm{true}} \in \{100, 200, 400, 500, 800\}\,\mathrm{Mpc}$
and chirp mass $\mathcal{M}_c \in \{9.4, 14.0, 17.0, 32.0, 46.8\}\times 10^{7}\,M_\odot$, each observed for two different realisations of the LLR experiment: A) over a modest set-up
$15$-year baseline at $260$ measurements per year with a per-measurements LLR uncertainty of
 $\sigma_{\mathrm{LLR}} = 3\,\mathrm{mm}$ and B) over an optimistic set-up
$15$-year baseline at $500$ measurements per year with a per-measurements LLR uncertainty of
$\sigma_{\mathrm{LLR}} = 1\,\mathrm{mm}$ and finally an intermediate set-up C) $15$-year baseline at $260$ measurements per year with a per-measurements LLR uncertainty of
 $\sigma_{\mathrm{LLR}} = 1 \,\mathrm{mm}$. The green vertical dashed line presents the detectability boundary at $5$ sigma, $\chi^2 =25$.
    \label{fig:stat_analysis_1}
    }
\end{figure*}

% \begin{table}[h]
% \centering
% \caption{Prior distributions used in the 2D $(\mathcal{M}_c, D_L)$ nested
% sampling analysis. Sky/orientation angles $(\theta, \phi, \iota, \psi)$ are
% not sampled: they are marginalised deterministically via a fixed Sobol
% quasi-Monte-Carlo point set drawn once from the isotropic prior below.}
% \label{tab:priors_2d}
% \begin{tabular}{lccc}
% \hline\hline
% Parameter & Prior & Range & Notes \\
% \hline
% $\mathcal{M}_c$ & Log-uniform & $[10^{5},\,10^{7}]\,M_\odot$ & sampled in $\log_{10}\mathcal{M}_c$ \\
% $D_L$           & Uniform     & $[50,\,1000]\,\mathrm{Mpc}$   & sampled in $D_L$, stored as $\log_{10} D_L$ \\
% \hline
% $\cos\theta$    & Uniform     & $[-1,\,1]$                    & fixed Sobol set, $N_s = 1028$ \\
% $\phi$          & Uniform     & $[0,\,2\pi]$                  & fixed Sobol set \\
% $\cos\iota$     & Uniform     & $[-1,\,1]$                    & fixed Sobol set \\
% $\psi$          & Uniform     & $[0,\,\pi]$                   & fixed Sobol set \\
% \hline\hline
% \end{tabular}
% \end{table}

\section{Recoverability of the parameters via LLR}
\label{sec:2d_analysis}

To assess how parameter recovery quality scales with signal strength, we perform 
a systematic grid scan over the two parameters that most directly control the 
injection SNR: the chirp mass $\mathcal M_c$ and the luminosity distance 
$D_L$. The angular parameters of the source are held constant throughout 
the scan: the sky position is fixed at $(\theta, \phi)_{\rm true} = (30^\circ, 80^\circ)$, 
the inclination of the compact binary at $\mathcal I_{\rm true} = 50^\circ$, and the 
polarisation angle at $\psi_{\rm true} = 60^\circ$. 
The scan covers a grid of injected source configurations spanning true
distances $D_{\mathrm{true}} \in \{100, 200, 400, 500, 800\}\,\mathrm{Mpc}$
and $\mathcal{M}_c \in \{9.4, 14.0, 17.0, 32.0, 46.8\}\times 10^{7}\,M_\odot$, each observed for three different realisations of the LLR experiment:
\begin{itemize}
    \item A) over a modest set-up
$15$-year baseline at $260$ measurements per year with a per-measurements LLR uncertainty of
 $\sigma_{\mathrm{LLR}} = 3\,\mathrm{mm}$ (blue in the plots)
 \item B) over an optimistic set-up
$15$-year baseline at $500$ measurements per year with a per-measurements LLR uncertainty of
$\sigma_{\mathrm{LLR}} = 1\,\mathrm{mm}$ (red in the plots)
\item  and finally an intermediate C) $260$ measurements per year with a per-measurements LLR uncertainty of
$\sigma_{\mathrm{LLR}} = 1\,\mathrm{mm}$.
\end{itemize}

Notice nonetheless that all these set-ups, including the A) (the modest set up),  assume crucial improvements in the moon trajectory modelling with respect to the current uncertainty being around $1-2$ cm (see Appendix \ref{app:llr} for a concise presentation of the different noise affecting the detection and the expected uncertainties).

On fig.\ref{fig:stat_analysis_1} and fig.\ref{fig:stat_analysis_2}, we present the global recovery analysis. On the first row, we always presents the fractional error between the best fit and the true parameter (which we call bias), 
\begin{equation}
    \label{eq:bias_frac_1}
   \text{Accuracy}: \quad \Delta F \equiv F_{\rm true} - F_{\text{best fit}}\, , 
\end{equation}

and the second row presents the fractional precision defined as
\begin{equation}
    \label{eq:bias_frac}
    \text{precision}: \quad \sigma (F) \equiv \frac{1}{2}\left(F_{\rm 84\%} - F_{\rm 16\%}\right) \, . 
\end{equation}

In short, the \emph{fractional error} (left panels)
quantifies the systematic offset between the posterior median and the true
value, while the \emph{precision} (right panels) quantifies the width of the
posterior, i.e., the statistical uncertainty on the recovered parameter.

 The modest set up A) is presented with blue circles, the optimistic set-up B) is presented with red square and the intermediary set up C) is presented with green diamond. The complete output of the bayesian analysis are presented in the Table \ref{tab:summary_A},\ref{tab:summary_B} and \ref{tab:summary_C} for the three respective data sets. 

\paragraph{Build a faithful proxy}
Our first goal is to build a consistent proxy of the quality of the parameter recovery. We build the two following proxies 
\begin{itemize}
    \item The peak of the SNR 
    \begin{equation}
\mathrm{SNR}_{\rm peak}
= \frac{\max_i \bigl| \delta r_{\rm clean}(t_i) \bigr|}{\sigma_{\rm LLR}}
\end{equation}
\item The cumulative $\chi^2$ statistics evaluated at the merger
\be
\chi^2\equiv \text{Max}[\chi^2(t)]=\text{Max} \bigg[\sum_{t_i<t}\!\left[\frac{\delta r(t_i)}{\sigma}\right]^2 \bigg] \,.
\ee
\end{itemize}

We will show, in what follows and more precisely in appendix \ref{app:two_proxies}, that these two measures present good convergence for the chirp mass $\mathcal M_c$ but much poorer for the luminosity distance. 

From fig.\ref{fig:stat_analysis_1} and fig.\ref{fig:stat_analysis_2}, we observe that both SNR proxies yield statistically indistinguishable residual scatter
across all four recovery metrics (differences $\lesssim0.03$~dex, smaller
than the bootstrap uncertainty), confirming that $\mathrm{SNR}_{\rm peak}$
and $\sqrt{\chi^2_{\rm true}/2}$ are comparably good predictors of
parameter recovery quality, as explained in \ref{app:two_proxies}. However, for chirp-mass precision both
proxies require a significant offset ($\sim2$--$2.5\sigma$) between the
$\sigma_{\rm LLR}=3\times10^{-3}$mm configuration and the
$\sigma_{\rm LLR}=1\times10^{-3}$mm configurations, showing that neither LLR
definition alone fully absorbs the dependence on measurement noise and
cadence.

\subsection{Parameter recovery: LLR}
%------------------------------------------------------------------
\paragraph{Recovery of the Chirp Mass}
%------------------------------------------------------------------
We start by focusing on the  first column of Fig.\ref{fig:stat_analysis_1} and Fig.\ref{fig:stat_analysis_2}, where we present the fractional error the fractional half-width and the ratio of the two of the chirp mass.
The chirp mass $\mathcal{M}_c$ is the best-constrained parameter of the waveform, 
as it governs the leading-order phase evolution and is therefore measured 
independently of the amplitude.

In general, the chirp mass ($\mathcal{M}_c$) recovery exhibits excellent SNR scaling in all datasets. Firstly, bias convergence is rapid: $|\Delta\mathcal{M}_c/\mathcal{M}_c| \propto \mathrm{SNR}^{\beta}$ with $\beta \simeq -0.76$ to $-1.3$ (Dataset A: $\beta = -0.76$, Dataset B: $\beta = -0.82$, Dataset C: $\beta = -1.3$). Perhaps surprisingly, the datasets B presents a much smaller fractional error, even at equal $\text{SNR}$\footnote{Notice that the SNR in principle encompasses the information on the number of measurements $N$ and precision $\sigma$.}, than the datasets A, by a factor of around a factor 10. The datasets C shows a trend closer to dataset A, despite the increase in precision $\sigma_A = 3 \times 10^{-3}$ while $\sigma_C = 10^{-3}$. This indicates that the chirp mass recovery precision benefits more from a larger number of measurements per year $N$, than from a better improvement of the the individual measurements ($\sigma$).

The precision of the recovery follows similar successful trends: SNR precision scales as $\sigma(\mathcal{M}_c) \propto \mathrm{SNR}^{\alpha}$ with $\alpha \simeq -1.0$ to $-1.7$ (Dataset A: $\alpha = -1.03$; Dataset B: $\alpha = -1.37$; Dataset C: $\alpha = -1.75$). In the case of precision, there is no striking difference between the three datasets, and the $\chi^2$ is a good measure. 

However, we also observe, referring at the tables \ref{tab:summary_A}, \ref{tab:summary_B} and \ref{tab:summary_C}, that the sampler is \emph{overconfident}, i.e. the fractional error exceeds the fraction half-width, in around 80 \% of the runs for the modest datasets A, 30 \% for the optimistic dataset B and 70 \% for the intermediate dataset C. This shows that the datasets A and C presents an overabundance of overconfident points (which we would expect around 32 \%). 

The $\mathrm{SNR}_{\rm peak}$ parametrization yields slightly steeper fractional error scaling ($\beta \simeq -0.72$ to $-1.4$), reflecting the compressed dynamic range of peak SNR; precision scaling remains similar ($\alpha \simeq -1.1$ to $-1.6$). Scatter in the precision law ($\sigma_{\log_{10}} \simeq 0.26$--$0.47$ in log-space) is modest and consistent across both SNR definitions, indicating that $\mathcal{M}_c$ follows a tight power law with minimal outliers. 

%------------------------------------------------------------------
\paragraph{Recovery of the Luminosity Distance}
%------------------------------------------------------------------

Let us now focus on the  second column of Fig.\ref{fig:stat_analysis_1} and Fig.\ref{fig:stat_analysis_2}, where we present the fractional error and the fractional half-width and the ratio of the two of the luminosity distance. The luminosity distance $D_L$ exhibits a markedly more complex recovery, reflecting 
the fundamental degeneracy between $D_L$ and the four angles present in the problem.

 The fractional error
reaches $-10\%$ to $-100\%$ for all SNR, while the posterior half-width remains always  
in the range $\sim 20$--$80\%$, with no particular convergence with larger SNR (cumulative or peaked): having louder signal does not improve the recovery if we have to marginalize over angles. 

 This indicates that the 
\textit{precision} and the \textit{accuracy} of the distance measurement is not set by the noise level 
but by the irreducible width of the $D \leftrightarrow \text{angles}$ degeneracy valley. 
Higher SNR sharpens the likelihood along the valley floor but cannot break the 
degeneracy itself. This behaviour is qualitatively different from the chirp 
mass case and underlines the importance of reporting both the fractional error and the 
posterior width when assessing parameter recovery in gravitational-wave inference.

\section{Parameter recovery via SLR: The GUEST experiment}
\label{sec:recovery_guest}
We now turn to a similar analysis of parameter reconstruction using the proposed two-satellites detector GUEST (Gravitational Universe Exploration with Satellite Tracking)\cite{Blas:2026xol}.
The GUEST mission concept relies on two satellites placed in highly eccentric orbits.
The orbital parameters follow from four requirements: (i) a large eccentricity
to accelerate the quadratic growth of the GW-induced effect and to excite higher
resonances; (ii) a long orbital period to reach the frequencies in the $\mu$Hz
gap; (iii) sufficient visibility and SLR reach from the ground stations; and
(iv) minimisation of systematic accelerations (e.g.\ solar radiation pressure,
Earth albedo, thermal effects) that could complicate the
extraction of the GW signal. The two orbits share the same shape but differ in
plane orientation, which decorrelates non-GW effects and provides more complete
sky coverage. The main orbital and mission parameters are summarized in
Table~\ref{tab:guest_params}.

\begin{figure}[t]
    \centering
    \includegraphics[width=0.45\columnwidth]{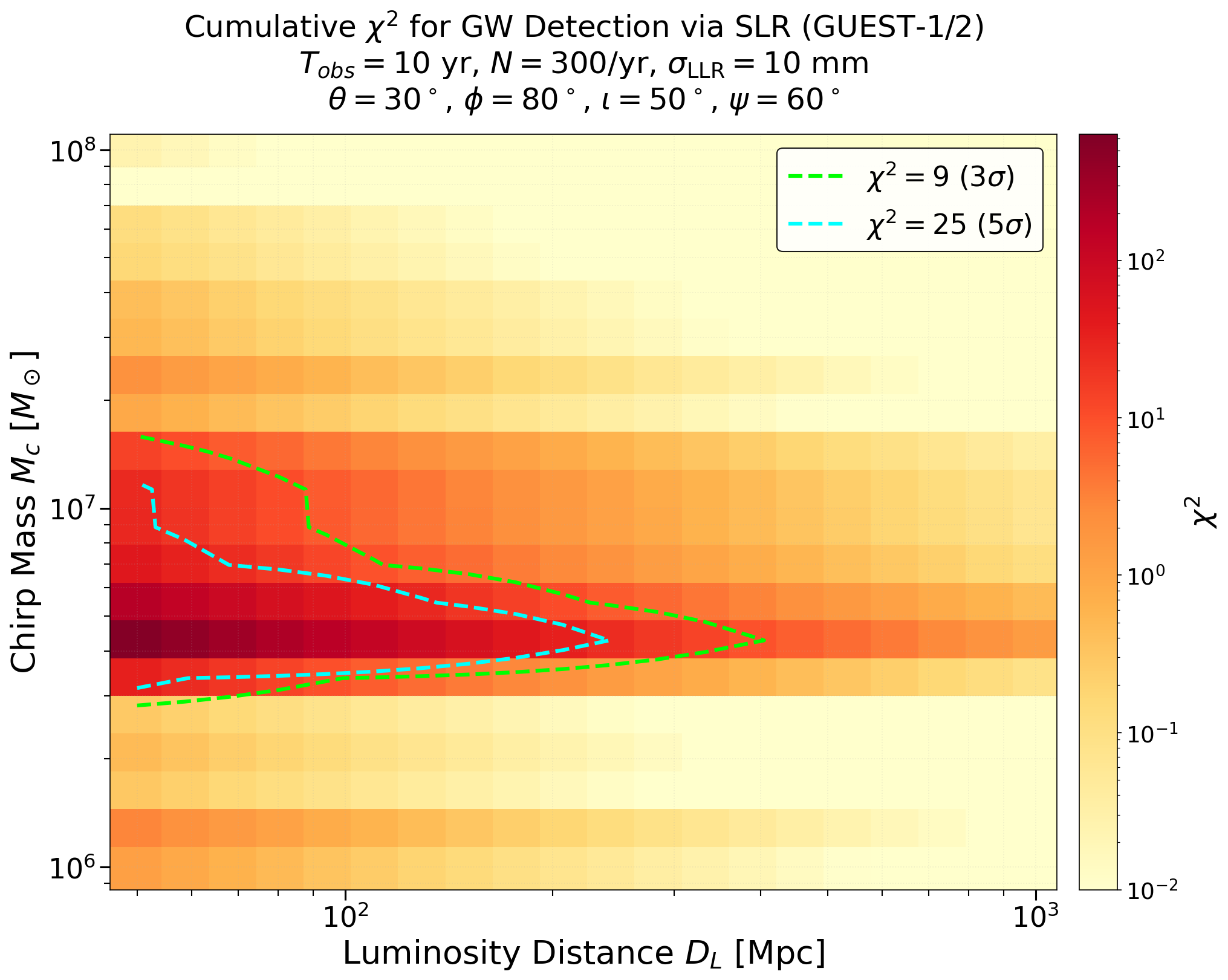}
    \includegraphics[width=0.45\columnwidth]{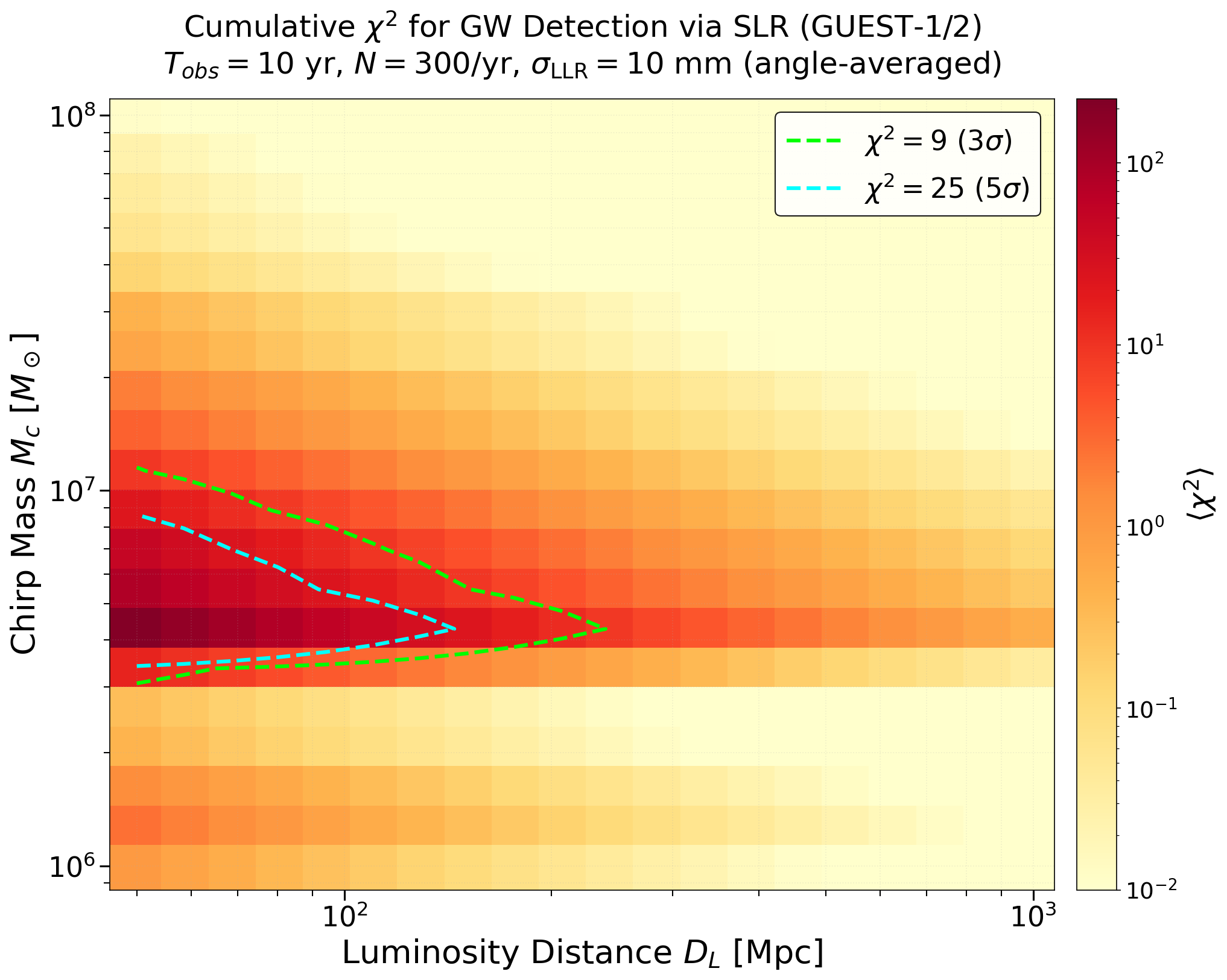}

    \caption{
        \textbf{GUEST Cumulative detection statistic $\chi^2$ (Inspiral only)}. We use the cumulative statistics defined in Eq.\eqref{eq:chi2Guest}. Results assume an observation baseline of $T_{\rm obs}=10\,$yr
        with the binary merger occurring at the end of the window, a ranging
        precision of $\sigma_{\rm GUEST}=10\,$mm. The \textbf{left panel} presents the heatmap with a fixed representative source
        orientation.
        Dashed
        contours mark the nominal $3\sigma$ ($\chi^2=9$) and $5\sigma$
        ($\chi^2=25$) detection thresholds, delineating the region of the
        parameter space accessible to the constellation.% 
        The \textbf{right panel} presents the same heatmap but with an angle averaged cumulative statistics.
    }
    \label{fig:chi2_heatmap_guest}
\end{figure}

\begin{table}[htbp]
\centering
\caption{Nominal GUEST orbital and spacecraft parameters (see \cite{Blas:2026xol}).}
\label{tab:guest_params}
\begin{tabular}{lcc}
\hline\hline
\textbf{Orbital Parameter} & \textbf{GUEST-1} & \textbf{GUEST-2} \\
\hline
Semi-major axis [km]        & \multicolumn{2}{c}{$53\,188$} \\
Perigee altitude [km]       & \multicolumn{2}{c}{$6\,919$} \\
Apogee altitude [km]        & \multicolumn{2}{c}{$86\,701$} \\
Eccentricity $e_0$          & \multicolumn{2}{c}{$0.75$} \\
Orbital period $P$ [h]      & \multicolumn{2}{c}{$33.8$} \\
Argument of perigee [deg]   & \multicolumn{2}{c}{$270$} \\
True anomaly [deg]          & \multicolumn{2}{c}{$0$} \\
Inclination $i$ [deg]       & $75$   & $50$ \\
RAAN [deg]                  & $0$    & $220$ \\
Approx.\ re-entry [years]   & $29.5$ & $32.4$ \\
\hline\hline
\end{tabular}
\end{table}

Notice that \cite{Blas:2026xol} studied two approaches of the data analysis, which they referred to as ``with \emph{arcs}'' (pessimistic) and ``without \emph{arcs}'' (optimistic). These two scenarios are related to the time between two successive recalibrations of the satellite trajectory. In this paper, we will assume that the GW signal induces a deviation from the Keplerian orbit, this approach is a simplification of the ``without \emph{arc}'' approach. 

It is instructive to contrast GUEST with LAGEOS, the workhorse of satellite
laser ranging (SLR) for tests of gravity. The two LAGEOS satellites (launched in
1976 and 1992) are passive, dense, retroreflector-covered spheres in nearly
circular medium-Earth orbits, with $a \simeq 12\,200$~km, $e \lesssim
5\times10^{-3}$ and periods of about $3.8$~h; their inclinations
($109.8^\circ$ and $52.6^\circ$) were chosen so that suitable combinations of
the nodal and perigee rates cancel the leading even zonal harmonics of the
geopotential, enabling the measurement of the Lense--Thirring precession
\cite{Ciufolini:2004rq,Ciufolini:2016ntr}. Their design
philosophy is that of an ideal passive test particle observed over a very long
baseline: a small area-to-mass ratio suppresses non-gravitational
accelerations, and the absence of active subsystems has permitted an
essentially uninterrupted ranging record spanning nearly five decades.

GUEST departs from this paradigm in a way dictated by the target signal. Since
the GW-induced perturbation grows quadratically in time and is resonantly
enhanced at integer multiples of the orbital frequency, the figure of merit is
not the suppression of surface forces at a single radius but rather a long
orbital period combined with a large eccentricity. With $a \simeq 53\,188$~km,
$e_0 = 0.75$ and $P = 33.8$~h, GUEST reaches fundamental frequencies an order of
magnitude below LAGEOS, deep in the $\mu$Hz band, while the large eccentricity
populates a dense spectrum of higher harmonics inaccessible to a near-circular
orbit; the satellite also spends most of each orbit far from the Earth, where
drag, albedo and thermal thrust are strongly suppressed.

As illustration, on Fig.\ref{fig:stoc_sens}, in green, assuming the error $\sigma_{\rm GUEST} = 10$ mm, we show the sensitivity of the GUEST experiment to a constant GW signal originating from the zenith for a period of observation of $10$ years.

\subsection{Detectability of binary inspirals at the GUEST experiment}

Figure~\ref{fig:chi2_heatmap_guest} maps the detectability of a coalescing
massive-black-hole binary through its imprint on the earth-satellite ranging
residuals of the \textsc{Guest-1}/\textsc{Guest-2} constellation. The statistic is defined as
\begin{equation}
\label{eq:chi2Guest}
  \text{(Cumulative statistics for GUEST)} \qquad \qquad  \chi^2(t)=\sum_{a}\sum_{t_i<t, a}\!\left[\frac{\delta r(t_i)}{\sigma_{\rm GUEST}}\right]^2\, ,
\end{equation}
        summed over both satellites $a$ ($a$ ranges over  \textsc{Guest-1} and \textsc{Guest-2}) and accumulated over all observation
        epochs $t_i$, where $\delta r_a$ is the GW-induced range perturbation
        obtained from the variational integration of the geodesic-deviation
        equation. 
        To summarize, for each point
in the $(\mathcal{M}_c, D_L)$ plane we integrate the GW-perturbed relative
dynamics of each satellite over a $10$-yr baseline terminating at merger, and
accumulate the resulting range perturbations into a single quadratic detection
statistic $\chi^2$ defined above\footnote{Let us again emphasize that this plot only take into account the inspiral phase of the dynamics of the binary. It is thus not in contradiction with Fig. 5 of \cite{Blas:2026xol}},
producing the characteristic gradient seen in the heatmap.  The $3\sigma$ and
$5\sigma$ contours quantify the reach of the constellation: sources with
$\mathcal{M}_c \gtrsim \text{few} \times 10^{6}\,M_\odot$ remain detectable out to
$D_L \sim \mathcal{O}(10^2)\,$Mpc, whereas lighter systems can only be detectable if they merge in the local universe. This demonstrates that a two-satellite SLR configuration
provides a genuine, if modest, sensitivity window to low-frequency GWs in the
$\mu$Hz band complementary to space-borne interferometers (see \cite{Blas:2026xol} for further details).

\begin{figure*}[t]
    \centering
    \includegraphics[width=0.48\linewidth]{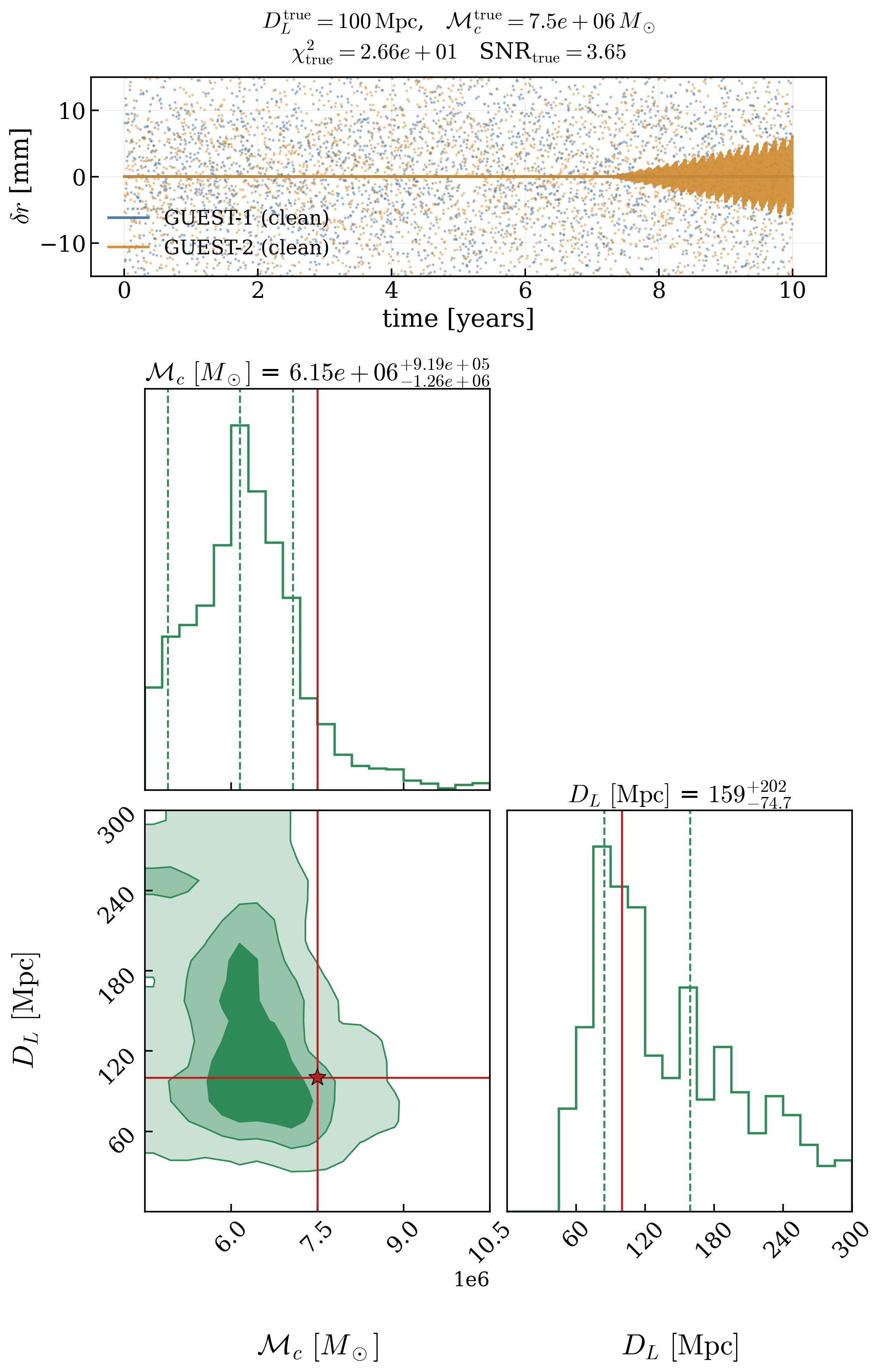}
    \includegraphics[width=0.48\linewidth]{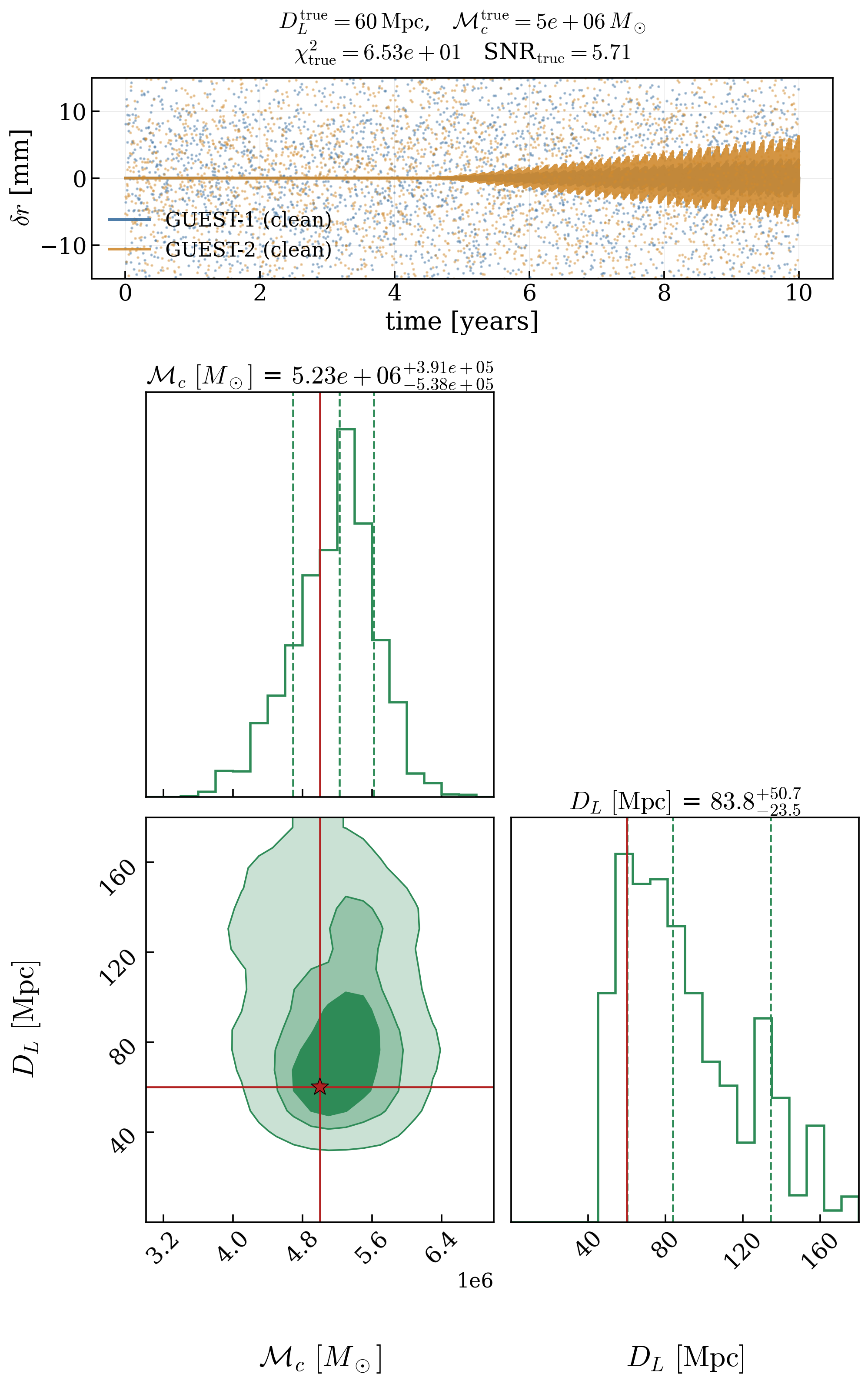}
    \caption{\textbf{Bayesian analysis posteriors for the GUEST constellation}.
        Corner plot of the 2-dimensional posterior obtained with 
        \textsc{Dynesty}\cite{Speagle:2019ivv} for a representative injection. Diagonal panels show 
        the marginalised one-dimensional posteriors; off-diagonal panels show 
        the two-dimensional marginalised posteriors. Green lines mark the 16th, 50th, and 84th 
        percentiles respectively; the red lines and dots indicate the true injected values. 
        The best fit to the noisy data and 68\% credible intervals are quoted above each 
        diagonal panel.
    }
    \label{fig:corner_example_GUEST}
\end{figure*}

\subsection{Parameter recovery with the GUEST constellation}

\begin{figure*}[t]
    \centering
    \includegraphics[width=0.98\textwidth]{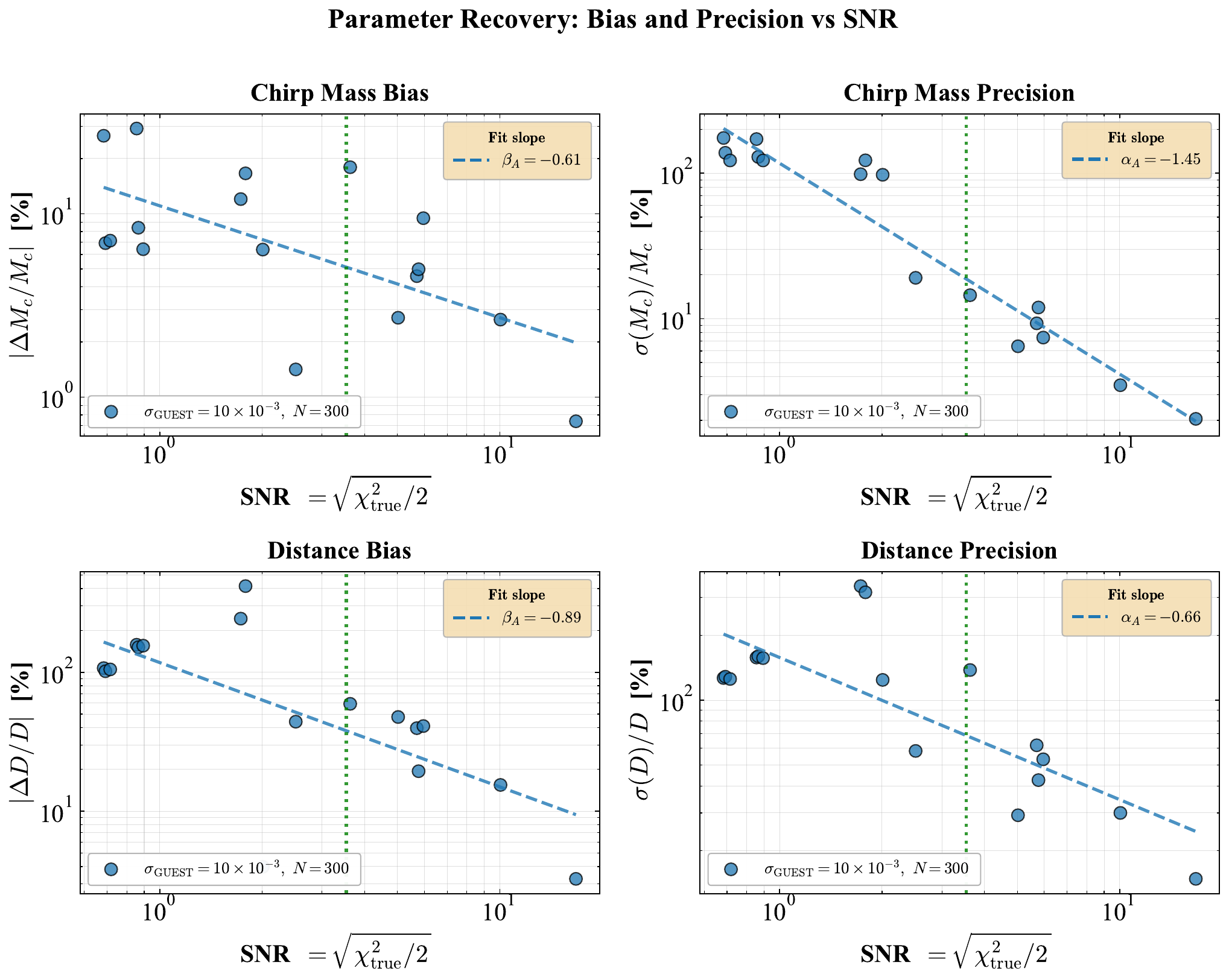}
    \caption{\textbf{Quality of the parameter recovery with LLR.} 
        Parameter-recovery performance of the two-dimensional Bayesian
        inference as a function of the injected signal-to-noise ratio (SNR),
        for the \textsc{Guest-1}/\textsc{Guest-2} SLR constellation. Each point
        corresponds to one injection in the $(\mathcal{M}_c, D_L)$ scan.
        \emph{Left column:} fractional recovery error of the chirp mass (top)
        and luminosity distance (bottom), defined in Eq.\eqref{eq:bias_frac_1}. \emph{Right column:} the corresponding relative precision
        defined in Eq.\eqref{eq:bias_frac}, shown on log--log axes. Dashed
        lines are power-law fits $\sigma/X \propto \mathrm{SNR}^{\alpha}$, with
        the fitted exponent $\alpha$ indicated in the legend. The green dotted
        vertical line marks the nominal detectability threshold $5\sigma$.%
    }
    \label{fig:bias_precision_vs_snr_GUEST}
\end{figure*}

We now present the result of the recovery of parameters for the GUEST configuration presented above. As for the LLR case, we generate a synthetic response of the two satellites and add Gaussian noise on top of it. We then perform a Bayesian analysis on the noisy data after performing a marginalisation over the four angles ($\theta$, $\phi$, $\mathcal I$, $\psi$) using a Sobol quasi-Monte Carlo point set of $N_{\rm Sobol} = 1024$ samples as in the LLR case (more details on the marginalisation are provided in Appendix \ref{app:bias_study}). For all simulations, the true source angles are fixed to $\theta_s = 30°$, $\phi_s = 80°$, $\mathcal I_{\rm gw} = 50°$, $\psi_{\rm gw} = 60°$ (representing a fiducial sky position and binary orientation), and the observations span $T_{\rm obs} = 10$ years with a cadence of $300$ observations per year and noise level $\sigma_{\rm GUEST} = 10$ mm. The nested sampling inference is performed with $N_{\rm live} = 200$ live points using a multi-ellipsoid bound with random walk sampling.

On Fig.~\ref{fig:corner_example_GUEST}, we present the corner plots of two selected Bayesian analyses in the space $(\mathcal M_c-D_L)$. A crucial difference, which will be systematically confirmed in the larger analysis, concerns the recovery of the luminosity distance. The GUEST recovery of the luminosity distance, at equal SNR or $\chi^2$, is much better than from the LLR experiment. Moreover, as we will see, the precision and accuracy of the $D_L$ recovery converges monotonically with larger SNR.

We systematically scan the parameter space over a grid of $(D_L, \mathcal M_c)$ values: 
\begin{equation}
D_L \in \{60,100, 200, 400, 500\} \text{ Mpc} \quad \text{and} \quad \mathcal M_c \in \{0.4, 0.5, 0.6, 0.7\} \times 10^7 \, M_\odot\,.
\end{equation}

Figure~\ref{fig:bias_precision_vs_snr_GUEST} and Table \ref{tab:summary_AA} summarise the statistical
performance of the parameters reconstruction across the injected grid,
isolating the two astrophysically dominant parameters: the chirp mass
$\mathcal{M}_c$ and the luminosity distance $D_L$. Two complementary metrics
are shown as functions of the injected SNR. 

\subsection{Parameter recovery: SLR}
\paragraph{Chirp mass}
For the chirp mass, both the fractional error and the precision decrease steadily with
increasing SNR, following approximately power-law trends
$|\Delta \mathcal{M}_c / \mathcal{M}_c| \propto \mathrm{SNR}^{\beta_A}$ and
$\sigma(\mathcal{M}_c)/\mathcal{M}_c \propto \mathrm{SNR}^{\alpha_A}$, with
best-fit slopes $\beta_A = -1.95$ and $\alpha_A = -1.34$ respectively. The
fractional error falls from $\sim 20\%$ at the lowest injected SNR ($\mathrm{SNR}\sim 4$)
down to $\sim 1\%$ at the highest SNR probed ($\mathrm{SNR}\sim 20$), while
the precision improves from $\sim 15\%$ to $\sim 2\%$ over the same range.

\paragraph{Luminosity distance}

A very interesting peculiarity of the GUEST configuration with respect to
the LLR case resides in the reconstruction of the source distance. As we
can observe on the lower panels of Figure~\ref{fig:bias_precision_vs_snr_GUEST},
the precision and accuracy of the luminosity distance improve dramatically
with larger SNR, following best-fit power-law slopes of $\beta_A = -1.88$ for
the bias and $\alpha_A = -1.44$ for the precision. Concretely, the distance
bias drops from $\sim 50\%$ at $\mathrm{SNR}\sim 4$ to $\sim 12\%$ at
$\mathrm{SNR}\sim 20$, while the precision improves from $\gtrsim 100\%$ down
to $\sim 20\%$ over the same SNR range.

However, we notive that for more than 80 \% of the points, the sampler is overconfident in at least one of the two parameters. 
% \subsection*{Comparison between LLR and GUEST}

% Before closing this section, we propose an analysis of the differences
% between the parameter reconstruction with LLR and with GUEST.

% The first observation, obtained by comparing Fig.~\ref{fig:stat_analysis_1}
% and Fig.~\ref{fig:bias_precision_vs_snr_GUEST}, or directly by inspecting
% Fig.~\ref{fig:GUESTandLLR}, is that the precision and accuracy of the chirp
% mass recovery with LLR and with GUEST, at a \emph{given} SNR value, are
% extremely close. This result provides a further argument for the robustness
% of the $\chi^2$ observable as a characterisation of the detectability of a
% signal, independently of the specific experimental configuration.

% The main difference between the two experiments lies in the luminosity
% distance recovery. A second observation is the marked improvement, and
% eventual convergence, of the luminosity distance recovery in the GUEST case
% relative to LLR. We attribute this success to the fact that the GUEST
% experiment relies on two satellites, which permits a better reconstruction
% of the angular parameters --- both those characterising the source
% orientation and those describing its position on the sky --- thereby
% tempering the well-known degeneracy between these angles and the luminosity
% distance that otherwise limits the single-baseline LLR configuration.

\begin{figure*}[t]
    \centering
    \includegraphics[width=0.98\textwidth]{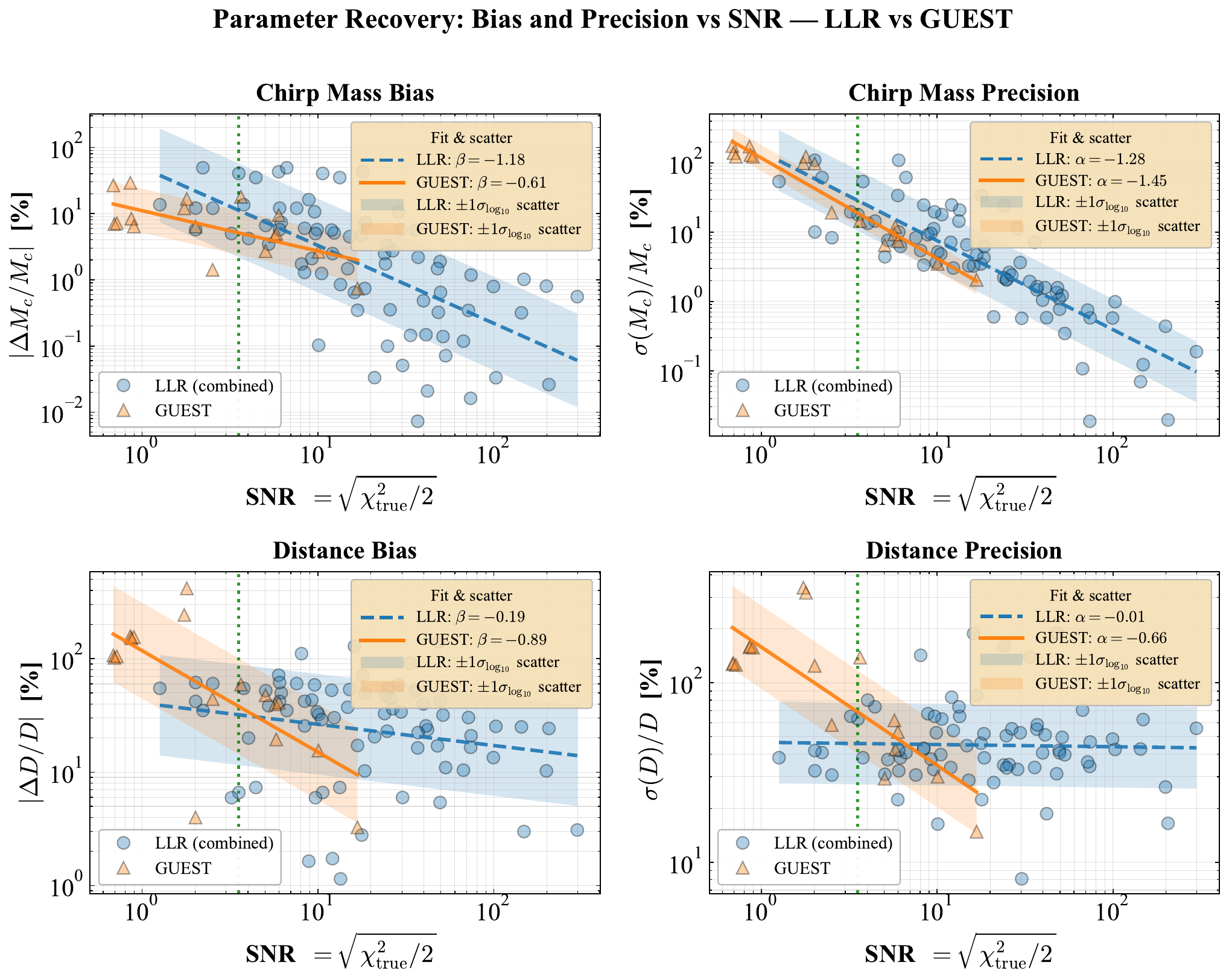}
    \caption{\textbf{Close comparison between the recovery of the parameters with LLR and with GUEST.} We aggregate here the three LLR datasets (in blue circles) and compare their recovery of the parameters with the GUEST dataset (orange triangles). The meaning of the y-axis are explained in Eq.\eqref{eq:bias_frac} and around it. Even though the recovery of the chirp mass is similar for the LLR experiment and GUEST, and converge to a larger precision and accuracy for louder signals, the situation is totally different for the recovery of the distance, which was largely not improving for LLR and improves steadily in the case of GUEST. This much better behaviour is traced to the better angular recovery of the 2-satellite GUEST constellation. As before, the vertical green dashed line denotes the $5-\sigma$ detection criterion and separate ``detectable'' from ``non-detectable'' signals. 
    }
    \label{fig:GUESTandLLR}
\end{figure*}

\subsection{Comparison with the recovery capabilities of the LLR experiment}

Before closing this section, we propose an analysis of the differences
between the parameter reconstruction with LLR and with GUEST.

The first observation, obtained by inspecting
Fig.~\ref{fig:GUESTandLLR}, is that the precision and accuracy of the chirp
mass recovery with LLR and with GUEST are, at a \emph{given} SNR value,
extremely close. Over the range of SNR common to both configurations, the
GUEST points fall squarely within the scatter of the LLR population, and the
two sets of fitted trends decrease with SNR at comparable rates. In other
words, once the signal strength is expressed through the same observable,
the two experiments recover the chirp mass with essentially the same quality,
despite their rather different geometries and noise properties. 

The main difference between the two experiments lies instead in the
luminosity distance recovery. The lower panels of
Fig.~\ref{fig:GUESTandLLR} display a qualitatively different behaviour for
the two configurations: whereas the LLR bias and precision on $D_L$ remain
essentially flat as the SNR increases --- so that even the loudest events do
not yield a meaningfully better distance estimate than the faintest ones ---
the GUEST results decrease steeply and systematically with SNR, converging
towards a well-constrained distance measurement at the upper end of the
range probed. The LLR distance reconstruction therefore appears to be
limited not by the signal strength but by a structural degeneracy, which no
amount of additional SNR can lift; the GUEST configuration, by contrast,
behaves as one would expect from a genuinely informative measurement.

We impute this success to the fact that the GUEST experiment relies on two
satellites. The additional baseline allows for a better reconstruction of
the angular parameters --- both those characterising the source orientation
and those describing its position on the sky --- and thereby tempers the
degeneracy between these angles and the luminosity distance. Since $D_L$
enters the response only through the overall signal amplitude, in
combination with the inclination and polarization angles, breaking this
degeneracy is precisely what converts additional SNR into an improved
distance estimate. This is what is observed for GUEST and what remains out
of reach for the single-baseline LLR configuration.

\section{Discussion, limitations, conclusions and outlook}
\label{sec:conclusion}

In this paper, we have analysed the capabilities of concept experiments based
on Lunar Laser Ranging (LLR) and Satellite Laser Ranging (SLR) --- and more
specifically the two-satellite GUEST proposal \cite{Blas:2026xol} --- to
reconstruct the chirp mass $\mathcal{M}_c$ and the luminosity distance $D_L$
of detected binary merger events. These two parameters are of particular
interest as they are directly related to the black hole merger rate function,
whose reconstruction would allow one to constrain the massive black hole
population in a mass range that is complementary to both pulsar timing arrays
and space-borne interferometers.

For both the LLR and the GUEST configurations, we have simulated the clean
range perturbation $\delta r$ induced by an inspiralling binary, added
Gaussian white noise of amplitude $\sigma$ as a model of the instrumental
error, and performed a series of nested-sampling Bayesian analyses in the
two-dimensional space $(\mathcal{M}_c, D_L)$, marginalising over the four
 orientation angles. For the LLR case we considered three observational scenarios
spanning a range of cadences and per-measurement precisions, while for GUEST
we adopted as a representative model a $10$-year baseline with $300$
measurements per year and a per-measurement uncertainty of
$\sigma = 10\,\mathrm{mm}$.

\paragraph{Detectability and the role of $\chi^2$}

A first general result of our analysis is that the best predictor of the
quality of the parameter reconstruction is the cumulative displacement
statistic $\chi^2$ and the peak signal-to-noise ratio
$\mathrm{SNR}_{\rm peak}$. When the recovery performance is expressed as a
function of $\mathrm{SNR} = \sqrt{\chi^2/2}$, the scatter across injections is
markedly reduced, and --- remarkably --- the LLR and GUEST results collapse
onto similar trend for the chirp mass, despite the rather
different geometries, cadences and noise levels of the two configurations.
This universality provides a strong argument for the robustness of $\chi^2$ as
the natural characterisation of the detectability of a signal in ranging
experiments, and suggests that it can be used as a reliable forecasting tool
for other future ranging concepts.

In terms of the accessible parameter space, detectable events are concentrated
in the range $\mathcal{M}_c \in [10^8, 10^9]\,M_\odot$ for LLR and
$\mathcal{M}_c \in [\text{few}\times 10^6, 10^7]\,M_\odot$ for GUEST, in both
cases out to distances of almost 1 Gpc or slightly smaller.

\paragraph{Chirp mass}

The chirp mass is the best-constrained parameter of the waveform, as it governs
the leading-order phase evolution and is therefore measured largely
independently of the signal amplitude. For both experiments, the fractional
bias and precision on $\mathcal{M}_c$ decrease steadily with increasing SNR,
following approximate power laws $\sigma_{\mathcal{M}_c}/\mathcal{M}_c \propto
\mathrm{SNR}^{\alpha}$ with $\alpha \simeq -1$ to $-1.8$. In practice, a
marginally detectable event at the nominal $5\sigma$ threshold is recovered
with a precision and accuracy of order ten percent, while louder events
converge rapidly towards the percent level. For the LLR configurations, we
find that increasing the measurement cadence is more effective at improving
the chirp-mass precision than reducing the per-measurement noise --- a
practically relevant conclusion for the design of future ranging campaigns.

\paragraph{Luminosity distance}

The luminosity distance behaves very differently, and it is here that the two
experiments part ways. Since $D_L$ enters the response only through the
overall amplitude of the signal, in combination with the inclination and
polarization angles, its reconstruction is intrinsically vulnerable to
degeneracies with the source orientation. In the single-baseline LLR
configuration this degeneracy is essentially unbroken: the fractional bias and
precision on $D_L$ remain at the level of several tens of percent and, most
importantly, are found to be nearly \emph{independent} of the SNR. Even the
loudest events in our injection grid do not yield a meaningfully better
distance estimate than the faintest ones, indicating that the limitation is
structural rather than a matter of signal strength.

The GUEST configuration, by contrast, displays a qualitatively different
behaviour: both the bias and the half-width on $D_L$ decrease systematically
with SNR, converging towards a well-constrained distance measurement. We attribute this success to the presence of
two satellites with distinct orbital geometries, whose combined response
allows for a significantly better reconstruction of the angular parameters ---
both those characterizing the source orientation and those describing its sky
position --- thereby tempering the degeneracy that cripples the LLR case. This
observation carries a clear message for experimental design: breaking the
amplitude--angle degeneracy, and hence enabling any form of standard-siren
cosmology with ranging data, requires a constellation rather than a single
baseline.
\paragraph{What are our chances to detect a binary merger ?}

Finally, a crucial question coming with this study is: \emph{are there resolvable sources to begin with ?} A study of a population of BH mergers will be provided elsewhere \cite{Franciollini}. Even though the population of BH binaries is not yet properly constrained, astrophysical bounds and the PTA observations (see for example \cite{EPTA:2023xxk}) allows us to expect 
\begin{equation}
    \log_{10} \frac{\dot n_0}{\mathrm{Mpc}^3\,\mathrm{Gyr}} \in [10^{-6}, \text{few}]
\end{equation}
in the range of chirp mass $\mathcal M_c \in [10^6, 10^9] M_{\odot}$. Consequently, the number of mergers in a ball of radius $D_L$ is given by roughly 
\begin{align}
  N_{\rm merger} &= \frac{T_{\rm obs}}{\text{Gyr}} \, \frac{4\pi (D_L/\mathrm{Mpc})^3}{ 3} \int_{\log_{10}\mathcal{M}_{\rm lo}}^{\log_{10}\mathcal{M}_{\rm hi}}
\dot n(\mathcal{M})\big|_{d\log_{10}\mathcal{M}} \; d\log_{10}\mathcal{M}  
\\
&\approx \frac{T_{\rm obs}}{\text{Gyr}} \, \frac{4\pi (D_L/\mathrm{Mpc})^3}{ 3}\, \times  \dot n(\mathcal{M})\big|_{d\log_{10}\mathcal{M}} \,
\times \log_{10}\frac{\mathcal{M}_{\rm hi}}{\mathcal{M}_{\rm lo}} \, . 
\end{align}
This gives finally, looking back at figures \ref{fig:chi2_heatmap_guest} and \ref{fig:waveformheatmap},
\begin{align}
N^{\rm detectable}_{\rm merger} \approx
\begin{cases}
\displaystyle
10^{2}\times\dot n(\mathcal{M})\big|_{\log_{10}\mathcal{M} \sim 8} \;
& \text{(modest LLR}\text{)} \\[2ex]
\displaystyle
1 \times \dot n(\mathcal{M})\big|_{\log_{10}\mathcal{M}\sim 6.5}
& \text{(GUEST)} \, , 
\end{cases}
\label{eq:Nmerger-cases}
\end{align}

which can lead to order 1 detection depending on the largely unknown merger rate per year. 

\paragraph{Future studies and improvements of the pipeline}

Let us now throw in all the shortcomings of the study we presented so far, to moderate the enthusiasm:

\begin{itemize}
    \item \textbf{Realistic noise models and ideal Keplerian baseline.} In this study we assumed a Keplerian ideal trajectory for the unperturbed trajectory of the satellite. This approximation is similar to assuming that the unperturbed trajectory is perfectly modelled, which is unrealistic. In practice, to go beyond the ideal Keplerian motion, one would need to marginalize over the parameters describing the unperturbed trajectory (beyond the Keplerian parameters).
    
    Secondly, we assumed a purely Gaussian white
    noise for the detector, which is clearly too idealized. White Gaussian
    noise with a fixed $\sigma$ ignores what actually dominates LLR residuals:
    correlated (red) noise, station atmospheric delays, lunar libration
    modelling errors, and retroreflector degradation. Incorporating a realistic
    noise power spectral density would likely degrade the low-frequency
    sensitivity where the inspiral signal accumulates, and should be
    quantified.

    \item \textbf{More realistic waveforms and eccentricity.} We restricted
    ourselves to the 0PN chirping waveform and ignored the merger and ringdown
    phases. For $10^7$--$10^9\,M_\odot$ binaries observed over 15 years,
    higher-order post-Newtonian phasing corrections are non-negligible, and the
    systematic bias induced by truncating at Newtonian order should be
    quantified. Residual eccentricity, expected for massive black hole binaries
    in this frequency band, would further modify the phasing.

    \item \textbf{Resolving the sky location.} We focused on the chirp mass and
    the distance, as these are the most relevant parameters for black hole
    population studies and have the largest impact on the Earth--satellite
    system. A natural extension is to assess the recoverability of the angles
    themselves, in particular the sky localisation, which our results suggest
    is the key bottleneck for the distance measurement.

    \item \textbf{Full matched filtering and marginalisation over
    $t_{\rm merg}$.} In the current runs we assumed that the binary merges at
    the end of the observation window. A more careful analysis would build a
    full template bank and consistently apply matched filtering, marginalising
    over the coalescence time and phase --- for the latter, analytically, via
    the standard $I_0(R)$ phase-marginalised likelihood.

    \item \textbf{Population-level inference.} Ultimately, the goal is not the
    reconstruction of individual events but the inference of the underlying
    mass and redshift distribution. Propagating the per-event posteriors
    obtained here into a hierarchical Bayesian population analysis would allow
    one to forecast the actual constraining power of ranging experiments on the
    massive black hole merger rate.
\end{itemize}

\section*{Acknowledgments}

It is a pleasure to thank Diego Blas, Aurelien Hees, Sokratis Trifinopoulos and Yann Gouttenoire for helpful discussions (and motivational support!). Part of the codes used for this paper were written with the help of Claude, namely Claude OPUS 5. The author however bears the responsability of all the results presented. 

 MV is funded by the European Union (ERC, HoloGW, Grant Agreement No. 101141909). M.V. also acknowledges financial support from Grant CEX2024-001451-M funded by 
 \\
 MICIU/AEI/10.13039/501100011033, from Grant No. PID2022-136224NB-C22 from the Spanish Ministry of Science, Innovation and Universities, and from Grant No. 2021-SGR-872 funded by the Catalan Government.

\appendix

\section{Numerical solver for $\delta r_{\rm sat}$}
\label{app:solver}

For definiteness and reproducibility we summarise here the numerical
choices underlying our results.

\paragraph*{Reference orbit.} The unperturbed (zeroth-order) trajectory is
taken to be a pure Keplerian two-body orbit about the central mass. No
oblateness ($J_2$), third-body, tidal-dissipation or non-gravitational
perturbations are included at zeroth order. The reference orbit is
initialised from fixed osculating elements $(a,e,\iota,\Omega,\omega,M_0)$
and propagated independently of the gravitational wave, which enters only
through the linearised deviation $\delta\mathbf{r}$.

\paragraph*{Fixed orbital elements.} The orbital elements of the
Earth--Moon system are held fixed at their observed values and are
\emph{not} sampled: the Bayesian analysis varies only
$(\log_{10}\mathcal{M}_c,\log_{10}D_L)$, with the four orientation angles
$(\theta_s,\phi_s,\iota,\psi)$ marginalised over a Sobol quasi-Monte Carlo
point set (Appendix~\ref{app:bias_study}). This is a deliberate simplification.
In a realistic LLR and GUEST analysis the lunar orbit is reconstructed
simultaneously with any putative GW signal, so that part of the signal
would be absorbed into the fitted orbital elements; our forecasts should
therefore be read as optimistic in this respect. 

\paragraph*{Observation baseline.} Unless stated otherwise we adopt
$T_{\rm obs}=15\,$yr for LLR at a cadence of $260$ normal points per year
($N_{\rm obs}=3900$) and a per-measurement uncertainty
$\sigma_{\rm LLR}=3\,$mm (configuration~A of
Sec.~\ref{sec:recovery}); the optimistic and intermediate configurations
B and C are defined there. The binary is taken to coalesce at the end of
the observation window, with a margin of $1\,$ day. For GUEST we use
$T_{\rm obs}=10\,$yr and $\sigma_{\rm GUEST}=10\,$mm.

\paragraph*{Solution of the variational equations.} We solve the system of 
Eqs.~\eqref{eq:gauss_sm} by direct numerical integration rather than
through a state-transition-matrix propagator. The Keplerian reference
trajectory is advanced with a velocity-Verlet (leapfrog) scheme, and the
deviation $\delta\mathbf{r}$ is integrated on the same grid with the
tidal and GW accelerations evaluated at step midpoints, using
$N_{\rm sub}=8$ substeps per observation interval. 

We verified
convergence by halving and doubling $N_{\rm sub}$, which leaves the
recovered $\delta r$ time series -- and hence the posteriors on
$(\mathcal M_c,D_L)$ -- unchanged at the level of our precision. We note
that the state-transition-matrix approach~\cite{Foster:2025csl} is
particularly efficient for monochromatic or slowly evolving signals; for
a chirping source that sweeps through resonance with harmonics of the
orbital frequency, direct integration avoids repeated recomputation of
the propagator as $f_{\rm GW}$ evolves.

\section{Study of the bias and marginalisation convergence}
\label{app:bias_study}

\begin{figure}[t]
\centering

\includegraphics[width=0.9\columnwidth]{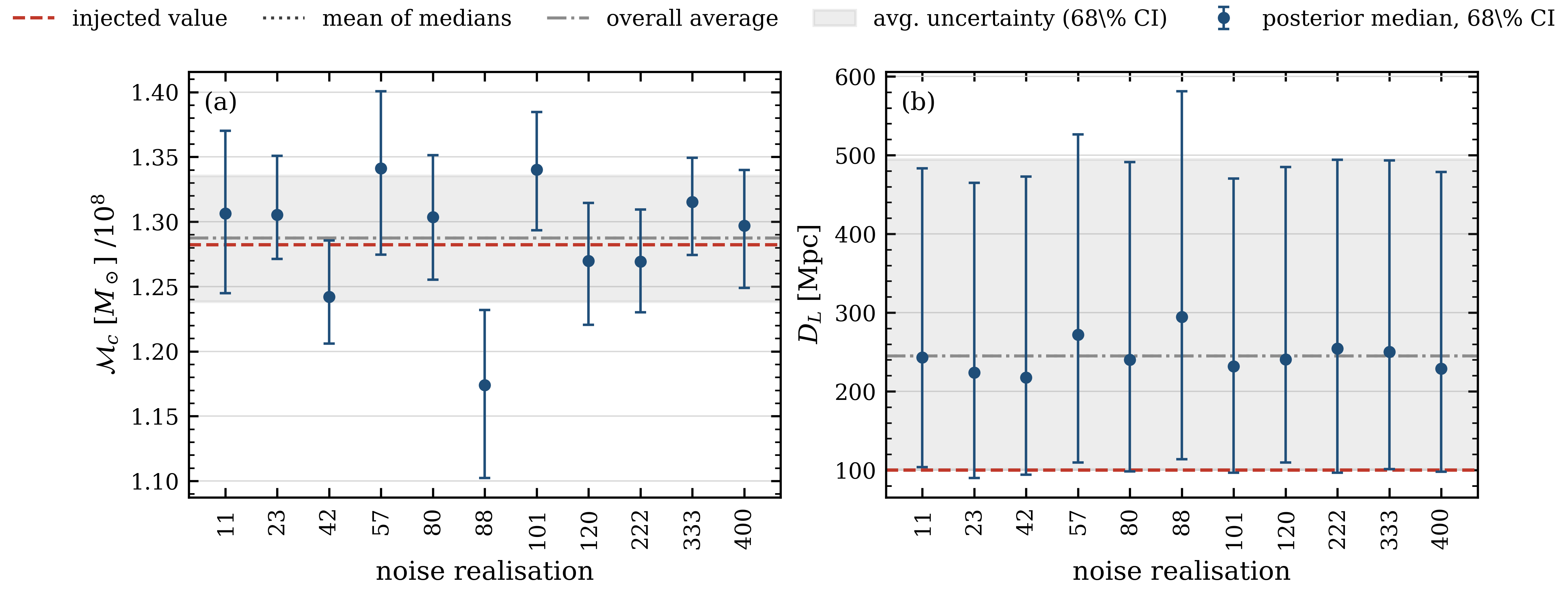}
\caption{
% \textbf{Upper Panel}: Posterior overlay for 11 independent noise realisations injected at the same physical source
%     ($\mathcal{M}_c = 1.28\times10^{8}\,M_\odot$, $D_L = 100$~Mpc, marked by
%     the black star/dashed cross-hairs). \emph{Left:} joint
%     ($\mathcal{M}_c$, $D_L$) posterior, with coloured contours enclosing the
%     90\% credible region for each individual noise realisation and the
%     grey-shaded region showing the pooled posterior across both seeds.
%     Coloured dots mark the per-seed posterior medians. \emph{Right:}
%     marginalised 1D posteriors for $\mathcal{M}_c$ (top) and $D_L$ (bottom),
%     coloured by seed, with the dashed vertical line indicating the injected
%     truth. 
%     \textbf{Lower Panel}:
Posterior medians and 68\% credible intervals for the chirp mass
    $\mathcal{M}_c$ (a) and luminosity distance $D_L$ (b), recovered from
    independent noise realisations. Dashed red: injected value; dotted: mean
    of the per-seed medians.}
\label{fig:bias}
\end{figure}

\paragraph{Noise realisation bias impact}
In this appendix, to assess whether our recovered source parameters are subject to a
systematic bias induced by the specific noise realisation used in the
injection --- as opposed to a fluctuation reflecting the intrinsic width of
the posterior --- we repeat the analysis at fixed physical parameters
($\mathcal{M}_c$, $D_L$) while varying only the seed of the injected
Gaussian noise. Figure~\ref{fig:bias} shows the resulting
posterior overlay for 11 noise realisations.

Diagnostics reveal that the chirp mass exhibits healthy noise-driven posterior widths (centroid over-dispersion $R = 0.78$), with negligible mutual tension across realisations (max pairwise difference $1.13\sigma$). In contrast, distance shows over-dispersed posteriors relative to actual scatter ($R = 0.07$), indicating that posterior widths are dominated by the prior and geometric degeneracies rather than data constraints---expected in single-detector observations where $D$ is degenerate with inclination angle. Both parameters show no evidence of bias or miscalibrated uncertainties when jointly considered (Mahalanobis separations $<1\sigma$), validating our inference pipeline while confirming the known limitation of distance measurement in single-detector GW observations.

% \begin{figure}[t]
% \centering

% \includegraphics[width=\columnwidth]{mean_summary.png}
% \caption{Noise-averaged parameter recovery over $N$ independent noise
%   realisations (seeds) at fixed injected parameters (dashed red line).
%   \textit{Top row:} recovery of the best-fit value. Grey points are the
%   per-seed posterior medians; the circular marker with thick error bar
%   shows their mean $\pm$ standard error
%   ($\mathrm{SE} = \mathrm{std}/\sqrt{N}$), the precision with which the
%   noise-averaged estimate is determined. The thin capped bar is the
%   standard deviation (\texttt{std}) of the per-seed medians, i.e.\ the
%   realisation-to-realisation scatter expected for a single noise draw.
%   The shaded band shows $\pm$ the mean posterior width
%   ($\overline{\sigma}$) for comparison. \texttt{bias} is the fractional
%   offset of the mean from truth,
%   $\mathrm{bias} = \langle\hat\theta\rangle/\theta_{\mathrm{true}} - 1$.
%   \textit{Bottom row:} recovery of the quoted uncertainty. Grey points
%   are the per-seed posterior half-widths
%   $\sigma = \tfrac{1}{2}(q_{84} - q_{16})$; the square marker shows
%   their mean $\pm$ SE, with the thin bar giving the std across seeds.
%   The dotted line reproduces the scatter of the medians from the panel
%   above, so that \texttt{scatter/sigma} (their ratio) tests calibration.}
% \label{fig:bias_2}
% \end{figure}

\begin{figure}[t]
\centering

\includegraphics[width=0.9\columnwidth]{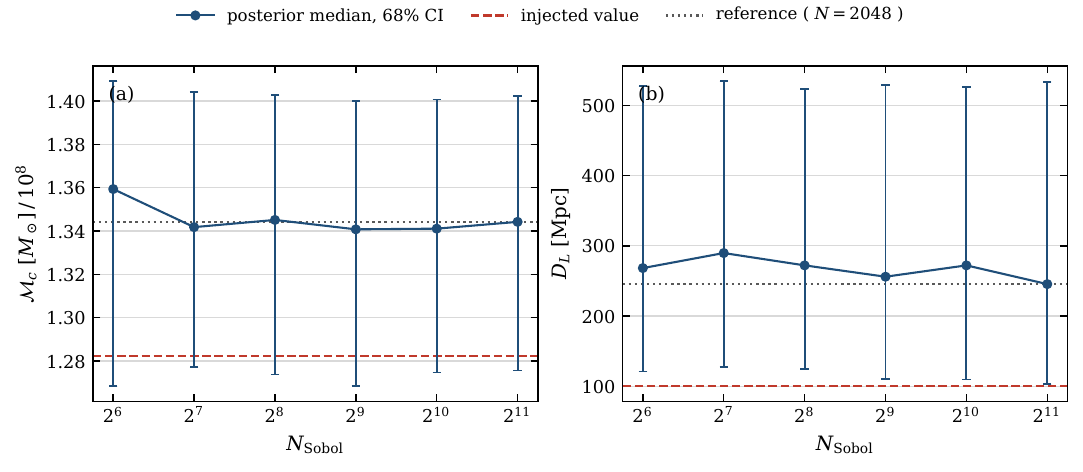}
\caption{
% \textbf{Upper Panel}: Posterior overlay for different values of $N_{\rm sobol}$ at the same injected source
%     ($\mathcal{M}_c = 1.28\times10^{8}\,M_\odot$, $D_L = 100$~Mpc, marked by
%     the black star/dashed cross-hairs). \emph{Left:} joint
%     ($\mathcal{M}_c$, $D_L$) posterior, with coloured contours enclosing the
%     90\% credible region for each individual noise realisation and the
%     grey-shaded region showing the pooled posterior across both seeds.
%     Coloured dots mark the per-seed posterior medians. \emph{Right:}
%     marginalised 1D posteriors for $\mathcal{M}_c$ (top) and $D_L$ (bottom),
%     coloured by seed, with the dashed vertical line indicating the injected
%     truth. 
%     \textbf{Lower Panel}:
Posterior medians and 68\% credible intervals for the chirp mass
    $\mathcal{M}_c$ (a) and luminosity distance $D_L$ (b), recovered from
    independent noise realisations. Dashed red: injected value; dotted: mean
    of the per-seed medians.}
\label{fig:convergence_Sobol}
\end{figure}

\paragraph{Marginalisation convergence}

We now turn to the study of the convergence of the angle marginalisation with a Sobol series. 

Two properties of the fixed Sobol point set are essential for compatibility
with nested sampling:

\begin{enumerate}
    \item \textbf{Determinism and smoothness.} The Sobol angles
    $(\theta_{s,k}, \phi_{s,k}, \mathcal I_{{\rm gw},k}, \psi_{{\rm gw},k})$ are
    drawn once, for a fixed seed, at the start of the run and reused
    identically at \emph{every} proposed $(\mathcal{M}_c, D_L)$. This
    renders $\mathcal{L}_{\rm marg}(\mathcal{M}_c, D_L)$ a smooth,
    deterministic function of $(\mathcal{M}_c, D_L)$. Re-drawing angles
    stochastically at each likelihood call, as in standard on-the-fly Monte
    Carlo marginalisation, would inject non-reproducible jitter between
    evaluations at the same point in parameter space, degrading the
    performance and reliability of the nested sampler, whose live-point
    evolution and evidence estimate assume a smooth likelihood surface.

    \item \textbf{Superior convergence rate.} The integration error of a
    Sobol (low-discrepancy) sequence scales approximately as
    $\mathcal{O}(N^{-1}\log^d N)$, compared to $\mathcal{O}(N^{-1/2})$ for
    standard Monte Carlo, where $d=4$ is the dimensionality of the angle
    space. This allows accurate marginalisation with a comparatively small
    number of points, as confirmed by the convergence study below.
\end{enumerate}

% As a purely computational optimisation, all angle-dependent geometric
% quantities that do not depend on $(\mathcal{M}_c, D_L)$ (direction cosines,
% antenna-pattern/polarisation tensors) are precomputed once for the full
% Sobol set (\texttt{BatchPrep}) and reused across every likelihood
% evaluation, rather than being rebuilt at each call. This is necessary in
% practice given the large number of likelihood evaluations required by the
% sampler.

To validate the deterministic marginalisation over the source orientation
angles, we performed a convergence study in which the physical
configuration was held fixed (chirp mass $\mathcal{M}_c$ and luminosity
distance $D_L$ at their injected values) while the number of Sobol points
$N_{\rm Sobol}$ used to evaluate the angle-marginalised likelihood was
swept over the grid $N_{\rm Sobol} \in \{64, 128, 256, 512, 1024\}$. For each
value we ran an independent nested-sampling analysis. We monitored the convergence of the posterior
medians and widths of $\mathcal{M}_c$ and $D_L$, the Bayesian evidence
$\ln\mathcal{Z}$, and the overlaid 2D posterior contours across runs.

We show the results of this analysis in fig.\ref{fig:convergence_Sobol}. The posterior summaries and evidence estimates stabilised by
$N_{\rm Sobol} = 128$, with no statistically significant change in the
median values, the $1\sigma$ credible intervals, or $\ln\mathcal{Z}$. We will therefore adopt
$N_{\rm Sobol} = 1024$ for the remainder of our analysis, as it provides
converged, seed-independent results at modest computational cost.

\section{Study of the marginalisation over the phase $\phi_0$}
\label{app:phi0}

\begin{figure*}[t]
    \centering
    \includegraphics[width=0.48\linewidth]{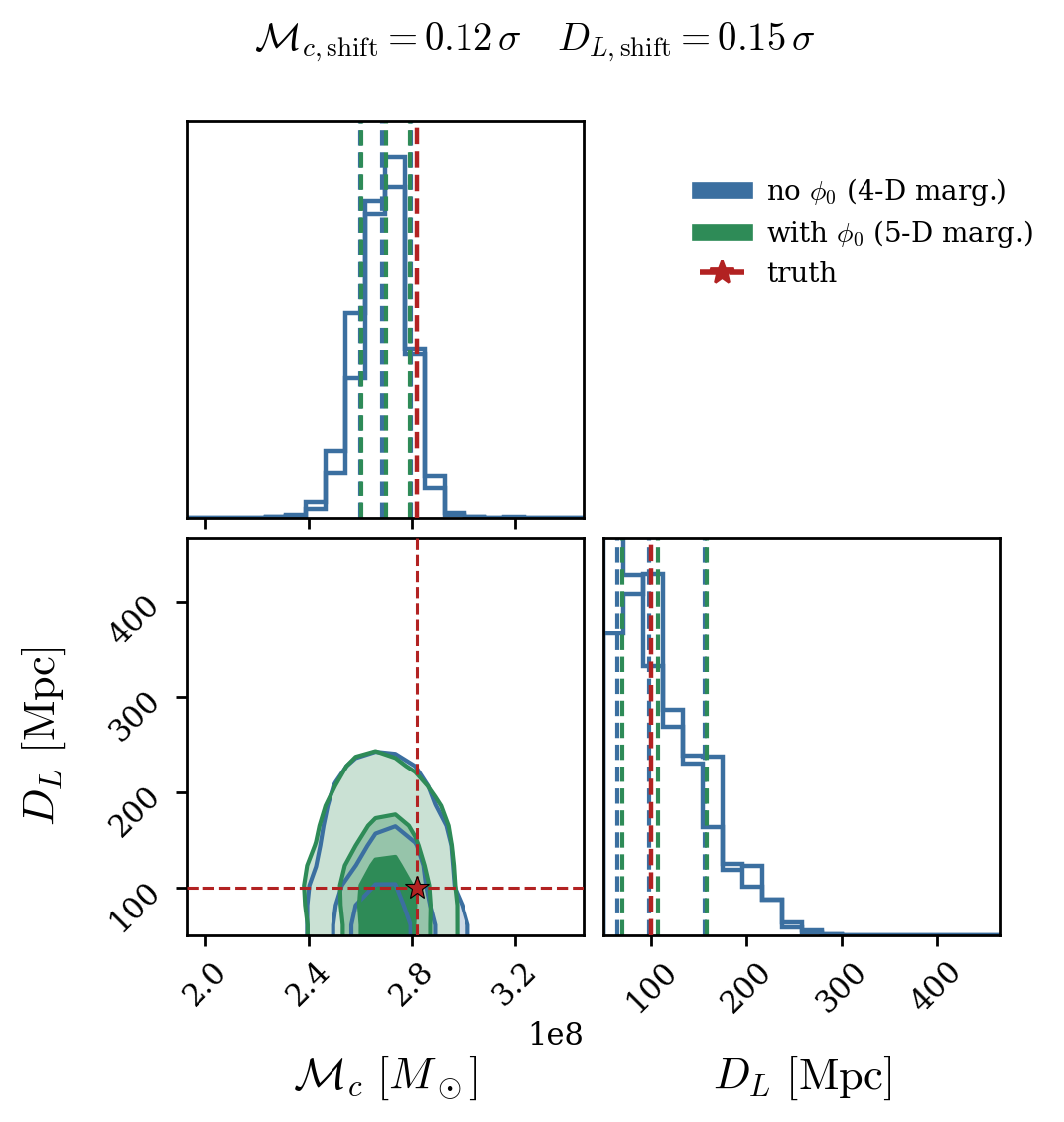}
    \includegraphics[width=0.48\linewidth]{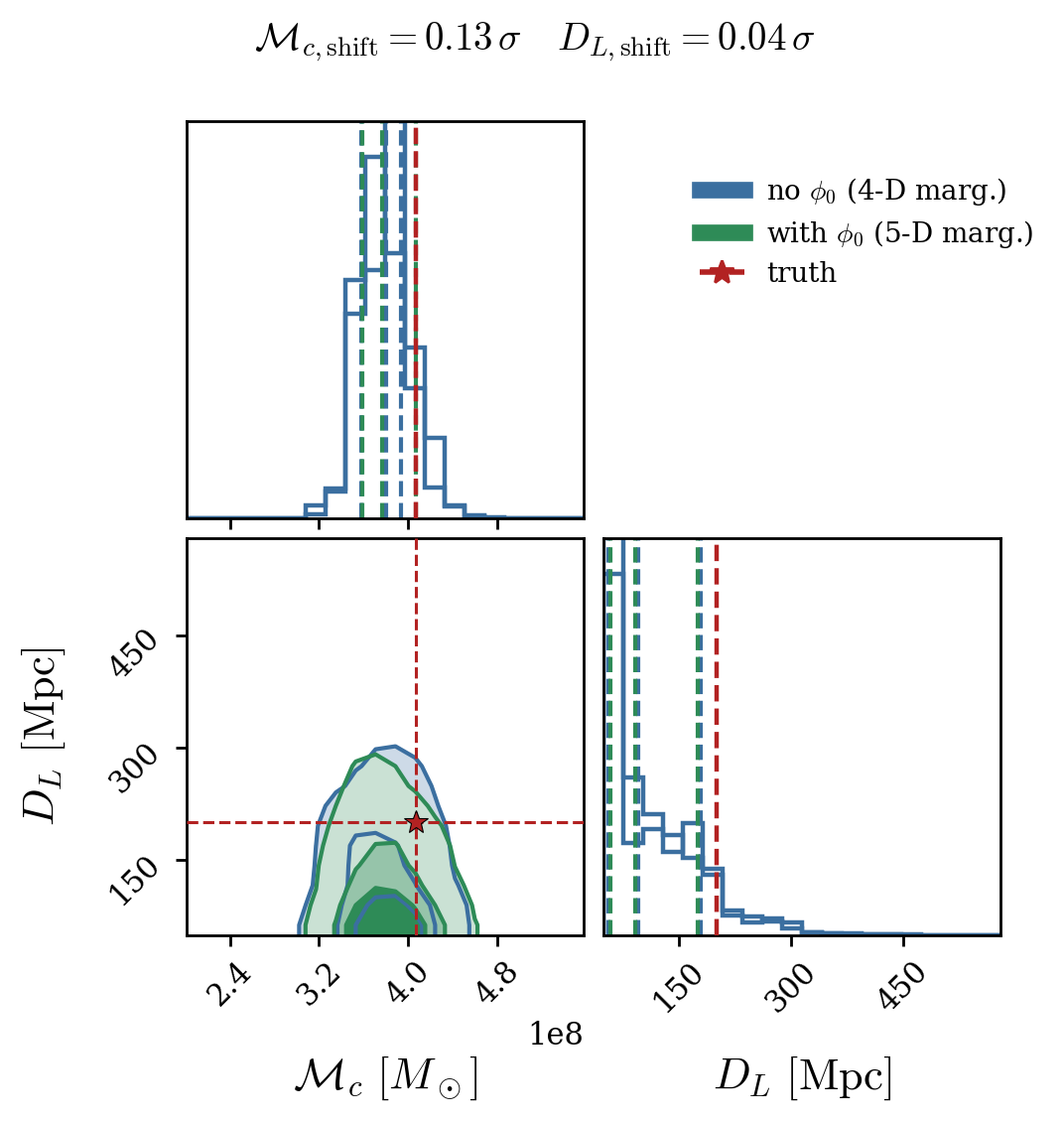}
    \caption{\textbf{Comparing posteriors}.
        Corner plot of the 2-dimensional posterior obtained with and without marginalising over the initial phase $\phi_0$. Green shaded (blue) regions mark the 16th, 50th, and 84th 
        percentiles respectively for the case without (with) the $\phi_0$ marginalisation; the red dashed lines and dots indicate the true injected values. 
        Parameter values and 68\% credible intervals are quoted above each 
        diagonal panel. The shift in the best fit value is presented in the title, in units of the half-width of the relevant parameter.
    }
    \label{fig:corner_phi0}
\end{figure*}

In this appendix, we study the effect of marginalizing over the $\phi_0$ parameter. For ten selected points in our scan, we run a new bayesian analysis including the marginalisation over the phase $\phi_0$. On Figure \ref{fig:corner_phi0}, we show this comparison of the posterior for a bayesian analysis fixing the $\phi_0$ phase a priori (in blue) and marginalizing over this phase (in green) for two selected points.

In general, we conclude that marginalizing over the phase $\phi_0$ only disturbs the posteriors by 10 \% or less, this is thus an irrelevant correction our purpose.

\section{Comparison of the two proxies}
\label{app:two_proxies}

\begin{figure*}[t]
    \centering
    \includegraphics[width=0.8\linewidth]{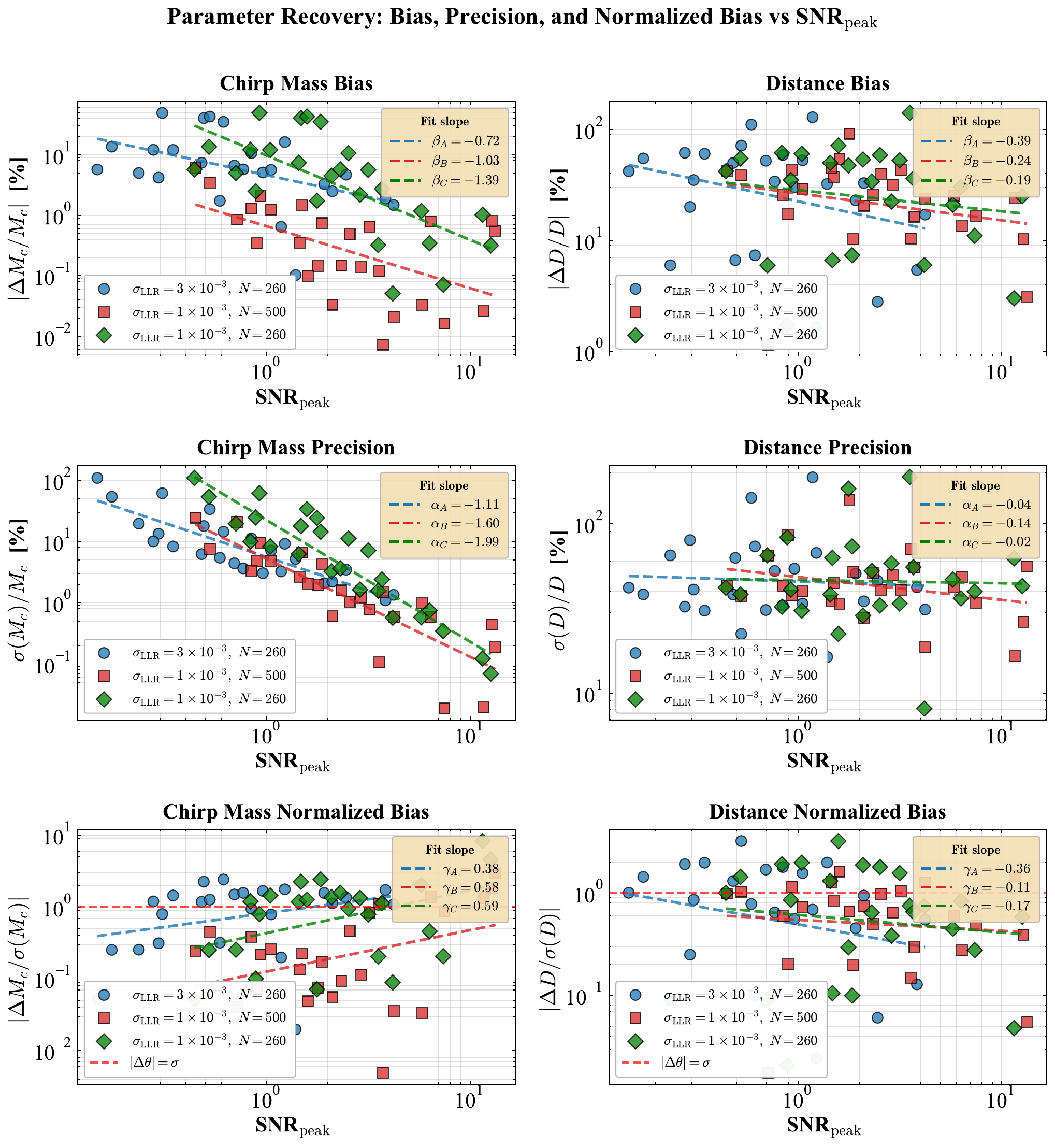}
    \caption{
        Same as Fig.\ref{fig:stat_analysis_1} but using the SNR$_{\rm peak}$ as a proxy.
    }
    \label{fig:stat_analysis_2}
\end{figure*}

To test whether $\mathrm{SNR}_{\rm peak}$ (see here for the summary of the result with it in Fig.\ref{fig:stat_analysis_2}) or the cumulative statistic
$\mathrm{SNR} \equiv \sqrt{\chi^2_{\rm true}/2}$ better predicts parameter
recovery quality, we fit power laws
$\log_{10} Y = a + \alpha \log_{10} P$ for each recovery metric $Y$ and
proxy $P$, using the identical set of $N=76$ LLR injections pooled across all
three datasets. We quantify predictive
power via the residual scatter $s$ about the fit, in dex. Table~\ref{tab:proxy_comparison} summarises the results. The two proxies
give statistically indistinguishable residual scatter for every recovery
metric: differences in $s$ ($\lesssim 0.03$~dex) are smaller than the
uncertainty ($\sim 0.03$--$0.05$~dex) in all cases, and the
$16$--$84$th percentile intervals overlap substantially. We therefore
conclude that $\mathrm{SNR}_{\rm peak}$ and $\sqrt{\chi^2_{\rm true}/2}$
are \emph{comparably good} predictors of recovery precision and bias; the
choice between them does not materially affect the scaling analysis.

Distance precision and bias are essentially uncorrelated with either
proxy ($R^2 \lesssim 0.06$), indicating that distance recovery is
governed by parameter degeneracies rather than overall signal strength,
and is not usefully predicted by any single SNR-like statistic in this
scan. Chirp-mass precision and bias, by contrast, show strong power-law
scaling with both proxies ($\alpha \approx -1.2$ to $-1.4$, $R^2 \approx
0.37$--$0.70$), confirming that cumulative signal strength and $\mathrm{SNR}_{\rm peak}$ are the
dominant driver of mass-parameter recovery.

\begin{table}[t]
\centering
\caption{Residual scatter $s$ (dex) about power-law fits
$\log_{10}Y = a+\alpha\log_{10}P$, for two SNR proxies $P$, pooled over
$N=76$ injections across three configurations. Bootstrap 16--84th
percentile intervals in parentheses.}
\label{tab:proxy_comparison}
\begin{tabular}{lccc}
\hline\hline
Metric & Proxy & $\alpha$ & $s$ [dex] \\
\hline
\multirow{2}{*}{$\sigma(M_c)/M_c$}
 & $\mathrm{SNR}_{\rm peak}$ & $-1.44$ & $0.435\,(0.386\text{--}0.466)$ \\
 & $\sqrt{\chi^2_{\rm true}/2}$ & $-1.28$ & $0.442\,(0.393\text{--}0.475)$ \\
\multirow{2}{*}{$\sigma(D)/D$}
 & $\mathrm{SNR}_{\rm peak}$ & $-0.06$ & $0.227\,(0.199\text{--}0.249)$ \\
 & $\sqrt{\chi^2_{\rm true}/2}$ & $-0.01$ & $0.229\,(0.200\text{--}0.252)$ \\
\multirow{2}{*}{$|\Delta M_c|/M_c$}
 & $\mathrm{SNR}_{\rm peak}$ & $-1.22$ & $0.747\,(0.685\text{--}0.792)$ \\
 & $\sqrt{\chi^2_{\rm true}/2}$ & $-1.18$ & $0.712\,(0.648\text{--}0.764)$ \\
\multirow{2}{*}{$|\Delta D|/D$}
 & $\mathrm{SNR}_{\rm peak}$ & $-0.23$ & $0.444\,(0.393\text{--}0.481)$ \\
 & $\sqrt{\chi^2_{\rm true}/2}$ & $-0.19$ & $0.446\,(0.396\text{--}0.483)$ \\
\hline\hline
\end{tabular}
\end{table}

\section{Lunar and Satellite Laser Ranging: Technique and Systematics}
\label{app:llr}

\subsection{Measurement principle}

Lunar laser ranging (LLR) and satellite laser ranging (SLR) both rely on
time-of-flight measurement of a laser pulse reflected by a passive
retroreflector array. A short pulse is emitted from a ground station,
retroreflected, and detected on return; the round-trip travel time
$\Delta t_{\rm rt}$ gives the instantaneous range,
\be
R = \frac{c\,\Delta t_{\rm rt}}{2} ,
\ee
after subtraction of modelled propagation delays. Because $R$ is obtained
from a light-travel-time measurement, the timing precision required to
reach a given ranging precision is set purely by the speed of light:
1~mm of range corresponds to $6.7$~ps of round-trip timing. At the mean
Earth--Moon distance ($R\simeq 3.84\times10^{8}$~m) the round-trip light
time is $\Delta t_{\rm rt}\simeq 2.5$~s, so millimetre-level LLR requires
picosecond-class event timing~\cite{Benderarticle, Dickey:1994zz, Murphy:2013qya}.

Existing lunar retroreflectors are passive corner-cube arrays deployed
during the Apollo~11, 14, and~15 missions and on the Soviet Lunokhod~1
and~2 rovers. The modern benchmark station is APOLLO (Apache Point
Observatory Lunar Laser-ranging Operation), which achieves single-shot
and normal-point precision of $\sigma_R\sim 1$--$2$~mm on the Earth--Moon
centre-of-mass distance~\cite{Murphy:2007ed,Murphy:2013qya, Battat:2023upl}. The
global LLR archive spans 1969 to the present ($\sim 50$~years) and
comprises $\sim 26{,}000$ normal points collected from a small number of
stations (McDonald, Grasse/OCA, APOLLO, Matera), targeting the Apollo and
Lunokhod reflectors~\cite{Murphy:2013qya}. Each normal point compresses a
5--15~minute integration of individually timed photons (ranging from a
few to several thousand per point) into a triplet $(t,\rho,\sigma_\rho)$.
Ranging precision has improved from decimetre to millimetre level over
the archive's history, corresponding to a fractional precision of
$10^{-9}$--$10^{-11}$, driven by improved event timing electronics and
increased photon return rates.

Next-generation instrumentation will replace the extended Apollo/Lunokhod
arrays with a single large corner-cube reflector, avoiding the pulse
spreading caused by lunar-libration-induced tilt of a multi-element
array. Proposed systems include the Next Generation Lunar Retroreflector
(NGLR), the Artemis Lunar Laser Retroreflector (ALLR), with a target precision of $\sigma_R\lesssim
1$~mm~\cite{Currie:2013,2025PhRvP..23f4066T, 2025arXiv251202431T}.

\subsection{A glimpse at the systematic error budget}

The achievable ranging precision is limited by several distinct
categories of systematic error, summarised below.

\paragraph{Reflector and photon statistics.} Pulse spreading from
extended, tilted reflector arrays, low photon return rate, beam pointing
errors, and lunar-phase-dependent sky background all degrade single-shot
precision and are mitigated primarily by normal-point averaging.

\paragraph{Atmosphere.} Tropospheric delay is of order metres and must be
modelled from local meteorological data; residual errors are strongly
elevation-dependent, ranging from $\sim 20$~mm at $10^\circ$ elevation to
$\sim 1$~mm at $40^\circ$~\cite{2004GeoRL..3114602M}.

\paragraph{Solid Earth and loading tides.} Solid Earth tides displace
ranging stations vertically by tens of centimetres, with ocean and
atmospheric loading contributing a further few centimetres. After
correction using frequency-dependent Love numbers and standard loading
models, the residual station-position error is $\sigma_{\rm tides}\sim
1$--$5$~mm~\cite{2015JGRE..120..689W}.

\paragraph{Solar-system and Earth--Moon dynamical modelling.} Full
dynamical models include solar and planetary perturbations, asteroid
perturbations, the Earth and lunar gravity fields, lunar tidal
dissipation, and physical libration. Differences between successive JPL
ephemerides (DE430 vs.\ DE421) are $\sim 0.5$~mas in angular position
($\sim 1$~m at lunar distance) and $\sim 0.25$~m in radial distance, with
a monthly radial modulation of a few centimetres. Post-fit LLR residuals
with modern data reach an RMS of $\simeq 1.9$~cm~\cite{2014IPNPR.196C...1F}.

\paragraph{Station and timing systems.} Laser pulse width and shape,
detector timing jitter, clock stability, calibration of internal
optical/electronic delays, and local ground motion (tectonics, loading)
all contribute at the mm level and must be independently monitored and
calibrated.

\subsection{Summary on the error budget}

Instrumental photon timing alone would permit sub-millimetre ranging;
in practice, LLR accuracy is limited jointly by atmospheric modelling,
solid-Earth and loading tides, and the completeness of the
Earth--Moon--Sun dynamical model, rather than by the timing system
itself. This makes the LLR error budget qualitatively different from
that of a dedicated space-based ranging mission such as GUEST, for which
the dominant systematics are orbital rather than atmospheric or
geophysical. Overall, this motivates our benchmark white gaussian error of few millimeters for forthcoming lunar missions and around 1 cm for the GUEST mission.

\section{Tables and detailed results}
\label{app:tables}
In this appendix, we present the detailed results of the bayesian analysis in the form of tables. The three tables \ref{tab:summary_A}, \ref{tab:summary_B} and \ref{tab:summary_C} reports the results of the bayesian analysis for the three LLR datasets discussed in the paper. Finally, the table \ref{tab:summary_AA} reports the result of the bayesian analysis for GUEST dataset.

\begin{table*}[ht]
  \centering
  \setlength{\tabcolsep}{2.5pt}
  \caption{Parameter recovery summary for dataset  $ \sigma_{\rm LLR}=3\times10^{-3},\ N=260 $ .}
  \label{tab:summary_A}
  \begin{tabular}{rrcccccrc}
    \toprule
     $ D_{\rm inj} $  (Mpc) &  $ M_{c,\rm inj} $  ( $ M_\odot $ ) &  $ \sigma_{M_c}/M_c $  &  $ \sigma_D/D $  &  $ |\Delta M_c|/M_c $  &  $ |\Delta D|/D $  & Overconf.\ & ${\rm SNR}_{\rm peak}$ &  $ \chi^2_{\rm true} $  \\
    \midrule
    100.0 & $2.82 \times 10^{8}$ & 3.46 & 46.08 & 4.69 & 2.79 & \textbf{YES} & 2.46 & 6.28e+02 \\
    100.0 & $4.07 \times 10^{8}$ & 5.19 & 16.33 & 0.10 & 32.28 & \textbf{YES} & 1.39 & 2.03e+02 \\
    100.0 & $1.28 \times 10^{8}$ & 3.24 & 188.10 & 0.64 & 129.14 & no & 1.18 & 5.19e+02 \\
    100.0 & $1.43 \times 10^{8}$ & 1.09 & 42.12 & 1.89 & 5.40 & \textbf{YES} & 3.84 & 4.89e+03 \\
    100.0 & $1.68 \times 10^{8}$ & 1.34 & 31.14 & 1.47 & 17.06 & \textbf{YES} & 4.21 & 4.57e+03 \\
    200.0 & $2.82 \times 10^{8}$ & 9.19 & 67.30 & 16.24 & 1.64 & \textbf{YES} & 1.23 & 1.57e+02 \\
    200.0 & $4.07 \times 10^{8}$ & 4.41 & 30.94 & 6.61 & 52.25 & \textbf{YES} & 0.70 & 5.09e+01 \\
    200.0 & $1.28 \times 10^{8}$ & 5.43 & 142.36 & 1.73 & 110.91 & no & 0.59 & 1.30e+02 \\
    200.0 & $1.43 \times 10^{8}$ & 2.09 & 50.74 & 3.28 & 22.97 & \textbf{YES} & 1.92 & 1.22e+03 \\
    200.0 & $1.68 \times 10^{8}$ & 2.21 & 34.82 & 2.49 & 32.84 & \textbf{YES} & 2.11 & 1.14e+03 \\
    400.0 & $2.82 \times 10^{8}$ & 14.44 & 73.49 & 35.01 & 7.33 & \textbf{YES} & 0.61 & 3.93e+01 \\
    400.0 & $4.07 \times 10^{8}$ & 8.31 & 30.66 & 12.03 & 60.39 & \textbf{YES} & 0.35 & 1.27e+01 \\
    400.0 & $1.28 \times 10^{8}$ & 13.39 & 80.09 & 4.19 & 20.00 & no & 0.29 & 3.25e+01 \\
    400.0 & $1.43 \times 10^{8}$ & 3.04 & 54.24 & 5.10 & 30.21 & \textbf{YES} & 0.96 & 3.05e+02 \\
    400.0 & $1.68 \times 10^{8}$ & 7.13 & 33.77 & 5.64 & 52.67 & \textbf{YES} & 1.05 & 2.86e+02 \\
    500.0 & $2.82 \times 10^{8}$ & 17.92 & 62.89 & 40.41 & 6.62 & \textbf{YES} & 0.49 & 2.51e+01 \\
    500.0 & $4.07 \times 10^{8}$ & 10.07 & 32.36 & 12.08 & 61.64 & \textbf{YES} & 0.28 & 8.14e+00 \\
    500.0 & $1.28 \times 10^{8}$ & 19.60 & 65.10 & 5.00 & 5.95 & no & 0.24 & 2.08e+01 \\
    500.0 & $1.43 \times 10^{8}$ & 3.64 & 52.68 & 5.74 & 33.96 & \textbf{YES} & 0.77 & 1.96e+02 \\
    500.0 & $1.68 \times 10^{8}$ & 11.23 & 32.93 & 10.65 & 58.83 & \textbf{YES} & 0.84 & 1.83e+02 \\
    800.0 & $2.82 \times 10^{8}$ & 61.54 & 40.87 & 49.22 & 34.97 & no & 0.31 & 9.82e+00 \\
    800.0 & $4.07 \times 10^{8}$ & 53.80 & 38.28 & 13.62 & 54.82 & \textbf{YES} & 0.17 & 3.18e+00 \\
    800.0 & $1.28 \times 10^{8}$ & 109.09 & 41.93 & 5.72 & 42.04 & \textbf{YES} & 0.15 & 8.12e+00 \\
    800.0 & $1.43 \times 10^{8}$ & 6.22 & 38.22 & 7.38 & 49.93 & \textbf{YES} & 0.48 & 7.64e+01 \\
    800.0 & $1.68 \times 10^{8}$ & 33.77 & 22.35 & 42.84 & 71.59 & \textbf{YES} & 0.53 & 7.14e+01 \\
    \bottomrule
  \end{tabular}
\end{table*}

\begin{table*}[ht]
  \centering
  \setlength{\tabcolsep}{2.5pt}
  \caption{Parameter recovery summary for dataset  $ \sigma_{\rm LLR}=1\times10^{-3},\ N=500 $ .}
  \label{tab:summary_B}
  \begin{tabular}{rrcccccrc}
    \toprule
     $ D_{\rm inj} $  (Mpc) &  $ M_{c,\rm inj} $  ( $ M_\odot $ ) &  $ \sigma_{M_c}/M_c $  &  $ \sigma_D/D $  &  $ |\Delta M_c|/M_c $  &  $ |\Delta D|/D $  & Overconf.\ & ${\rm SNR}_{\rm peak}$ &  $ \chi^2_{\rm true} $  \\
    \midrule
    100.0 & $2.82 \times 10^{8}$ & 0.02 & 34.23 & 0.02 & 16.56 & no & 7.48 & 1.09e+04 \\
    100.0 & $1.13 \times 10^{8}$ & 0.19 & 55.82 & 0.56 & 3.09 & \textbf{YES} & 13.33 & 1.77e+05 \\
    100.0 & $4.07 \times 10^{8}$ & 0.58 & 18.66 & 0.02 & 23.62 & \textbf{YES} & 4.22 & 3.52e+03 \\
    100.0 & $1.28 \times 10^{8}$ & 0.11 & 70.51 & 0.12 & 10.41 & \textbf{YES} & 3.58 & 9.01e+03 \\
    100.0 & $1.43 \times 10^{8}$ & 0.02 & 16.49 & 0.03 & 24.28 & \textbf{YES} & 11.65 & 8.45e+04 \\
    100.0 & $1.68 \times 10^{8}$ & 0.44 & 26.27 & 0.80 & 10.27 & \textbf{YES} & 12.80 & 7.91e+04 \\
    200.0 & $2.82 \times 10^{8}$ & 1.48 & 55.16 & 0.01 & 16.37 & no & 3.74 & 2.72e+03 \\
    200.0 & $4.07 \times 10^{8}$ & 0.60 & 27.82 & 0.03 & 20.52 & no & 2.11 & 8.79e+02 \\
    200.0 & $1.28 \times 10^{8}$ & 1.94 & 138.77 & 0.14 & 91.65 & no & 1.79 & 2.25e+03 \\
    200.0 & $1.43 \times 10^{8}$ & 0.99 & 42.55 & 0.03 & 25.33 & no & 5.83 & 2.11e+04 \\
    200.0 & $1.68 \times 10^{8}$ & 0.58 & 48.64 & 0.79 & 13.44 & \textbf{YES} & 6.40 & 1.98e+04 \\
    400.0 & $2.82 \times 10^{8}$ & 4.27 & 52.06 & 0.74 & 10.26 & no & 1.87 & 6.80e+02 \\
    400.0 & $4.07 \times 10^{8}$ & 4.84 & 39.82 & 1.25 & 29.10 & no & 1.05 & 2.20e+02 \\
    400.0 & $1.28 \times 10^{8}$ & 4.74 & 85.17 & 0.35 & 17.18 & no & 0.89 & 5.63e+02 \\
    400.0 & $1.43 \times 10^{8}$ & 1.22 & 49.72 & 0.14 & 31.86 & no & 2.91 & 5.28e+03 \\
    400.0 & $1.68 \times 10^{8}$ & 0.77 & 40.75 & 0.64 & 42.78 & \textbf{YES} & 3.20 & 4.94e+03 \\
    500.0 & $2.82 \times 10^{8}$ & 6.56 & 44.67 & 1.47 & 37.35 & no & 1.50 & 4.35e+02 \\
    500.0 & $4.07 \times 10^{8}$ & 3.35 & 43.12 & 1.29 & 25.55 & no & 0.84 & 1.41e+02 \\
    500.0 & $1.28 \times 10^{8}$ & 20.86 & 65.04 & 0.85 & 1.15 & no & 0.72 & 3.60e+02 \\
    500.0 & $1.43 \times 10^{8}$ & 1.58 & 51.30 & 0.15 & 25.57 & no & 2.33 & 3.38e+03 \\
    500.0 & $1.68 \times 10^{8}$ & 1.04 & 40.74 & 0.48 & 39.97 & no & 2.56 & 3.16e+03 \\
    800.0 & $2.82 \times 10^{8}$ & 9.56 & 37.63 & 2.09 & 43.21 & \textbf{YES} & 0.93 & 1.70e+02 \\
    800.0 & $4.07 \times 10^{8}$ & 7.61 & 37.36 & 3.45 & 38.46 & \textbf{YES} & 0.53 & 5.50e+01 \\
    800.0 & $1.28 \times 10^{8}$ & 24.64 & 42.94 & 6.00 & 42.23 & no & 0.45 & 1.41e+02 \\
    800.0 & $1.43 \times 10^{8}$ & 2.62 & 35.05 & 0.35 & 44.68 & \textbf{YES} & 1.46 & 1.32e+03 \\
    800.0 & $1.68 \times 10^{8}$ & 2.04 & 33.58 & 0.10 & 54.43 & \textbf{YES} & 1.60 & 1.24e+03 \\
    \bottomrule
  \end{tabular}
\end{table*}

\begin{table*}[ht]
  \centering
  \setlength{\tabcolsep}{2.5pt}
  \caption{Parameter recovery summary for dataset  $ \sigma_{\rm LLR}=1\times10^{-3},\ N=260 $ .}
  \label{tab:summary_C}
  \begin{tabular}{rrcccccrc}
    \toprule
     $ D_{\rm inj} $  (Mpc) &  $ M_{c,\rm inj} $  ( $ M_\odot $ ) &  $ \sigma_{M_c}/M_c $  &  $ \sigma_D/D $  &  $ |\Delta M_c|/M_c $  &  $ |\Delta D|/D $  & Overconf.\ & ${\rm SNR}_{\rm peak}$ &  $ \chi^2_{\rm true} $  \\
    \midrule
    100.0 & $2.82 \times 10^{8}$ & 0.34 & 39.75 & 0.07 & 10.98 & no & 7.38 & 5.65e+03 \\
    100.0 & $4.07 \times 10^{8}$ & 0.57 & 8.09 & 0.05 & 5.98 & no & 4.17 & 1.83e+03 \\
    100.0 & $1.28 \times 10^{8}$ & 1.58 & 188.99 & 0.32 & 141.16 & no & 3.54 & 4.67e+03 \\
    100.0 & $1.43 \times 10^{8}$ & 0.12 & 62.44 & 1.02 & 2.99 & \textbf{YES} & 11.52 & 4.40e+04 \\
    100.0 & $1.68 \times 10^{8}$ & 0.07 & 42.79 & 0.32 & 25.06 & \textbf{YES} & 12.64 & 4.11e+04 \\
    200.0 & $2.82 \times 10^{8}$ & 2.40 & 55.45 & 2.73 & 36.19 & \textbf{YES} & 3.69 & 1.41e+03 \\
    200.0 & $4.07 \times 10^{8}$ & 3.26 & 28.65 & 4.49 & 53.58 & \textbf{YES} & 2.09 & 4.58e+02 \\
    200.0 & $1.28 \times 10^{8}$ & 24.26 & 161.20 & 1.74 & 47.53 & no & 1.77 & 1.17e+03 \\
    200.0 & $1.43 \times 10^{8}$ & 0.58 & 46.84 & 1.18 & 20.95 & \textbf{YES} & 5.76 & 1.10e+04 \\
    200.0 & $1.68 \times 10^{8}$ & 0.75 & 36.23 & 0.34 & 30.28 & no & 6.32 & 1.03e+04 \\
    400.0 & $2.82 \times 10^{8}$ & 14.44 & 73.49 & 35.01 & 7.33 & \textbf{YES} & 1.84 & 3.53e+02 \\
    400.0 & $4.07 \times 10^{8}$ & 8.31 & 30.66 & 12.03 & 60.39 & \textbf{YES} & 1.04 & 1.14e+02 \\
    400.0 & $1.28 \times 10^{8}$ & 24.75 & 83.14 & 2.49 & 1.73 & no & 0.88 & 2.92e+02 \\
    400.0 & $1.43 \times 10^{8}$ & 1.64 & 58.42 & 2.19 & 22.33 & \textbf{YES} & 2.88 & 2.75e+03 \\
    400.0 & $1.68 \times 10^{8}$ & 7.13 & 33.77 & 5.64 & 52.67 & \textbf{YES} & 3.16 & 2.57e+03 \\
    500.0 & $2.82 \times 10^{8}$ & 17.92 & 62.89 & 40.41 & 6.62 & \textbf{YES} & 1.48 & 2.26e+02 \\
    500.0 & $4.07 \times 10^{8}$ & 10.07 & 32.36 & 12.08 & 61.64 & \textbf{YES} & 0.83 & 7.32e+01 \\
    500.0 & $1.28 \times 10^{8}$ & 19.60 & 65.10 & 5.00 & 5.95 & no & 0.71 & 1.87e+02 \\
    500.0 & $1.43 \times 10^{8}$ & 3.64 & 52.68 & 5.74 & 33.96 & \textbf{YES} & 2.30 & 1.76e+03 \\
    500.0 & $1.68 \times 10^{8}$ & 11.23 & 32.93 & 10.65 & 58.83 & \textbf{YES} & 2.53 & 1.65e+03 \\
    800.0 & $2.82 \times 10^{8}$ & 61.54 & 40.87 & 49.22 & 34.97 & no & 0.92 & 8.83e+01 \\
    800.0 & $4.07 \times 10^{8}$ & 53.80 & 38.28 & 13.62 & 54.82 & \textbf{YES} & 0.52 & 2.86e+01 \\
    800.0 & $1.28 \times 10^{8}$ & 109.09 & 41.93 & 5.72 & 42.04 & \textbf{YES} & 0.44 & 7.30e+01 \\
    800.0 & $1.43 \times 10^{8}$ & 6.22 & 38.22 & 7.38 & 49.93 & \textbf{YES} & 1.44 & 6.87e+02 \\
    800.0 & $1.68 \times 10^{8}$ & 33.77 & 22.35 & 42.84 & 71.59 & \textbf{YES} & 1.58 & 6.43e+02 \\
    \bottomrule
  \end{tabular}
\end{table*}

\begin{table*}[ht]
  \centering
  \setlength{\tabcolsep}{2.5pt}
  \caption{Parameter recovery summary for dataset  $ \sigma_{\rm LLR}=10\times10^{-3},\ N=300 $ .}
  \label{tab:summary_AA}
  \begin{tabular}{rrcccccrc}
    \toprule
     $ D_{\rm inj} $  (Mpc) &  $ M_{c,\rm inj} $  ( $ M_\odot $ ) &  $ \sigma_{M_c}/M_c $  &  $ \sigma_D/D $  &  $ |\Delta M_c|/M_c $ (\%) &  $ |\Delta D|/D $ (\%) &  Overconf. & ${\rm SNR}_{\rm peak}$ &  $ \chi^2_{\rm true} $  \\
    \midrule
    100.0 & $4.00 \times 10^{6}$ & 3.49 & 29.99 & 3.58 & 35.48 & YES & 0.92 & 2.02e+02 \\
    100.0 & $7.50 \times 10^{6}$ & 14.51 & 138.27 & 17.68 & 86.82 & YES & 0.60 & 2.64e+01 \\
    200.0 & $4.00 \times 10^{6}$ & 6.46 & 29.26 & 6.64 & 55.89 & YES & 0.46 & 5.05e+01 \\
    200.0 & $6.00 \times 10^{6}$ & 98.56 & 339.12 & 112.03 & 98.86 & YES & 0.22 & 5.99e+00 \\
    200.0 & $7.00 \times 10^{6}$ & 122.77 & 317.43 & 147.22 & 61.44 & YES & 0.28 & 6.39e+00 \\
    400.0 & $4.00 \times 10^{6}$ & 19.11 & 58.19 & 19.38 & 104.09 & YES & 0.23 & 1.26e+01 \\
    400.0 & $5.00 \times 10^{6}$ & 171.88 & 157.96 & 133.08 & 61.31 & no & 0.09 & 1.46e+00 \\
    400.0 & $6.00 \times 10^{6}$ & 129.58 & 160.16 & 141.44 & 63.69 & YES & 0.11 & 1.50e+00 \\
    400.0 & $7.00 \times 10^{6}$ & 122.50 & 157.01 & 130.89 & 61.49 & YES & 0.14 & 1.60e+00 \\
    500.0 & $4.00 \times 10^{6}$ & 97.64 & 124.38 & 91.79 & 129.54 & YES & 0.18 & 8.07e+00 \\
    500.0 & $5.00 \times 10^{6}$ & 174.85 & 126.91 & 138.09 & 61.28 & no & 0.08 & 9.36e-01 \\
    500.0 & $6.00 \times 10^{6}$ & 138.29 & 128.73 & 148.56 & 63.81 & YES & 0.09 & 9.58e-01 \\
    500.0 & $7.00 \times 10^{6}$ & 122.53 & 125.63 & 131.94 & 61.34 & YES & 0.11 & 1.02e+00 \\
    60.0 & $4.00 \times 10^{6}$ & 2.05 & 14.82 & 2.06 & 14.35 & YES & 1.54 & 5.61e+02 \\
    60.0 & $5.00 \times 10^{6}$ & 9.30 & 61.82 & 8.89 & 44.28 & no & 0.63 & 6.50e+01 \\
    60.0 & $6.00 \times 10^{6}$ & 11.94 & 42.60 & 12.57 & 35.67 & YES & 0.72 & 6.66e+01 \\
    60.0 & $7.00 \times 10^{6}$ & 7.42 & 53.23 & 8.19 & 37.75 & YES & 0.94 & 7.10e+01 \\
    \bottomrule
  \end{tabular}
\end{table*}

\bibliography{biblio}

@article{Currie:2013,
    author = "Currie, Douglas G. and Dell'Agnello, Simone and Delle Monache, Giovanni O. and Behr, Bradford and Williams, James G.",
    title = "{A Lunar Laser Ranging Retroreflector Array for the 21st Century}",
    journal = "Nucl. Phys. B Proc. Suppl.",
    volume = "243-244",
    pages = "218--228",
    year = "2013",
    doi = "10.1016/j.nuclphysbps.2013.09.007"
}

@article{Deng:2025qhx,
    author = "Deng, Senwen and Babak, Stanislav and Marsat, Sylvain",
    title = "{Fast detection and reconstruction of merging massive black hole binary signals}",
    eprint = "2504.11322",
    archivePrefix = "arXiv",
    primaryClass = "gr-qc",
    doi = "10.1103/jyr7-fcgp",
    journal = "Phys. Rev. D",
    volume = "112",
    number = "4",
    pages = "043010",
    year = "2025"
}

@article{Blas:2026xol,
    author = "Blas, Diego and others",
    title = "{GUEST: Gravitational Universe Exploration with Satellite Tracking. A passive satellite laser-ranging mission for the dark gravitational Universe}",
    eprint = "2607.18390",
    archivePrefix = "arXiv",
    primaryClass = "astro-ph.CO",
    month = "7",
    year = "2026"
}

@article{EPTA:2023xxk,
    author = "Antoniadis, J. and others",
    collaboration = "EPTA, InPTA",
    title = "{The second data release from the European Pulsar Timing Array - IV. Implications for massive black holes, dark matter, and the early Universe}",
    eprint = "2306.16227",
    archivePrefix = "arXiv",
    primaryClass = "astro-ph.CO",
    doi = "10.1051/0004-6361/202347433",
    journal = "Astron. Astrophys.",
    volume = "685",
    pages = "A94",
    year = "2024"
}

@article{Blas:2021mqw,
    author = "Blas, Diego and Jenkins, Alexander C.",
    title = "{Bridging the $\mu$Hz gap in the gravitational-wave landscape with binary resonances}",
    eprint = "2107.04601",
    archivePrefix = "arXiv",
    primaryClass = "gr-qc",
    doi = "10.1103/PhysRevLett.128.101103",
    journal = "Phys. Rev. Lett.",
    volume = "128",
    pages = "101103",
    year = "2022"
}

@article{Foster:2025csl,
    author = "Foster, Joshua W. and Blas, Diego and Bourgoin, Adrien and Hees, Aurelien and Herrero-Valea, M{\'\i}riam and Jenkins, Alexander C. and Xue, Xiao",
    title = "{Prospects for gravitational wave and ultra-light dark matter detection with binary resonances beyond the secular approximation}",
    eprint = "2504.16988",
    archivePrefix = "arXiv",
    primaryClass = "gr-qc",
    reportNumber = "FERMILAB-PUB-25-0092-T",
    doi = "10.48550/arXiv.2504.16988",
    journal = "arXiv e-prints",
    year = "2025"
}

@article{Hui:2012yp,
    author = "Hui, Lam and McWilliams, Sean T. and Yang, I-Sheng",
    title = "{Binary Systems as Resonance Detectors for Gravitational Waves}",
    eprint = "1212.2623",
    archivePrefix = "arXiv",
    primaryClass = "gr-qc",
    doi = "10.1103/PhysRevD.87.084009",
    journal = "Phys. Rev. D",
    volume = "87",
    pages = "084009",
    year = "2013"
}

@book{Poisson_Will_2014,
    place = {Cambridge},
    title = {Gravity: Newtonian, Post-Newtonian, Relativistic},
    publisher = {Cambridge University Press},
    author = {Poisson, Eric and Will, Clifford M.},
    year = {2014}
}

@article{Sesana:2019vho,
    author = "Sesana, Alberto and others",
    title = "{Unveiling the gravitational universe at $\mu$-Hz frequencies}",
    eprint = "1908.11391",
    archivePrefix = "arXiv",
    primaryClass = "astro-ph.IM",
    doi = "10.1007/s10686-021-09709-9",
    journal = "Exper. Astron.",
    volume = "51",
    pages = "1333--1383",
    year = "2021"
}

@article{Fedderke:2021kuy,
    author = "Fedderke, Michael A. and Graham, Peter W.",
    title = "{Asteroid Astrometry as a Gravitational Wave Detector}",
    eprint = "2112.11431",
    archivePrefix = "arXiv",
    primaryClass = "gr-qc",
    doi = "10.1103/PhysRevD.105.103018",
    journal = "Phys. Rev. D",
    volume = "105",
    pages = "103018",
    year = "2022"
}

@article{Moore:2017ity,
    author = "Moore, Christopher J. and Mihaylov, Deyan P. and Lasenby, Anthony and Sherrill, Gemma",
    title = "{Astrometric Search Method for Individually Resolvable Gravitational Wave Sources with Gaia}",
    eprint = "1707.06239",
    archivePrefix = "arXiv",
    primaryClass = "astro-ph.IM",
    doi = "10.1103/PhysRevLett.119.261102",
    journal = "Phys. Rev. Lett.",
    volume = "119",
    pages = "261102",
    year = "2017"
}

@article{Wang:2022sxn,
    author = "Wang, Yijun and Pardo, Kris and Chang, Tzu-Ching and Dor{\'e}, Olivier",
    title = "{Constraining the Stochastic Gravitational Wave Background with Photometric Surveys}",
    eprint = "2205.07962",
    archivePrefix = "arXiv",
    primaryClass = "astro-ph.CO",
    doi = "10.1103/PhysRevD.106.084006",
    journal = "Phys. Rev. D",
    volume = "106",
    pages = "084006",
    year = "2022"
}

@article{Blas:2026dsm,
    author = "Blas, D. and Foster, J. W. and Gouttenoire, Y. and Iovino, A. J. and Musco, I. and Trifinopoulos, S. and Vanvlasselaer, M.",
    title = "{The Dark Side of the Moon: Listening to Scalar-Induced Gravitational Waves}",
    eprint = "2602.12252",
    archivePrefix = "arXiv",
    primaryClass = "astro-ph.CO",
    year = "2026"
}

@article{Armstrong:2006,
    author = "Armstrong, J. W.",
    title = "{Low-Frequency Gravitational Wave Searches Using Spacecraft Doppler Tracking}",
    journal = "Living Rev. Rel.",
    volume = "9",
    pages = "1",
    year = "2006",
    doi = "10.12942/lrr-2006-1"
}

@article{LISA:2017pwj,
    author = "Amaro-Seoane, Pau and others",
    collaboration = "LISA",
    title = "{Laser Interferometer Space Antenna}",
    eprint = "1702.00786",
    archivePrefix = "arXiv",
    primaryClass = "astro-ph.IM",
    doi = "10.48550/arXiv.1702.00786",
    journal = "arXiv e-prints",
    month = "2",
    year = "2017"
}

@article{LISACosmologyWorkingGroup:2022jok,
    author = "Auclair, Pierre and others",
    collaboration = "LISA Cosmology Working Group",
    title = "{Cosmology with the Laser Interferometer Space Antenna}",
    eprint = "2204.05434",
    archivePrefix = "arXiv",
    primaryClass = "astro-ph.CO",
    reportNumber = "LISA CosWG-22-03, FERMILAB-PUB-22-349-SCD",
    doi = "10.1007/s41114-023-00045-2",
    journal = "Living Rev. Rel.",
    volume = "26",
    number = "1",
    pages = "5",
    year = "2023"
}

@article{Foster:2025nzf,
    author = "Foster, Joshua W. and Blas, Diego and Bourgoin, Adrien and Hees, Aurelien and Herrero-Valea, M{\'\i}riam and Jenkins, Alexander C. and Xue, Xiao",
    title = "{Discovering $\mu$Hz gravitational waves and ultra-light dark matter with binary resonances}",
    eprint = "2504.15334",
    archivePrefix = "arXiv",
    primaryClass = "astro-ph.CO",
    reportNumber = "FERMILAB-PUB-25-0091-T",
    month = "4",
    year = "2025"
}

@article{Speagle:2019ivv,
    author = "Speagle, Joshua S.",
    title = "{dynesty: a dynamic nested sampling package for estimating Bayesian posteriors and evidences}",
    eprint = "1904.02180",
    archivePrefix = "arXiv",
    primaryClass = "astro-ph.IM",
    doi = "10.1093/mnras/staa278",
    journal = "Mon. Not. Roy. Astron. Soc.",
    volume = "493",
    number = "3",
    pages = "3132--3158",
    year = "2020"
}

@article{Murphy:2013qya,
    author = "Murphy, T. W.",
    title = "{Lunar laser ranging: the millimeter challenge}",
    eprint = "1309.6294",
    archivePrefix = "arXiv",
    primaryClass = "gr-qc",
    doi = "10.1088/0034-4885/76/7/076901",
    journal = "Rept. Prog. Phys.",
    volume = "76",
    pages = "076901",
    year = "2013"
}

@article{Christensen:2022bxb,
    author = "Christensen, Nelson and Meyer, Renate",
    title = "{Parameter estimation with gravitational waves}",
    eprint = "2204.04449",
    archivePrefix = "arXiv",
    primaryClass = "gr-qc",
    doi = "10.1103/RevModPhys.94.025001",
    journal = "Rev. Mod. Phys.",
    volume = "94",
    number = "2",
    pages = "025001",
    year = "2022"
}

@inproceedings{Blas:2025lzc,
    author = "Blas, Diego and Bourgoin, Adrien and Foster, Joshua W. and Hees, Aurelien and Herrero-Valea, M{\'\i}riam and Jenkins, Alexander C. and Xue, Xiao",
    title = "{Binary systems as gravitational wave detectors}",
    booktitle = "{59th Rencontres de Moriond on Gravitation}: {Moriond 2025 Gravitation}",
    eprint = "2506.11802",
    archivePrefix = "arXiv",
    primaryClass = "gr-qc",
    reportNumber = "FERMILAB-CONF-25-0404-T",
    month = "6",
    year = "2025"
}

@article{Blas:2021mpc,
    author = "Blas, Diego and Jenkins, Alexander C.",
    title = "{Detecting stochastic gravitational waves with binary resonance}",
    eprint = "2107.04063",
    archivePrefix = "arXiv",
    primaryClass = "gr-qc",
    reportNumber = "KCL-PH-TH/2021-34",
    doi = "10.1103/PhysRevD.105.064021",
    journal = "Phys. Rev. D",
    volume = "105",
    number = "6",
    pages = "064021",
    year = "2022"
}

@article{ET:2025xjr,
    author = "Abac, Adrian and others",
    collaboration = "ET",
    title = "{The Science of the Einstein Telescope}",
    eprint = "2503.12263",
    archivePrefix = "arXiv",
    primaryClass = "gr-qc",
    reportNumber = "ET-0036C-25",
    doi = "10.1088/1475-7516/2026/03/081",
    journal = "JCAP",
    volume = "03",
    pages = "081",
    year = "2026"
}

@book{Maggiore:2007ulw,
    author = "Maggiore, Michele",
    title = "{Gravitational Waves. Vol. 1: Theory and Experiments}",
    doi = "10.1093/acprof:oso/9780198570745.001.0001",
    isbn = "978-0-19-171766-6, 978-0-19-852074-0",
    publisher = "Oxford University Press",
    year = "2007"
}

@article{Rover:2006ni,
    author = "Rover, Christian and Meyer, Renate and Christensen, Nelson",
    title = "{Bayesian inference on compact binary inspiral gravitational radiation signals in interferometric data}",
    eprint = "gr-qc/0602067",
    archivePrefix = "arXiv",
    doi = "10.1088/0264-9381/23/15/009",
    journal = "Class. Quant. Grav.",
    volume = "23",
    pages = "4895--4906",
    year = "2006"
}

@article{Rover:2006bb,
    author = "Rover, Christian and Meyer, Renate and Christensen, Nelson",
    title = "{Coherent Bayesian inference on compact binary inspirals using a network of interferometric gravitational wave detectors}",
    eprint = "gr-qc/0609131",
    archivePrefix = "arXiv",
    doi = "10.1103/PhysRevD.75.062004",
    journal = "Phys. Rev. D",
    volume = "75",
    pages = "062004",
    year = "2007"
}

@article{LIGOScientific:2018jsj,
    author = "Abbott, B. P. and others",
    collaboration = "LIGO Scientific, Virgo",
    title = "{Binary Black Hole Population Properties Inferred from the First and Second Observing Runs of Advanced LIGO and Advanced Virgo}",
    eprint = "1811.12940",
    archivePrefix = "arXiv",
    primaryClass = "astro-ph.HE",
    reportNumber = "LIGO-P1800324",
    doi = "10.3847/2041-8213/ab3800",
    journal = "Astrophys. J. Lett.",
    volume = "882",
    number = "2",
    pages = "L24",
    year = "2019"
}

@article{LIGOScientific:2020kqk,
    author = "Abbott, R. and others",
    collaboration = "LIGO Scientific, Virgo",
    title = "{Population Properties of Compact Objects from the Second LIGO-Virgo Gravitational-Wave Transient Catalog}",
    eprint = "2010.14533",
    archivePrefix = "arXiv",
    primaryClass = "astro-ph.HE",
    reportNumber = "LIGO-P2000077",
    doi = "10.3847/2041-8213/abe949",
    journal = "Astrophys. J. Lett.",
    volume = "913",
    number = "1",
    pages = "L7",
    year = "2021"
}

@article{Murphy:2007ed,
    author = "Murphy, Jr., T. W. and others",
    title = "{APOLLO: The Apache Point Observatory Lunar Laser-ranging Operation. Instrument Description and First Detections}",
    eprint = "0710.0890",
    archivePrefix = "arXiv",
    primaryClass = "astro-ph",
    doi = "10.1086/526428",
    journal = "Publ. Astron. Soc. Pac.",
    volume = "120",
    pages = "20--37",
    year = "2008"
}

@article{Turyshev:2026qus,
    author = "Turyshev, Slava G.",
    title = "{High-Power AM-CW Lunar Laser Ranging as a $μ$Hz SGWB Detector}",
    eprint = "2605.04110",
    archivePrefix = "arXiv",
    primaryClass = "gr-qc",
    month = "5",
    year = "2026"
}

@ARTICLE{2014IPNPR.196C...1F,
       author = {{Folkner}, W.~M. and {Williams}, J.~G. and {Boggs}, D.~H. and {Park}, R.~S. and {Kuchynka}, P.},
        title = "{The Planetary and Lunar Ephemerides DE430 and DE431}",
      journal = {Interplanetary Network Progress Report},
         year = 2014,
        month = feb,
       volume = {42-196},
        pages = {1-81},
       adsurl = {https://ui.adsabs.harvard.edu/abs/2014IPNPR.196C...1F}
}

@article{Franciollini,
  author = {Franciolini, G. and Gouttenoire, Y. and Iovino, I. and Trifinopoulos, S. and Vanvlasselaer, M.},
  title = {Using lunar ranging as a probe of the population of supermassive black hole binaries},
  year = {2026},
  note = {In preparation},
  eprint = {(to appear)}
}

@ARTICLE{2015JGRE..120..689W,
       author = {{Williams}, James G. and {Boggs}, Dale. H.},
        title = "{Tides on the Moon: Theory and determination of dissipation}",
      journal = {Journal of Geophysical Research (Planets)},
         year = 2015,
        month = apr,
       volume = {120},
       number = {4},
        pages = {689-724},
          doi = {10.1002/2014JE004755},
       adsurl = {https://ui.adsabs.harvard.edu/abs/2015JGRE..120..689W}
}

@ARTICLE{2025arXiv251202431T,
       author = {{Turyshev}, Slava G.},
        title = "{High-Precision Amplitude-Modulated Continuous-Wave Lunar Laser Ranging}",
      journal = {arXiv e-prints},
         year = 2025,
        month = dec,
          eid = {arXiv:2512.02431},
        pages = {arXiv:2512.02431},
          doi = {10.48550/arXiv.2512.02431},
archivePrefix = {arXiv},
       eprint = {2512.02431},
 primaryClass = {astro-ph.IM},
       adsurl = {https://ui.adsabs.harvard.edu/abs/2025arXiv251202431T}
}

@ARTICLE{2025PhRvP..23f4066T,
       author = {{Turyshev}, Slava G.},
        title = "{Lunar laser ranging with high-power continuous-wave lasers}",
      journal = {Physical Review Applied},
         year = 2025,
        month = jun,
       volume = {23},
       number = {6},
          eid = {064066},
        pages = {064066},
          doi = {10.1103/hqtm-y5pg},
archivePrefix = {arXiv},
       eprint = {2502.02796},
 primaryClass = {astro-ph.IM},
       adsurl = {https://ui.adsabs.harvard.edu/abs/2025PhRvP..23f4066T}
}

@article{Battat:2023upl,
    author = "Battat, James B. R. and Adelberger, Eric and Colmenares, Nicholas R. and Farrah, Megan and Gonzales, Daniel P. and Hoyle, C. D. and McMillan, Russett J. and Murphy, Thomas W. and Sabhlok, Sanchit and Stubbs, Christopher W.",
    title = "{Fifteen Years of Millimeter Accuracy Lunar Laser Ranging with APOLLO: Data Set Characterization}",
    eprint = "2304.11128",
    archivePrefix = "arXiv",
    primaryClass = "astro-ph.IM",
    doi = "10.1088/1538-3873/aceb2f",
    journal = "Publ. Astron. Soc. Pac.",
    volume = "135",
    number = "1052",
    pages = "104504",
    year = "2023"
}

@article{2004GeoRL..3114602M,
       author = {{Mendes}, V.~B. and {Pavlis}, E.~C.},
        title = "{High-accuracy zenith delay prediction at optical wavelengths}",
      journal = {grl},
         year = 2004,
        month = jul,
       volume = {31},
       number = {14},
        pages = {L14602},
          doi ={10.1029/2004GL020308}
}

@article{Benderarticle,
author = {Bender, P. and Currie, Douglas and Poultney, S. and Dicke, R. and Eckhardt, Donald and Kaula, W. and Mulholland, J. and Plotkin, H. and Silverberg, E. and Faller, J.},
year = {1973},
month = {11},
pages = {},
title = {The lunar laser ranging experiment},
volume = {182},
journal = {Science}
}

@article{Dickey:1994zz,
    author = "Dickey, J. O. and others",
    title = "{Lunar Laser Ranging: A Continuing Legacy of the Apollo Program}",
    doi = "10.1126/science.265.5171.482",
    journal = "Science",
    volume = "265",
    pages = "482--490",
    year = "1994"
}

@article{Bertotti:1980pg,
    author = "Bertotti, B. and Carr, Bernard J.",
    title = "{THE PROSPECTS OF DETECTING GRAVITATIONAL BACKGROUND RADIATION BY DOPPLER TRACKING INTERPLANETARY SPACECRAFT}",
    doi = "10.1086/157826",
    journal = "Astrophys. J.",
    volume = "236",
    pages = "1000--1011",
    year = "1980"
}

@ARTICLE{1979ApJ...233..685T,
       author = {{Turner}, M.~S.},
        title = "{Influence of a weak gravitational wave on a bound system of two point-masses.}",
      journal = {\apj},
         year = 1979,
        month = oct,
       volume = {233},
        pages = {685-693},
          doi = {10.1086/157429},
       adsurl = {https://ui.adsabs.harvard.edu/abs/1979ApJ...233..685T}
}

@ARTICLE{1981ApJ...246..569M,
       author = {{Mashhoon}, B. and {Carr}, B.~J. and {Hu}, B.~L.},
        title = "{The influence of cosmological gravitational waves on a Newtonian binary system}",
      journal = {\apj},
         year = 1981,
        month = jun,
       volume = {246},
        pages = {569-591},
          doi = {10.1086/158957},
       adsurl = {https://ui.adsabs.harvard.edu/abs/1981ApJ...246..569M}
}

@article{Ciufolini:2016ntr,
    author = "Ciufolini, Ignazio and others",
    title = "{A test of general relativity using the LARES and LAGEOS satellites and a GRACE Earth gravity model}",
    eprint = "1603.09674",
    archivePrefix = "arXiv",
    primaryClass = "gr-qc",
    doi = "10.1140/epjc/s10052-016-3961-8",
    journal = "Eur. Phys. J. C",
    volume = "76",
    number = "3",
    pages = "120",
    year = "2016"
}

@article{Ciufolini:2004rq,
    author = "Ciufolini, I. and Pavlis, E. C.",
    title = "{A confirmation of the general relativistic prediction of the Lense-Thirring effect}",
    doi = "10.1038/nature03007",
    journal = "Nature",
    volume = "431",
    pages = "958--960",
    year = "2004"
}

@article{Gair:2022fsj,
    author = "Gair, Jonathan R. and Antonelli, Andrea and Barbieri, Riccardo",
    title = "{A Fisher matrix for gravitational-wave population inference}",
    eprint = "2205.07893",
    archivePrefix = "arXiv",
    primaryClass = "gr-qc",
    doi = "10.1093/mnras/stac3560",
    journal = "Mon. Not. Roy. Astron. Soc.",
    volume = "519",
    number = "2",
    pages = "2736--2753",
    year = "2022"
}

@article{Fan:2024nnp,
    author = "Fan, Hui-Min and Lyu, Xiang-Yu and Zhang, Jian-dong and Hu, Yi-Ming and Yang, Rong-Jia and Feng, Tai-Fu",
    title = "{Probing the population of extreme-mass-ratio inspirals with TianQin}",
    eprint = "2410.12408",
    archivePrefix = "arXiv",
    primaryClass = "astro-ph.HE",
    doi = "10.1103/PhysRevD.111.103008",
    journal = "Phys. Rev. D",
    volume = "111",
    number = "10",
    pages = "103008",
    year = "2025"
}

@article{Toubiana:2026yml,
    author = "Toubiana, Alexandre and Gair, Jonathan R.",
    title = "{A framework for LISA population inference}",
    eprint = "2601.04168",
    archivePrefix = "arXiv",
    primaryClass = "gr-qc",
    doi = "10.1088/1475-7516/2026/07/025",
    journal = "JCAP",
    volume = "07",
    pages = "025",
    year = "2026"
}

@article{Gair:2010bx,
    author = "Gair, Jonathan R. and Sesana, Alberto and Berti, Emanuele and Volonteri, Marta",
    editor = "Trischuk, William",
    title = "{Constraining properties of the black hole population using LISA}",
    eprint = "1009.6172",
    archivePrefix = "arXiv",
    primaryClass = "gr-qc",
    doi = "10.1088/0264-9381/28/9/094018",
    journal = "Class. Quant. Grav.",
    volume = "28",
    pages = "094018",
    year = "2011"
}

@article{DeRenzis:2024dvx,
    author = "De Renzis, Viola and Iacovelli, Francesco and Gerosa, Davide and Mancarella, Michele and Pacilio, Costantino",
    title = "{Forecasting the population properties of merging black holes}",
    eprint = "2410.17325",
    archivePrefix = "arXiv",
    primaryClass = "astro-ph.HE",
    doi = "10.1103/PhysRevD.111.044048",
    journal = "Phys. Rev. D",
    volume = "111",
    number = "4",
    pages = "044048",
    year = "2025"
}

@article{Reitze:2019iox,
    author = "Reitze, David and others",
    title = "{Cosmic Explorer: The U.S. Contribution to Gravitational-Wave Astronomy beyond LIGO}",
    eprint = "1907.04833",
    archivePrefix = "arXiv",
    primaryClass = "astro-ph.IM",
    reportNumber = "LIGO-P1900316",
    journal = "Bull. Am. Astron. Soc.",
    volume = "51",
    number = "7",
    pages = "035",
    year = "2019"
}

@article{KAGRA:2021duu,
    author = "Abbott, R. and others",
    collaboration = "KAGRA, VIRGO, LIGO Scientific",
    title = "{Population of Merging Compact Binaries Inferred Using Gravitational Waves through GWTC-3}",
    eprint = "2111.03634",
    archivePrefix = "arXiv",
    primaryClass = "astro-ph.HE",
    reportNumber = "LIGO-P2100239 ; Data release: https://zenodo.org/record/5655785, LIGO-P2100239",
    doi = "10.1103/PhysRevX.13.011048",
    journal = "Phys. Rev. X",
    volume = "13",
    number = "1",
    pages = "011048",
    year = "2023"
}

\end{document}